\input amssym.def
\input amssym.tex

\parindent=10pt
\def\Z{\Bbb Z}\def\R{\Bbb R}\def\C{\Bbb C}\def\Q{\Bbb Q}
\def\la{{\lambda}}\def\al{{\alpha}}\def\be{{\beta}}\def\ga{{\gamma}}
\def\ome{{\omega}}\def\vp{{\varphi}}
\def\Tr{{\rm Tr}\,}\def\ad{{\rm ad}\,}\def\dim{{\rm dim}\,}\def\id{{\rm id}}
\def\ch{{\rm ch}}\def\mult{{\rm mult}\,}
\def\Ppkb{\overline{P_+^k}}
\def\zb{{\bar z}}
\def\pzp{{\{z\}}}
\def\di{{\partial}}
\def\h{\hat}
\def\>{{\rangle}}
\def\<{{\langle}}
\def\K{{\cal K}}
\def\T{{\cal T}}
\def\Tn{{\cal T}_\#}
\def\Nk{{{ }^{(k)}N}}\def\Vk{{{ }^{(k)}V}}
\def\Sk{{S^{(k)}}}\def\Tk{{T^{(k)}}}
\def\LR{{Littlewood-Richardson}}
\def\ie{{\rm i.e.}}\def\card{{\rm card}\,}
\def\Sp{{\rm Span}\,}

\def\b#1{\kern-0.25pt\vbox{\hrule height 0.2pt\hbox{\vrule
width 0.2pt \kern2pt\vbox{\kern2pt \hbox{#1}\kern2pt}\kern2pt\vrule
width 0.2pt}\hrule height 0.2pt}}
\def\ST#1{\matrix{\vbox{#1}}}
\def\STrow#1{\hbox{#1}\kern-1.35pt}
\def\bv{\b{\phantom{1}}}

\font\huge=cmr10 scaled \magstep2
\font\small=cmr8

\def\boxit#1{\leavevmode\kern5pt\hbox{
	\vrule width.2pt\vtop{\vbox{\hrule height.2pt\kern5pt
        \hbox{\kern5pt{#1}\kern5pt}}
      \kern5pt\hrule height.2pt}\vrule width.2pt}\kern5pt}
\def\boxEq#1{\boxit{$\displaystyle #1$}}

\def\ubrackfill#1{$\mathsurround=0pt
	\kern2.5pt\vrule depth#1\leaders\hrule\hfill\vrule depth#1\kern2.5pt$}
\def\contract#1{\mathop{\vbox{\ialign{##\crcr\noalign{\kern3pt}
	\ubrackfill{3pt}\crcr\noalign{\kern3pt\nointerlineskip}
	$\hfil\displaystyle{#1}\hfil$\crcr}}}\limits
}

\def\ubrack#1{$\mathsurround=0pt
	\vrule depth#1\leaders\hrule\hfill\vrule depth#1$}
\def\dbrack#1{$\mathsurround=0pt
	\vrule height#1\leaders\hrule\hfill\vrule height#1$}
\def\ucontract#1#2{\mathop{\vbox{\ialign{##\crcr\noalign{\kern 4pt}
	\ubrack{#2}\crcr\noalign{\kern 4pt\nointerlineskip}
	$\hskip #1\relax$\crcr}}}\limits
}
\def\dcontract#1#2{\mathop{\vbox{\ialign{##\crcr
	$\hskip #1\relax$\crcr\noalign{\kern0pt}
	\dbrack{#2}\crcr\noalign{\kern0pt\nointerlineskip}
	}}}\limits
}

\def\ucont#1#2#3{^{\kern-#3\ucontract{#1}{#2}\kern #3\kern-#1}}
\def\dcont#1#2#3{_{\kern-#3\dcontract{#1}{#2}\kern #3\kern-#1}}


\input harvmac
\input epsf

\baselineskip20pt

\centerline{\bf{\huge Affine Kac-Moody Algebras}}
 
\centerline{\bf{\huge and the}}
 
\centerline{\bf{\huge Wess-Zumino-Witten
Model}}

\vskip.75cm
\baselineskip15pt

\centerline{Mark Walton}
\centerline{\it Physics Department, University of Lethbridge}
\centerline{\it Lethbridge, Alberta, Canada\ \ T1K 3M4}
\centerline{\rm walton@uleth.ca}

\vskip.75cm

\noindent {\bf Abstract:}\ \ These lecture notes are a brief introduction
to Wess-Zumino-Witten models, and their current algebras, the affine
Kac-Moody algebras. After reviewing the general background, we focus
on the application of representation theory to
the computation of 3-point functions and fusion rules.

\newsec{Introduction}

In 1984, Belavin, Polyakov and Zamolodchikov \ref\bpz{A.A. Belavin, A.M.
Polyakov, A.B. Zamolodchikov, Nucl. Phys. {\bf B241} (1984) 333}\ showed how
an infinite-dimensional field theory problem could effectively be reduced to a 
finite problem, by the presence of an infinite-dimensional symmetry. The
symmetry algebra was the Virasoro algebra, or two-dimensional conformal
algebra, and the field theories studied were examples of two-dimensional
conformal field theories. The authors showed how to solve
the minimal models of conformal field theory, so-called because they
realise just the Virasoro algebra, and they do it in a minimal fashion. All
fields in these models could be grouped into a discrete, finite  set
of conformal families, each associated with a representation of the Virasoro
algebra. 

This strategy has since been extended to a large class of conformal field
theories with similar structure, the rational conformal field theories
(RCFT's) \ref\msrev{G. Moore, N. Seiberg, in the proceedings of the Trieste
Spring School Superstrings, 1989, M. Green et al, eds. (World
   Scientific, 1990);\hfill\break 
and in the proceedings of the NATO Advanced Summer Institute and Banff Summer
School, 1989, H.C. Lee, ed. (Plenum Press, 1990)}. The new feature is that the theories realise infinite-dimensional
algebras that contain the Virasoro algebra as a subalgebra. The larger
algebras are known as $W$-algebras \ref\bousch{P. Bouwknegt, K. Schoutens, 
Phys. Rep. {\bf 223} (1993) 183}\ in the physics literature. 

Thus the study of conformal field theory (in two dimensions) is intimately
tied to infinite-dimensional algebras. The rigorous framework for such
algebras is the subject of vertex (operator) algebras \ref\Kacva{V. Kac, 
{\it Vertex Algebras for Beginners}, 2nd ed. (Amer. Math. Soc.,
1998)} \ref\gann{T. Gannon, lecture notes, this volume}. A related, more
physical approach is called meromorphic conformal field theory \ref\gabgo{M.
Gaberdiel, P. Goddard, lecture notes, this volume}.

Special among these infinite-dimensional algebras are the affine Kac-Moody
algebras (or their enveloping algebras), realised in the Wess-Zumino-Witten
(WZW) models  \ref\wzwit{E. Witten, Comm. Math.
Phys. {\bf 92} (1984) 455}.
They are the simplest infinite-dimensional extensions of ordinary
semi-simple Lie algebras. Much is known about them, and so also about the
WZW models. The affine Kac-Moody algebras are the subject of these lecture
notes, as are their applications in conformal field theory. For brevity we
restrict  consideration to the WZW models; the goal will be to indicate how the
affine Kac-Moody algebras allow the solution of WZW models, in the same way
that the Virasoro algebra allows the solution of minimal models, and
$W$-algebras the solution of other RCFT's. We will also give a couple of 
examples of remarkable mathematical properties that find an
``explanation'' in the WZW context.  

One might think that focusing on the special examples of affine Kac-Moody
algebras is too restrictive a strategy. There are good counter-arguments
to this criticism. Affine Kac-Moody algebras can
tell us about many other RCFT's: the coset construction \ref\gko{P.
Goddard, A. Kent,  D. Olive, Phys. Lett. {\bf 152B} (1985) 88}\ builds a
large class of new theories as differences of WZW models, roughly  speaking. 
Hamiltonian reduction  \ref\balogW{ J. Balog, L. Feher, L.
O'Raifeartaigh, P. Forgacs , A. Wipf, Annals Phys. {\bf 203} (1990) 76} 
constructs $W$-algebras from the affine Kac-Moody algebras. In addition, many 
more conformal field theories can be
constructed from WZW and coset models by the orbifold procedure \ref\dhvw{L.
Dixon, J. Harvey, C. Vafa, E. Witten, Nucl. Phys. {\bf B261} (1985) 620; {\bf
274} (1986) 285} \ref\dvvv{R. Dijkgraaf, C. Vafa, E. Verlinde, H. Verlinde,
Commun. Math. Phys. {\bf 123} (1989) 485}. Incidentally, all three 
constructions can be understood in the context of
gauged WZW models.  

Along the same lines, the question ``Why study two-dimensional conformal
field theory?'' arises. First, these field theories are solvable
non-perturbatively, and so are toy models that hopefully prepare us to
treat the non-perturbative regimes of physical field theories. Being conformal,
they also describe statistical systems at criticality  \ref\car{J. Cardy, in Les
Houches session XILX, Fields, Strings, and Critical Phenomena, eds. E.
Br\'ezin, J. Zinn-Justin (Elsevier, 1989)}. Conformal field theories have found
application in condensed matter physics \ref\aff{I. Affleck, Acta Phys. Polon.
{\bf B26} (1995) 1869}.  Furthermore, they  are vital components of string
theory \ref\gsw{M. Green, J. Schwarz, E. Witten, {\it Superstring Theory,
Vols.  1 \& 2} (Cambridge U. Press, 1987)}, a candidate theory of quantum
gravity, that also provides a consistent framework for unification of all the
forces.   

The basic subject of these lecture notes is close to that of \ref\gaw{K.
Gawedzki, lecture notes, this volume}. It is hoped, however, that this
contribution will complement that of Gawedzki, since our emphases are quite
different.

The layout is as follows. Section 2 is a brief introduction to the WZW model,
including its current algebra. Affine Kac-Moody algebras are reviewed in
Section 3, where some background on simple Lie algebras is also provided. Both
Sections 2 and 3 lay the foundation for Section 4: it discusses applications,
especially 3-point functions and fusion rules. We indicate how {\it a
priori} surprising mathematical properties of the algebras find a natural
framework in WZW models, and their duality as rational
conformal field theories.

\newsec{Wess-Zumino-Witten Models}

\subsec{Action}

Let $G$ denote a compact connected Lie group, and $g$ its simple Lie
algebra\foot{For an excellent, elementary introduction to Lie algebras, with
physical motivation, see \ref\georgi{H. Georgi, {Lie Algebras in Particle
Physics} (Benjamin/Cummings, 1982)}.}. Suppose $\ga$ is a $G$-valued field on
the complex plane. The Wess-Zumino-Witten (WZW) action is written as \wzwit
\ref\novi{S.P. Novikov, Usp. Mat. Nauk. {\bf 37} (1982) 3}
\eqn\action{\boxEq{S_k(\ga)\ =\ -{k\over{8\pi}}\,\int{\cal
K}(\ga^{-1}\di^\mu\ga,\ga^{-1}\di_\mu\ga)\,d^2x\ +\ 2\pi k\,\tilde S(\ga)\
,}} where $\di_\mu=\di/\di x^\mu$, the summation convention is used with
Euclidean metric, and ${\cal K}$ denotes the Killing form of $g$, which is
nondegenerate for $g$ simple,  \eqn\killing{{\cal K}(x,y)\ =\
{{\Tr(\ad_x\,\ad_y)}\over{2h^\vee}}\ \ ,\ x,y\in g\ .}  Here $h^\vee$ is an
integer fixed by the algebra $g$, called the dual Coxeter number of $g$, and
$\ad_x(z):=[x,z]$. The second term is the Wess-Zumino action. To describe it,
imagine that the complex plane (plus the point at $\infty$) is a large 2-sphere
$S^2$. $\ga$ then maps $S^2$ into the group manifold of $G$.

\midinsert
\vskip-0cm
\epsfxsize=8cm
\centerline{\epsfbox{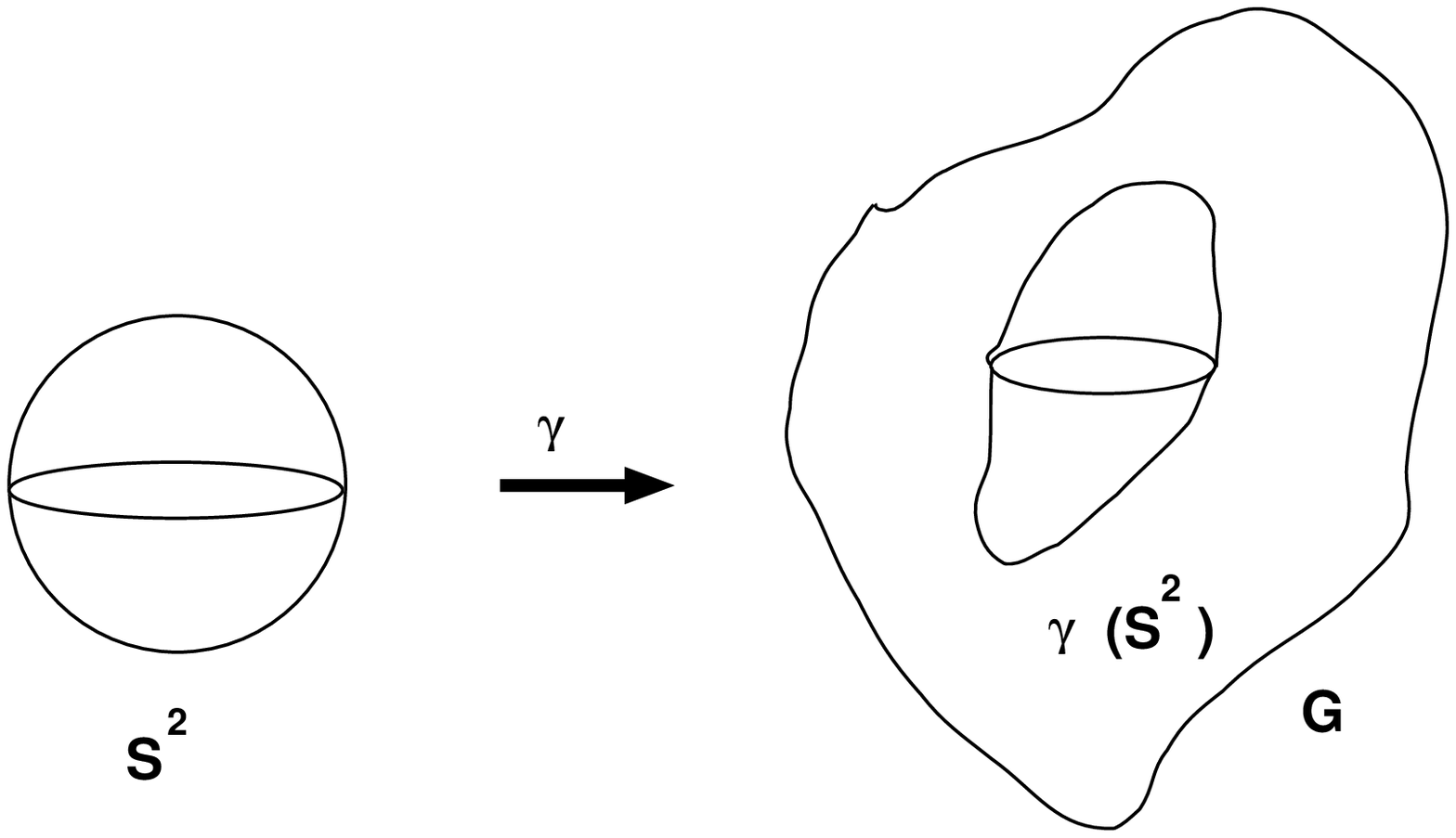}}
\smallskip\smallskip
\leftskip=1cm
\rightskip=1cm
\noindent
\baselineskip=12pt
{\it Figure 1}. The map $\ga:\ S^2\rightarrow
G$. $B$ is the 3-dimensional solid ball with $S^2$
as its boundary, $\di B=S^2$.\bigskip
\leftskip=0cm\rightskip=0cm
\baselineskip=15pt  
\endinsert 

The homotopy groups $\pi_n(G)$ thus enter consideration
(see \ref\nak{M. Nakahara, {\it Geometry, Topology and Physics} (IOP, 1990)},
for example). The elements of $\pi_n(G)$ are the equivalence classes of
continuous maps of the $n$-sphere $S^n$ into (the group manifold of) $G$. Two
such maps are equivalent if their images can be continuously deformed into
each other. If all images of $S^n$ in $G$ are contractible to a point, then the
$n$-th homotopy group of $G$ is trivial, $\pi_n(G)=0$. A non-trivial $\pi_n(G)$
indicates the presence of non-contractible $n$-cycles in $G$. 
(A cycle is a $n$-dimensional submanifold without boundary; a
non-contractible one is also not a boundary itself.) So, homotopy is quite a
fine measure of the topology of a group manifold $G$. For example, 
$\pi_n(G)=\Z$ implies there is a non-contractible  $n$-cycle ${\cal C}_n$ in $G$
that generates $\pi_n(G)$, and a map $S^n\rightarrow G$ can ``wind around'' this
cycle any number ($\in\Z$) of times. 

By a generalisation of Stokes' theorem, the existence of non-contractible
cycles in $G$ has to do with the existence of harmonic $n$-forms on $G$. A
harmonic form $h_n$ is a differential form that is closed ($dh_n = 0$), but not
exact (no form $p$ exists such that $h_n = dp$). Recall that if ${\cal C}_n$ is
a non-contractible cycle, its boundary vanishes ($\di {\cal C}_n = 0$),  and
${\cal C}_n$ is not a boundary itself. In the case $\pi_n(G)=\Z$ just
mentioned, there exists a harmonic $n$-form $h_n$ on $G$, that can be
identified with the volume form on ${\cal C}_n$, and can be normalised so that
the volume of ${\cal C}_n$ computed with it is 1:\ $\int_{{\cal C}_n}\,h_n\,
=\, 1$.  

Getting back to the Wess-Zumino term, since $\pi_2(G)=0$ for any compact
connected Lie group, $\ga$ can be extended to a map ($\tilde\ga$ when we want
to emphasise this) of $B$ into $G$, where $\di B=S^2$. The Wess-Zumino action
can be written as \eqn\wz{\boxEq{\tilde S(\ga)\ =\ {{-1}\over{48\pi^2}}\ \int_B
\, \epsilon^{ijk}\,{\cal K}\left(\tilde\ga^{-1}{{\di\tilde\ga}\over{\di y^i}},
\left[  \tilde\ga^{-1}{{\di\tilde\ga}\over{\di y^j}}, 
\tilde\ga^{-1}{{\di\tilde\ga}\over{\di y^k}} \right]\right)\,d^3y\ }}
where $y^i$ ($i=1,2,3$) denote the coordinates of $B$. 

Let $t^a$ denote the
elements of a basis of $g$, \ie\ $g=\Sp \{\,  t^a\,:\,a=1,\ldots,\dim g\,
\}$. For Hermitian $t^a$, $(t^a)^\dagger = t^a$, the commutation
relations of $g$ can be written as 
\eqn\gcomm{[t^a, t^b]\ =\ \sum_{c}\, if^{abc}\, t^c\ \ ,}
where the structure constants $f^{abc}$ are real. Normalising so that
$\K(t^a,t^b) = \delta^{ab}$,  we get  \eqn\Kfabc{{\cal
K}\left(t^a,[t^b,t^c]\right)\ =\ i\, f^{abc}\ .} 
Since $\tilde\ga^{-1} {{\di\tilde\ga}\over {\di y^i}}$ is an element of $g$,
we see by \wz\  that the structure constants $f^{abc}$ enter the Wess-Zumino
action. 

Now the
totally antisymmetric structure constants $f^{abc}$ of $g$ define a harmonic
3-form $h_3$ on the group manifold of $G$. $\tilde S$ is an  integral over
the pull-back of this harmonic 3-form $h_3$ to the space $B$: \eqn\wzpb{\tilde
S(\ga)_B\ :=\ \int_B\,\tilde\ga^*h_3\ =\  \int_{\di^{-1}S^2}\,\tilde\ga^*h_3\ .}
By the discussion of the previous paragraph, this points to a relation between
the Wess-Zumino term and the homotopy of $G$. This will be made explicit soon. 

If the WZW action is to describe a local theory on $S^2$, then the formal
expression $B=\di^{-1}S^2$ should indicate that the physics is independent of
which 3-dimensional extension $B$ of $S^2$ is used. Picture $S^2$ as a
circle, in order to draw a simple diagram. $\ga:\ S^2\,\rightarrow\,G$ can be
depicted as in Fig. 2.

\midinsert
\vskip-0cm
\epsfxsize=6cm
\centerline{\hskip-2cm\epsfbox{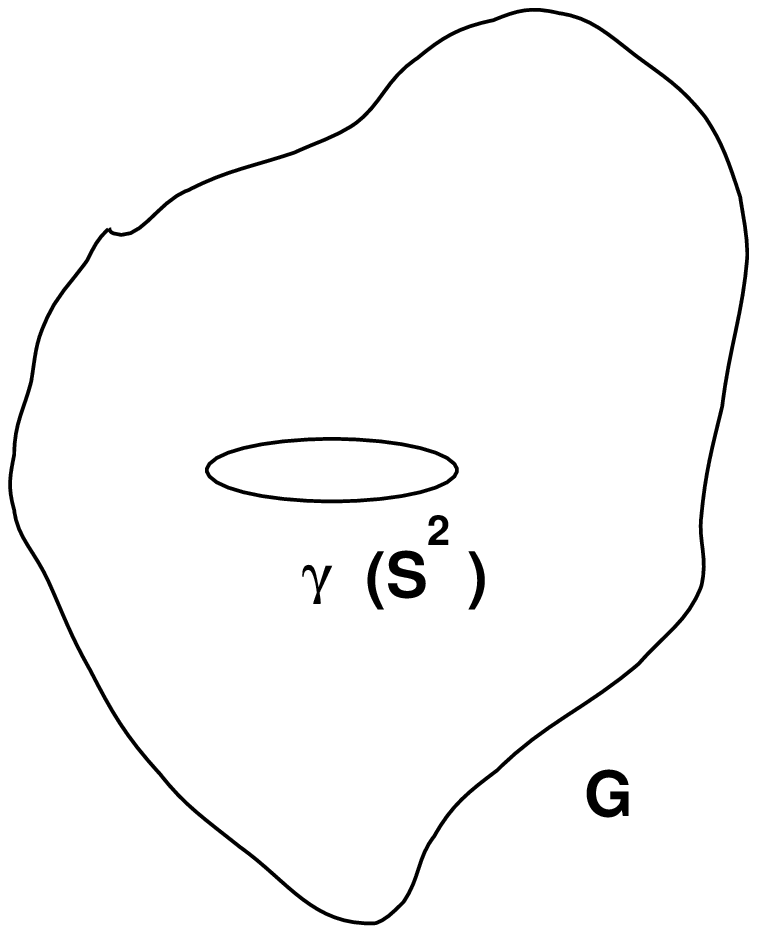}}
\smallskip\smallskip
\leftskip=1cm
\rightskip=1cm
\noindent
\baselineskip=12pt
{\it Figure 2}. The map $\gamma\ :\ S^2\rightarrow G$ depicted in one
lower dimension ($S^1$ replaces $S^2$).\bigskip \leftskip=0cm\rightskip=0cm
\baselineskip=15pt  
\endinsert

In Fig. 3, the images by $\gamma$ of two different extensions $B,B'$
of $S^2$ are pictured. In order that the physics described by $B$ is equivalent to that described
by $B'$, we require \eqn\eBBp{\exp\left[2\pi ik\tilde
S(\gamma)_B\right]\ =\  \exp\left[2\pi ik\tilde S(\gamma)_{B'}\right]\ }
or 
\eqn\BmBp{\exp\left[2\pi ik\tilde S(\gamma)_{B'-B}\right]\ =\ 1\ .}
$\gamma(B'-B)$ is homotopically equivalent to $S^3$ (depicted as $S^2$ in Fig.
4). 

\midinsert
\vskip-0cm
\epsfxsize=8cm
\centerline{\hskip-2cm\epsfbox{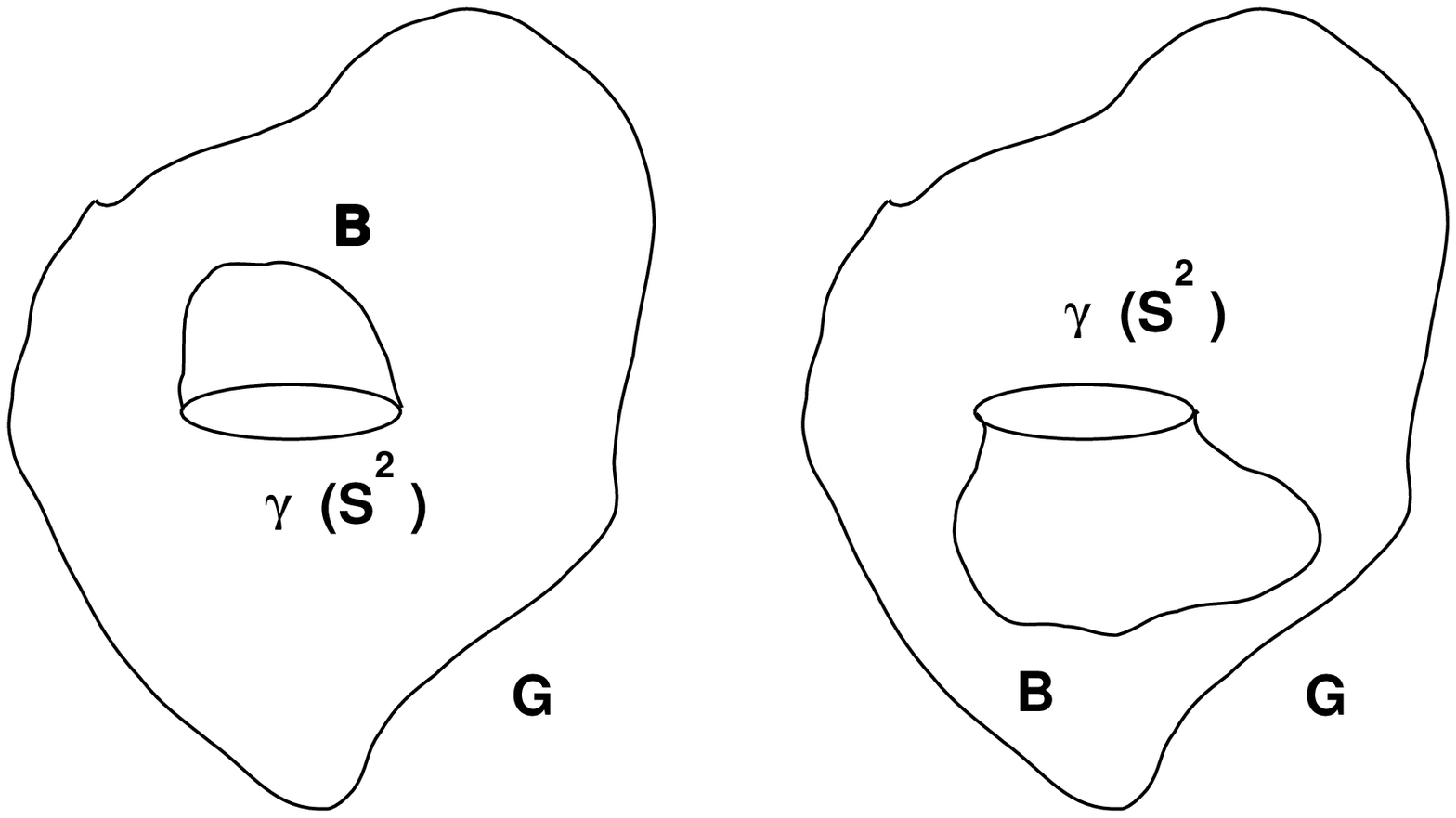}}
\smallskip\smallskip
\leftskip=1cm
\rightskip=1cm
\noindent
\baselineskip=12pt
{\it Figure 3}. The $\ga$-images of two different extensions $B,B'$ of
$S^2$.\bigskip \leftskip=0cm\rightskip=0cm
\baselineskip=15pt  
\endinsert 

\midinsert
\vskip-0cm
\epsfxsize=6cm
\centerline{\hskip-2cm\epsfbox{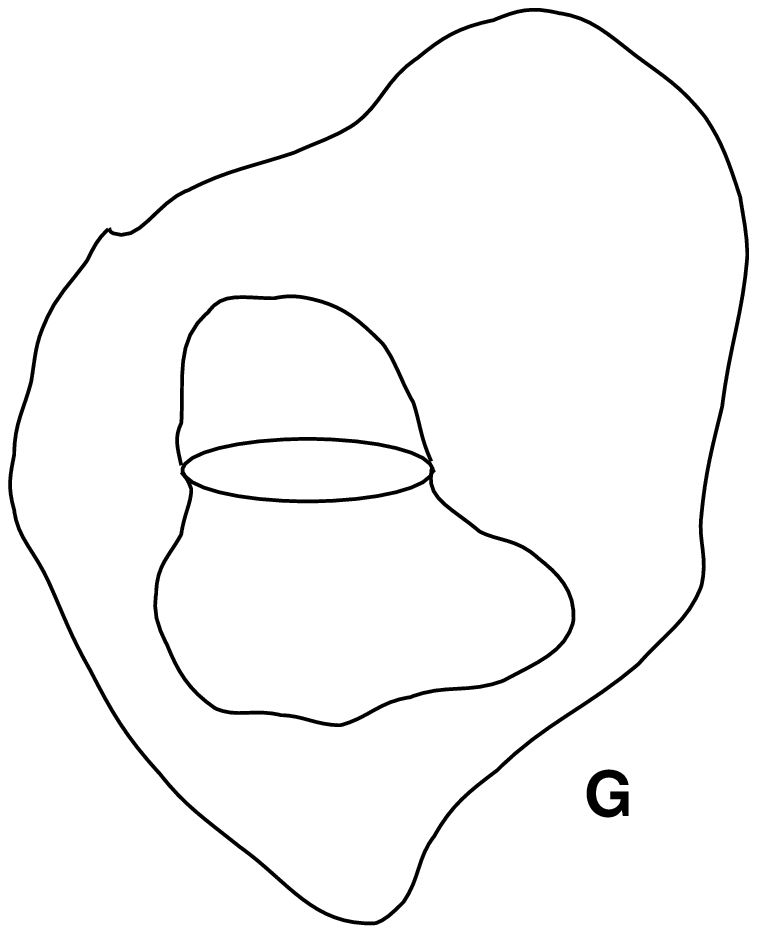}}
\smallskip\smallskip
\leftskip=1cm
\rightskip=1cm
\noindent
\baselineskip=12pt
{\it Figure 4}. $\gamma(B-B')$ (see Fig. 3). \bigskip
\leftskip=0cm\rightskip=0cm
\baselineskip=15pt  
\endinsert

Now to the homotopic significance of the WZ term: $\tilde S(\gamma)_{S^3}=N$
is the winding number of the map $\tilde \gamma\,:\,S^3\rightarrow G$. Since
$\pi_3(G)=\Z$ (for $G$ any compact connected simple Lie group), we have $N\in
\Z$. Therefore \BmBp\ requires $k\in\Z$, and since $k$ and $-k$ yield
indistinguishable physics, we use $k\in\Z_{\geq 0}$, which will be the so-called
{\it level} of the affine Kac-Moody algebra realised by the WZW model. 

The quantisation of the WZ term can also be understood by its relation to
anomalies, which have topological significance (see \nak, Chapter 13, for
example). Consider first a fermionic model with Lagrangian density
\eqn\Lferm{{\cal L} \ =\ {1\over 2}\,{\overline \Psi}\Gamma^\mu(\di_\mu+{\Bbb
A}_\mu) \Psi\ =\ \psi^\dagger(\partial_\zb+\bar A)\psi + 
\bar\psi^\dagger(\partial_z+ A)\bar\psi\ .} 
Here $z=x_1+ix_2$, $\bar z=x_1-ix_2$, and $\Gamma^\mu$ are the Dirac (gamma)
matrices, with anti-commutation relations $\{\Gamma^\mu,\Gamma^\nu\} =
2\delta^{\mu,\nu}$.  In the first expression, $\Psi$ is a Dirac spinor, 
${\overline \Psi} = \Psi^\dagger\Gamma^1$, and
${\Bbb A}_\mu$ is the gauge potential. In the second, the chiral components
$\psi=(1+\Gamma)\Psi/2$,  $\bar\psi=(1-\Gamma)\Psi/2$ appear, where
$\Gamma:=i\Gamma^1\Gamma^2$ is the chirality operator, and $A={\Bbb A}_1+
i{\Bbb A}_2$, $\bar A={\Bbb A}_1 - 
i{\Bbb A}_2$. The Lagrangian is invariant under the gauge
transformations \eqn\gauge{\eqalign{A\,\rightarrow\, &UAU^{-1}+U\di_zU^{-1}\ ,\
\ \psi\,\rightarrow\, U\psi\ \ ;\cr
\bar A\,\rightarrow\, &\bar U\bar A\bar U^{-1}+\bar U\di_\zb\bar U^{-1}\ ,\ \
\bar\psi\,\rightarrow\, \bar U\bar\psi\ \ ,\cr}}
where $U$ and $\bar U$ are two independent elements of the gauge group
$G$. This chiral $G\otimes G$ gauge invariance is the result of the 
vector and axial-vector gauge invariance of the Dirac Lagrangian: $U=\bar U$
specifies a vector gauge transformation, and when $U=\bar U^{-1}$, we get an  
axial-vector transformation. 

In a spacetime of $N$ dimensions, a gauge boson has $N-2$ degrees of
freedom. In $N=2$, what this means is that all ${\Bbb A}_\mu$ can be obtained
by applying gauge transformations to some fixed ${\Bbb A}_\mu$, 0, say. So we
can parametrise \eqn\AaAbb{A=\alpha^{-1}\partial_z\alpha\ \ ,\ \ \ \bar
A=\be^{-1}\di_\zb\be\ \ ,} with $\alpha,\beta\in G$. Then, the gauge
transformations \gauge\ become
\eqn\gUbUb{\al\ \rightarrow\ \al U\ \ ,\ \ \ \ \be\ \rightarrow\ \be\bar U\ \
.} 
Equivalently, we can say that the fields $A,\bar A$ are subsidiary fields that
could have been eliminated in \Lferm, giving a four-fermion interaction (and
so a Thirring model). 

The path integral \eqn\intf{\int\,[d\psi][d\bar\psi]\,\exp\left(\, -\int {\cal
L}\, d^2x\,  \right)\ =\ e^{-S_{\rm eff}}\  }
defines an effective action $S_{\rm eff}$ from which the fermions have been
eliminated. Because of the form of the Lagrangian \Lferm\ in \intf, one often
writes 
\eqn\Seffd{S_{\rm eff}\ =\ \log\det\,\left[\Gamma^\mu(\partial_\mu +
{\Bbb A}_\mu)\right]\ .} 
In simple cases, these path integrals can be computed
explicitly.  

Suppose there are extra flavour indices for the fermions, suppressed in
\Lferm, running over a number $N_L$ of
(flavours of) left-handed fermions $\psi$, and a number $N_R$ of right-handed
fermions $\bar\psi$. If $N_L\not= N_R$, the axial-vector gauge invariance is
destroyed when quantum corrections are
taken into account. There is a chiral anomaly, proportional to $N_L-N_R$. In
the path-integral formalism, this happens because the integration measure for
the chiral fermions cannot be regularised in a way that preserves the 
invariance \ref\fujik{K. Fujikawa, Phys. Rev. Lett. {\bf 42} (1979) 1195}.   

Following Polyakov and Wiegmann \ref\polwie{A.M. Polyakov, P.B.
Wiegmann, Phys. Lett. {\bf 141B} (1984) 223; Phys. Lett. {\bf 131B} (1983) 121}
(see also \ref\ddp{P. Di Vecchia, B. Durhuus, J.L. Petersen, Phys. Lett. {\bf
144B} (1984) 245}), consider a fermionic model  with $N_L$ flavours of
left-handed fermions $\psi$ 
and similarly $N_R$ right-handed fermions $\bar\psi$: \eqn\thirr{{\cal L}\ =\
\psi^\dagger(\partial_\zb+ \bar A)\psi +  \bar\psi^\dagger(\partial_z+
A)\bar\psi +  v{\cal K}(A,\bar A)\ ,} where  $v$ is a constant. 

First consider the case $N_L=N_R=1$. In the
gauge $\bar A=0$, 
integrating out the fermions gives \polwie\ 
\eqn\logd{\log\det\,\left[\Gamma^\mu(\partial_\mu+{\Bbb A}_\mu)\right]\ =\ 
S_1(\alpha)\ .} This contribution comes from the left-handed fermion 
$\psi$.  In general gauge, one would expect terms 
$S_1(\al)+S_1(\be^{-1})$, the second  term coming from the left-handed
fermion $\bar\psi$. This is not the complete answer, however, since by vector
gauge invariance, we expect a result that depends only on $\al\be^{-1}$. This
is where the $\K(A,\bar A)$ term comes in. One finds  
\eqn\logdi{\log\det\,\left[\Gamma^\mu(\partial_\mu+{\Bbb A}_\mu)\right]\ =\ 
S_1(\al\be^{-1})\ ,} and the constant $v$ of \thirr\ is adjusted so
that  \eqn\pwid{S_1(\alpha\be^{-1})\ =\ S_1(\alpha) +
S_1(\be^{-1}) - {1\over{4\pi}}\,\int d^2x\,{\cal
K}\left((\alpha^{-1}\partial_z\alpha),(\be^{-1}\di_\zb\be)\right)\ .} This
is the {\it
Polyakov-Wiegmann identity}. As we'll see, the affine current
algebra of the WZW model can be derived from it. 

Now suppose $N_L\not=N_R$, so that the theory has a chiral anomaly.
Eliminating fermions gives  \eqn\seff{
\eqalign{S_{\rm eff}&(A,\bar A)\ =\ N_LS_1(\al) + N_RS_1(\be^{-1}) - 
{1\over{8\pi f}}\,\int
d^2x\,{\cal K}(\gamma^{-1}\di^\mu\gamma, \gamma^{-1}\di_\mu\gamma)\cr 
           =&\ {1\over 2}(N_L+N_R)\big[\, S_1(\al)+S_1(\be^{-1})\,\big] + 
               {1\over 2}(N_L-N_R)\big[\, S_1(\al)-S_1(\be^{-1}) \,\big] \cr
&\ \ \ \ \ \ \ \ \  
-{1\over{8\pi f}}\,\int
d^2x\,{\cal K}(\gamma^{-1}\di^\mu\gamma, \gamma^{-1}\di_\mu\gamma)\ ,\cr}}
where $f$ is a constant, related to $v$. It is not fixed by gauge invariance 
here. Now take the limit $N_L+N_R\rightarrow\infty$, with
$N_L-N_R$ fixed. The term ${1\over 2}(N_L+N_R)\big[\,
S_1(\al)+S_1(\be^{-1})\,\big]$ is forced to vanish. This implies pure gauge
$A,\bar A$: $A=\ga^{-1}\di_z\ga$, $\bar A=\gamma^{-1}\di_\zb\gamma$ (compare
to  \AaAbb).  So we have  \eqn\swzwf{S_{\rm eff}\ =\ -{1\over{8\pi f}}\,\int
d^2x\,{\cal K}(\gamma^{-1}\di_\mu\gamma, \gamma^{-1}\di_\mu\gamma) +  2\pi
k\tilde S(\gamma)\ ,} with $k=N_L-N_R$. From this point of view then, $k$ is
quantised because it is the difference in the number of left-handed and
right-handed fermions. 

This last action is not quite that of the WZW model, with $f$ being an arbitrary
constant. It's that of a (two-dimensional) principal chiral $\sigma$-model,
with WZ term. Such a sigma model is asymptotically free, as is the
$\sigma$-model without the WZ term. Without the WZ term, the sigma model is 
strongly interacting in the infrared. But with the WZ term
present in the action, there is an infrared fixed point, at ${1\over f}=k$. The
WZW model describes the dynamics of this fixed point. We'll remain at ${1\over
f}=k$ henceforth. 

\subsec{Current algebra}

Let's rewrite the Polyakov-Wiegmann identity \pwid\ as 
\eqn\pw{\boxEq{S_k(\gamma\vp)\ =\ S_k(\gamma)\ +\ S_k(\varphi)\ +\
C_k(\gamma,\vp)\ ,}}
putting $\vp=\be^{-1}$, and using $S_k(\ga)=kS_1(\ga)$, and 
\eqn\CkC{C_k(\ga,\vp)\ =\ k\,C_1(\ga,\vp)\ =\ -
{k\over{4\pi}}\,\int
d^2x\,{\cal K}\left((\gamma^{-1}\partial_z\gamma),(\vp\di_\zb\vp^{-1})\right)\
.} The term $C_k(\gamma,\vp)$ is a cocycle:
\eqn\Ccoc{C_k(\gamma\vp,\sigma)\ +\ C_k(\gamma,\vp)\ =\ 
C_k(\gamma,\vp\sigma)\ +\ C_k(\vp,\sigma)\ .}
The presence of this cocycle indicates a projective
representation, of the loop group $LG$ of $G$ \ref\preseg{A.
Pressley, G. Segal, {\it Loop Groups} (Oxford U. Press, 1986)}. 
Alternatively, we can say that the group 
$\widehat{LG}$, an extension of $LG$, is represented non-projectively. This
extension $\widehat{LG}$ has as its Lie algebra the (untwisted) affine
Kac-Moody algebra $\hat g$, the central extension of the loop algebra of $g$.
We'll call $\hat g$ an affine algebra, for short. Let's see how the WZW model
realises $\hat g\oplus \hat g$ as a current algebra \wzwit \ref\knizam{V.G.
Knizhnik, A.B. Zamolodchikov, Nucl. Phys. {\bf B247} (1984) 83}. Then conformal
invariance can be established. 

Since
\eqn\Ck{C_k(\Omega,\bar\Omega^{-1})\ =\
-{k\over{4\pi}}\,\int d^2x\,{\cal K}\big(\Omega^{-1}\partial_\zb\Omega,
\bar\Omega^{-1}\partial_z\bar\Omega\big)\ ,} 
if either $\di_\zb\Omega=0$ or $\di_z\bar\Omega=0$, then 
$C_k(\Omega,\bar\Omega^{-1})=0$, and also $S_k(\Omega)= S_k(\bar\Omega) = 0$.
\pw\ thus establishes the local $G\otimes G$ invariance of the WZW model:
\eqn\GGwzw{S_k \left(\Omega(z)\gamma(z,\zb)\bar\Omega^{-1}(\zb) \right) \ =\ 
S_k\left(\gamma(z,\zb)\right)\ ,}
sometimes called the ``$G(z)\otimes G(\zb)$ invariance''. 

For infinitesimal transformations $\Omega=\id + \omega(z)$,
$\bar\Omega(\zb)=\id +\bar\omega(\zb)$, the WZW field $\gamma$ transforms as 
\eqn\gaOm{\delta_\omega\,\gamma\ =\ \omega\gamma,\ \ \
\delta_{\bar\omega}\,\gamma\ =\  -\gamma\bar\omega\ .}
With $\delta\gamma = \delta_\omega\ga + \delta_{\bar\omega}\ga$, we find
\eqn\deSk{\delta S_k(\gamma)\ =\ -{k\over{\pi}}\, \int d^2x\,
\left\{\, {\cal K}\left(\omega,\di_\zb(\di_z\gamma\gamma^{-1})\right)\ -\ 
{\cal K}\left(\bar\omega,\di_z(\gamma^{-1}\di_\zb\gamma)\right) \right\}\ .}

The equations of motion of the WZW model are
\eqn\wzwem{\di^\mu(\gamma^{-1}\di_\mu\gamma)\ +\
i\epsilon_{\mu\nu}\di^\mu(\gamma^{-1}\di^\nu\gamma)\ =\ 0\ .}
Switching to the complex coordinates $z,\zb$, and using $\di^z=2\di_z$,
$\epsilon_{z\zb}=i/2$, etc., these give
$\di_z(\gamma^{-1}\di_\zb\gamma)=0$, with hermitian conjugate
$-\di_\zb(\di_z\gamma\gamma^{-1})=0$. Defining  \eqn\JJb{\boxEq{J\ :=\
-k\di_z\gamma\gamma^{-1}\ ,\ \ \bar J\ :=\ k\gamma^{-1}\di_\zb\gamma\ \ ,}}
we have
\eqn\holJ{\di_\zb J\ =\ 0\ ,\ \ \ \di_z\bar J\ =\ 0\ .}
So the currents $J,\bar J$ are purely holomorphic, antiholomorphic,
respectively; \ie\ $J=J(z)$, $\bar J=\bar J(\zb)$. These currents will realise
two copies of the affine algebra $\hat g$. 

First we must explain the quantisation scheme. We consider the Euclidean
time direction to be the radial direction, so that constant time surfaces are
circles centred on the origin. More explicitly, the (conformal) transformation 
\eqn\cyl{z=e^{\tau+i\sigma}\, ,\ \bar z=e^{\tau-i\sigma}\ }
maps the complex plane (punctured at $z=0,\infty$) to a cylinder, with Euclidean
time coordinate $\tau\in\R$ running along its length, and a periodic space
coordinate $\sigma\equiv \sigma+2\pi$. The origin $z=0$ then  corresponds to
the distant past $\tau=-\infty$, and the distant future $\tau=+\infty$ is at
$|z|=\infty$. 

This is called {\it radial quantisation}. In (3+1)-dimensional QFT the 
$n$-point
functions are vacuum-expectation-values of time-ordered products of
fields. Similarly, in radial quantisation one needs to consider radially-ordered
products of fields:
\eqn\Rord{R\left(\,A(z)B(w)\,\right)\ :=\ \left\{\matrix{ A(z)B(w)\, ,\ &\
|z|>|w|\ ,\cr
B(w)A(z)\, ,\ & \ |z|<|w|\ .\cr}\right.\ }
Define the correlation functions as vacuum-expectation-values of radially 
ordered products of fields, \ie\
\eqn\corrAB{\langle A(z)B(w) \rangle\ :=\ \langle 0|\,R\left( A(z)B(w)\right)
\,|0\rangle\ .}

We make the
operator product expansion 
\eqn\ope{R\left(A(z)B(w)\right)\ =\ \sum_{n=-n_0}^\infty\,(z-w)^nD_{(n)}(w)\,,\
\ (n_0\ge0)\ .} 
We are also assuming $n_0<\infty$, \ie\ that there is no essential
singularity at $z=w$. Break this product up by defining the contraction 
\eqn\contr{\contract{A(z) B}(w)\ :=\
\sum_{n=-n_0}^{-1}\,(z-w)^nD_{(n)}(w)}
or singular part, and the normal-ordered product
\eqn\nprod{N\left(A(z)B(w)\right)\ :=\ \sum_{n\geq 0}\,(z-w)^n\,D_{(n)}(w)\ .}
If we also define 
\eqn\npzero{N\left(AB\right)(w)\ :=\ D_{(0)}(w)\
,}
then we can write  
\eqn\RNO{\eqalign{R\left(A(z)B(w)\right)\ =&\ \contract{A(z)B}(w)\ +\
N\left(A(z)B(w)\right)\ \cr
=&\ \contract{A(z)B}(w)\ +\ N\left(AB\right)(w)\ +O(z-w)\ .}}

Radial ordering will be assumed henceforth. Often it is only the singular
parts of operator product expansions (OPE's) that are relevant. We write 
\eqn\sope{A(z)B(w)\ \sim\ \sum_{n=-n_0}^{-1}(z-w)^nD_{(n)}(w)\ =\
\contract{A(z)B}(w)\ ,} \ie\ $\sim$ indicates that only the singular
terms are written. 

To show that the currents realise two copies of $\hat g$, we integrate the 
right hand side  of \deSk\ by parts, using counter-clockwise integration
contours, to  get
\eqn\deSz{{i\over{4\pi}}\,\oint_0 dz\;\K\left(\,\omega(z),J(z)\right)\
-\  {i\over{4\pi}}\,\oint_0 d\zb\;\K\left(\,\bar\omega(\zb),\bar J(\zb)\right)\
.} ($\oint_w dz$ will indicate integration around a contour enclosing the point
$z=w$.) Expanding $\omega=\sum_a \omega^at^a$, $J=\sum_aJ^at^a$, (and
$\bar\omega,\ \bar  J$ similarly), using $\K(t^a,t^b)=\delta^{ab}$, we get 
\eqn\deSa{\delta S_k(\gamma)\ =\ {{-1}\over{2\pi i}}\oint_0
dz\,\sum_a\omega^aJ^a\ +\ {1\over{2\pi i}}\oint_0 d\zb\,\sum_a\bar\omega^a\bar
J^a\ .} This transformation rule leads to the $\hat g\oplus \hat g$ current
algebra. In the Euclidean path integral formulation,  a correlation function
of the product $X$ of fields is given by \eqn\cX{\langle X\rangle\ =\
{{\int[d\Phi]\,X\,e^{-S[\Phi]}}\over {\int[d\Phi]\,e^{-S[\Phi]}}}\ ,}
where $[d\Phi]$ indicates path integration over the fields $\Phi$ of the
theory. If the action $S$ transforms with $\delta S\ =\ -\oint_0\,dz\,\delta
s(z)$, then 
\eqn\deX{\delta\langle X\rangle\ =\ -\oint_0 dz\;\langle(\delta s)X\rangle\ .} 
So \deSa\ implies 
\eqn\deomX{\delta_\omega\langle X\rangle\ =\ {{1}\over{2\pi i}}\oint_0 dz\,
\sum_a\omega^a(z)\,\langle J^a(z) X \rangle\ ,}
where we have put $\bar\omega=0$ for simplicity. 
Put $X=J^b(w)$. With $J(w)=-k(\partial_z
\ga)\ga^{-1}$ and $\delta_\ome \ga=\ome\ga$, we get  \eqn\dwJ{\delta_\ome J\ =\
[\ome,J]\ -\ k\partial_w\ome\ .} More explicitly this is 
\eqn\dwJb{\delta_\ome J^b(w)\ =\ \sum_{c,d}if^{bcd}\ome^c(w)J^d(w)\ -\
k\partial_w\ome^b(w)\ .} 

In \deomX\ this gives
\eqn\dzomf{\eqalign{{1\over{2\pi i}}\oint_w dw\,\ome^c(w)\,\langle
if^{cbd}&{{J^d(w)}\over{z-w}}\ +\ {{k\delta_{bc}}\over{(z-w)^2}}\rangle\ \cr
=&\ {1\over{2\pi i}}\oint_w dw\,\ome^a(w)\,\langle J^a(z)J^b(w)\rangle\ .\cr}}
This relation determines the singular part of the (radially-ordered) operator
product of two currents: 
\eqn\JaJb{\boxEq{J^a(z)J^b(w)\ \sim\ {{k\delta_{ab}}\over{(z-w)^2}}\ +\
{{if^{abc}J^c(w)}\over{z-w}}\ .}}
A similar OPE holds for the currents $\bar J^a(\bar z)$. This OPE is
equivalent to an affine algebra. 

The Laurent expansion of a current about $z=0$ is $J^a(z)=\sum_{n\in\Z}
J^a_nz^{-1-n}$,  or equivalently, $J^a_n\ =\ (1/2\pi i)\,\oint_0 dz\,
z^nJ^a(z)$. We can translate this expansion, so that  
\eqn\JLaur{J^a(z)\ =\ \sum_{n\in\Z}\,(z-w)^{-1-n}\, J^a_n(w)\ }
is the Laurent expansion about the point $z=w$, and $J_n^a(0)=J^a_n$.  
Of course, we also have
\eqn\JintJ{J^a_n(w)\ =\ {1\over{2\pi i}}\,\oint_w dz\,(z-w)^nJ^a(z)\ .}  

This allows us to write
\eqn\JJcomm{\eqalign{[J^a_m,J^b_n]\ =\ {1\over{2\pi i}}\oint_0dw\,&w^n\,
{1\over{2\pi i}}\oint_{|z|>|w|}dz\,z^m\, J^a(z)J^b(w)\ \cr -&\ 
{1\over{2\pi i}}\oint_0 dw\, w^m\, {1\over{2\pi i}}\oint_{|z|<|w|}dz\, z^m\,
J^b(w)J^a(z)\ ,\cr}}
where here radial ordering is not implicit in the operator products. Both
operator products in the integrands are $R\left(J^a(z)J^b(w)\right)$, however.
So, by subtraction of contours, we obtain \eqn\JJci{[J^a_m,J^b_n]\ =\
{1\over{2\pi i}}\oint_0 dw\, {1\over{2\pi i}}\oint_w dz\, z^m w^n\, R\left(
J^a(z)J^b(w) \right)\ , } as indicated in Fig. 5.

\midinsert
\vskip1cm
\epsfxsize=10cm
\centerline{\epsfbox{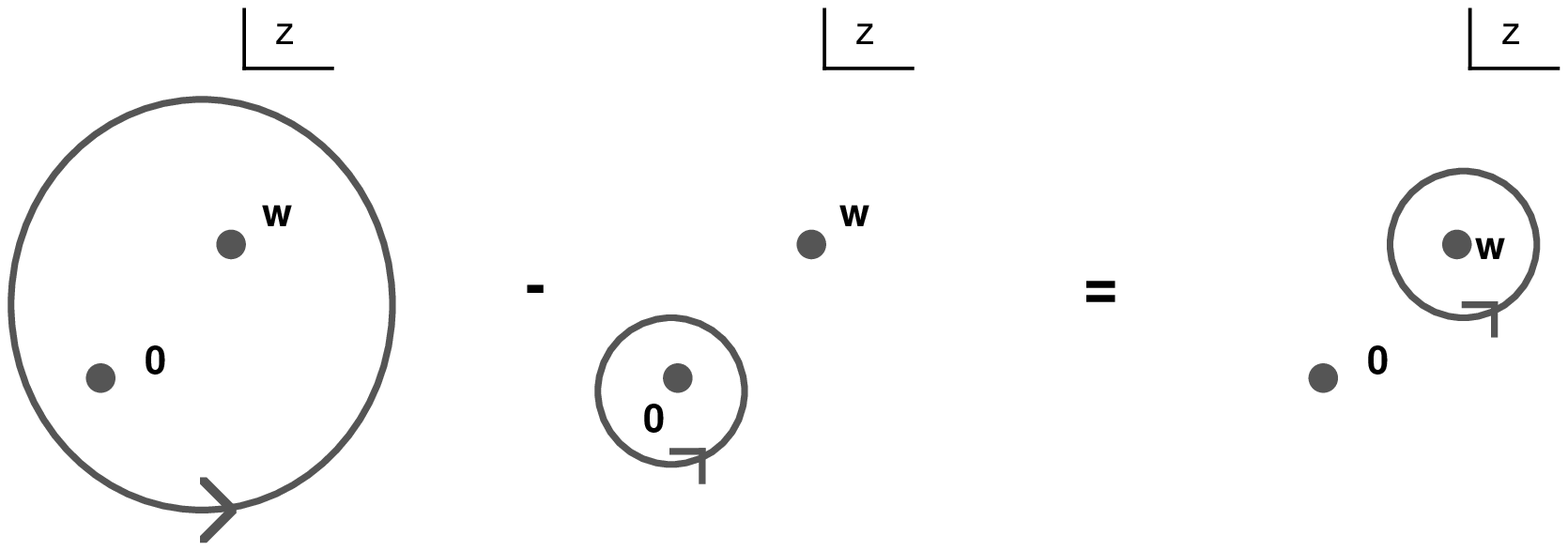}}
\smallskip\smallskip
\leftskip=1cm
\rightskip=1cm
\noindent
\baselineskip=12pt
{\it Figure 5}. Subtraction of contours for \JJci. \bigskip
\leftskip=0cm\rightskip=0cm
\baselineskip=15pt  
\endinsert 

\noindent After substituting \JaJb\ into the last result, residue
calculus then gives 
 \eqn\JJmn{\boxEq{ [J^a_n,J^b_m]\ =\ \sum_c\,if^{abc}J^c_{n+m}\ +\
kn\delta^{ab}\delta_{m+n,0}\ .}} Identical commutation relations hold for
the current modes $\bar J_m^a$. 

These are the commutation relations of $\hat g\oplus \hat g$. It is easy to
see that \JJmn\ is a central extension of the loop algebra of $g$. Consider
$J^a\otimes s^n$, with $s$ on the unit circle in the complex plane, and
$n\in\Z$. The loop algebra of $g$ is generated by the $J^a\otimes s^n$, since
they are $g$-valued functions on $S^1$ (the loop). Now
\eqn\Witt{[J^a\otimes s^m,J^b\otimes s^n]\ =\ [J^a,J^b]\otimes s^{m+n}\ =\
if^{abc}J^c\otimes s^{m+n}\ .} 
So only the central extension term ($\propto k$) is missing. 

The central extension term is known as a Schwinger term. \JaJb\ is not the 
usual form in quantum field
theory, because radial
quantisation is not typical. If we switch variables using $z=\exp(2\pi ix/L)$,
then Laurent series become Fourier series, and we recover the more familiar 
form 
\eqn\scwh{[\tilde J^a(x),\tilde J^b(y)]\ =\ if^{abc}\tilde J^c(x)\delta(x-y)
\ +\ {1\over{2\pi}}\,\delta^{ab}k\delta'(x-y)\ ,}
where we have put $\tilde J(x):=zJ^a(z)/L$. 
The Schwinger term is a quantum effect (as powers of $\hbar$ would show) and
is related to  chiral anomalies, as the  presence of $k$ suggests (recall that
$k=N_L-N_R$ in the derivation of the WZW model as an effective theory with
fermions integrated out).\foot{For more detail on the relation between chiral
anomalies and Schwinger terms, see \ref\mic{J. Mickelsson, {\it Current
Algebras and Groups}  (Plenum, 1989)}, Chapter 5.} 

The conformal invariance of the model can now be established in a
straightforward way. The Sugawara construction expresses the stress-energy
tensor in terms of normal-ordered products of currents $J^a(z)$. The
normal-ordered product \nprod\ of two operators $X(w),Y(w)$ can be rewritten as 
\eqn\NXY{N(XY)(w)\ =\ {1\over{2\pi i}}\,\oint_w {{dz}\over{z-w}}\,X(z)Y(w)\
.}
With this form of normal ordering, the appropriate version of Wick's theorem is
\eqn\wick{\eqalign{\contract{X(z)\,N(Y}Z)(w)\ =\ {1\over{2\pi
i}}\oint_w{{dx}\over{x-w}}
\Big\{\contract{X(z)Y}&(x)\,Z(w)\ \cr &+\ Y(x)\,\contract{X(z)Z}(w)\Big\}\
.\cr}} 

Using this we calculate 
\eqn\JJJ{\contract{\,J^a(z)\,\sum_bN(J^b}J^b)(w)\ =\
2(k+h^\vee)\,{{J^a(w)}\over{(z-w)^2}}\ ,}
using \JaJb. Here $h^\vee=\sum_{a,b,c}f^{abc}f^{abc}/(2\dim g)$ is the dual
Coxeter number of $g$ (this is consistent with \killing, \gcomm, \Kfabc). So
\eqn\JJJi{\sum_b N(J^b\contract{ J^b)(z)\, J^a}(w)\ =\ 2(k+h^\vee)
\left\{ {{J^a(w)}\over{(z-w)^2}}\ +\ {{\partial J^a(w)}\over{z-w}}\right\}\ .}
The Sugawara stress-energy tensor  is 
\eqn\sug{\boxEq{T(z)\ =\ {1\over{2(k+h^\vee)}}\,\sum_a N(J^aJ^a)(z)\ .}}
Using \wick\ and \JJJi\ then gives 
\eqn\TTope{\boxEq{T(z)T(w)\ \sim\ {{c/2}\over{(z-w)^4}}\ +\
{{2T(w)}\over{(z-w)^2}}\ +\ {{\di T(w)}\over{z-w}}\ ,}}
with the {\it central charge} 
\eqn\cwzw{c\ =:\ c(g,k)\ =\ {{k\,\dim g}\over{k+h^\vee}}\ .}

This last result is the conformal algebra with central extension, or {\it
Virasoro algebra}, in OPE form. So conformal invariance is established.
Substituting $T(z)=\sum_{n\in\Z} z^{-2-n}L_n$ yields the usual form of $Vir$
(the Virasoro algebra): \eqn\Vir{\boxEq{[L_m,L_n]\ =\ (m-n)L_{m+n}\ +\
{c\over{12}}(m^3-m)\delta_{m+n,0}\ .}}

For completeness, we also write
\eqn\TJope{\boxEq{T(z)J^a(w)\ \sim\ {{J^a(w)}\over{(z-w)^2}}\ +\ {{\di
J^a(w)}\over{z-w}}\ ,}}
which corresponds to 
\eqn\LmJn{\boxEq{[L_m,J_n^a]\ =\ -nJ^a_{m+n}\ .}}
This shows that $\hat g$ and $Vir$ extend to a semi-direct product in
the theory. Furthermore, the full {\it chiral algebra} of the WZW model is
$Vir\ltimes\hat g$, with commutation relations \Vir,\JJmn\ and \LmJn. 

\subsec{Factorisation and primary fields}

A factorised form for $\ga(z,\zb)$ solves the classical equation of motion:
\eqn\fact{\ga(z,\zb)\ =\ \ga_L(z)\ga_R(\zb)\ \ \Rightarrow\ \ 
\di_z\big(\ga^{-1}\di_\zb\ga\big)\ =\ 0\ .}
This factorisation survives in the following form in the quantum theory. As
already mentioned, under an infinitesimal  $G(z)\otimes G(z)$ transformation,
we have 
\eqn\GGinf{\delta_\ome\ga\ =\ \ome\ga\ \ ,\ \ \ \ \delta_{\bar\ome}\ga\ =\
-\ga\bar\ome\ \ .}
The currents $J^a(z),\bar J^a(\zb)$ generate the infinitesimal transformations
of the fields, so we have
\eqn\Jga{\eqalign{J^a(z)\ga(w,\bar w)\ \sim&\
{{-1}\over{z-w}}\,t^a_\ga\ga(w,\bar w)\ ,\cr
\bar J^a(z)\ga(w,\bar w)\ \sim&\ {{1}\over{\zb-\bar w}}\,\ga(w,\bar
w)t^a_\ga\ ,\cr
 }}
where $t_\ga^a$ is the $g$-generator $t^a$ in the representation appropriate
to $\ga(z,\zb)$. 

The WZW model also contains other fields, besides $\ga(z,\zb)$, that transform
in similar fashion. These are the so-called {\it primary fields}
$\Phi_{\la,\mu}(z,\zb)$:
\eqn\JPhi{\eqalign{J^a(z)\Phi_{\la,\mu}(w,\bar w)\ \sim&\
{{-1}\over{z-w}}\,t^a_\la\Phi_{\la,\mu}(w,\bar w)\ ,\cr
\bar J^a(z)\Phi_{\la,\mu}(w,\bar w)\ \sim&\ {{1}\over{\zb-\bar
w}}\,\Phi_{\la,\mu}(w,\bar w)t^a_\mu\ ,\cr
 }}
Here $\la,\mu$ are two highest weights of integrable unitary irreducible
representations $L(\la),L(\mu)$ of $g$, and $t^a_\la,t^a_\mu$ are the
generators in those representations. 

To find the action of the ``current modes'' $J^a_n$ that generate $\hat
g$, we  use $J^a(z)=\sum_{n\in \Z} z^{-1-n}J_n^a$ in 
\eqn\JPhiz{J^a(z)\Phi_{\la,\mu}(0,0)\ \sim\
-{1\over{z}}\,{t^a_\la}\Phi_{\la,\mu}(0,0)\ }
to get 
\eqn\JnPhi{\eqalign{[J_0^a,\Phi_{\la,\mu}(0,0)]\ =&\
-t^a_\la\Phi_{\la,\mu}(0,0)\ ,\cr [J_n^a,\Phi_{\la,\mu}(0,0)]\ =&\ 0\ 
,\ {\rm for\ }n>0\
.\cr}} This implies that the primary field $\Phi_{\la,\mu}$ transforms as a
highest-weight representation $L(\h\la)$ of the affine algebra $\hat g$.
Similar considerations work for the right action, so $\Phi_{\la,\mu}$
transforms as $L(\h\la)\otimes L(\h\mu)$.\foot{As we'll see, a highest weight
$\h\la$ for $\h g$ determines a highest weight $\la$ for $g$.} 

For most purposes, it suffices to consider only the left {\it or} right action
of $\hat g$. So we will write, instead of \JPhi, 
\eqn\Jphi{\boxEq{J^a(z)\phi_\la(w)\ \sim\ {{-t^a_\la\phi_\la(w)}\over{z-w}}\
,}} and similarly for $\bar J^a(\zb)$ and $\bar \phi_\mu(\bar z)$, if need be.
We must emphasise, however, that $\phi_\la(z),\bar\phi_\mu(\zb)$ are {\it not}
sensible local fields; they are only the holomorphic (left-moving) and
antiholomorphic (right-moving) parts of the primary field
$\Phi_{\la,\mu}(z,\zb)$. If you like, 
$\Phi_{\la,\mu}(z,\zb)=\phi_\la(z)\bar\phi_\mu(\zb)$. 

To see this, first note that the primary field $\Phi_{\la,\mu}(z,\zb)$
transforms nicely under conformal transformations. That's because of the
Sugawara construction \sug, expressing the stress-energy tensor as a
normal-ordered product of the currents. In terms of modes, the Sugawara
construction gives
\eqn\sugm{L_n\ =\ {1\over{2(k+h^\vee)}}\,\sum_a\sum_{m\in\Z}\,
N(J^a_{n-m}J^a_m)\ .}
where 
\eqn\Nmode{N(J^a_pJ^b_q)\ =\ \left\{\matrix{J^a_pJ^b_q\ ,\ \ &\ p\leq q\ ;\cr
                             J^b_qJ^a_p\ ,\ \ &\ p> q\ .}\right.}  
We get 
\eqn\Lnphi{[L_n,\phi_\la(0)]\ =\ \left\{\matrix{0 ,\ \ &\ n>0\ ;\cr
                             h_\la\phi_\la(0)\ ,\ \ &\ n= 0\ .}\right.}
where 
\eqn\hla{h_\la\ =\ {{\sum_a\, \Tr (t^a_\la t^a_\la)}\over{2(k+h^\vee)\,\Tr
(\id_\la)}}\ =\ {{(\la,\la+2\rho)}\over{2(k+h^\vee)}}\ }
is the {\it conformal weight} of the ``primary field'' $\phi_\la$, and $\rho$
is the Weyl vector of $g$ (the half-sum of the positive roots of $g$). 

In OPE language, this is
\eqn\Tphi{\boxEq{T(z)\phi_\la(0)\ \sim\ {{h_\la\phi_\la(0)}\over{z^2}}\ +\
{{\di\phi_\la(0)}\over z}\ .}}
Similarly,
\eqn\Tphib{\bar T(\zb)\bar\phi_\mu(0)\ \sim\ {{
h_{\mu}\bar\phi_{\mu}(0)}\over{\zb^2}}\ +\
{{\bar\di\bar\phi_\mu(0)}\over \zb}\ .}

Now, the generator of infinitesimal scaling is $L_0+\bar L_0$, as we'll show
below. So $h_\la+h_\mu$ is the scaling dimension of the primary field
$\Phi_{\la,\mu}$. (In radial quantisation, scaling = time-translation, so the
Hamiltonian ${\bf H}= L_0+\bar L_0$.) $L_0-\bar L_0$ generates rotations, so
that $h_\la-h_\mu$ is the {\it spin} of $\Phi_{\la,\mu}$. For a single-valued
(local) field, we therefore  require $h_\la-h_\mu\in\Z$. This is a highly
nontrivial constraint on pairs $(\la,\mu)$, since $h_\la,h_\mu\in\Q$. It is in
this sense that $\phi_\la(z)$ cannot be considered a sensible local field in
its own right. 

The fields $\Phi_{\la,\mu}$ are primary because all others are in the span of 
operator products of currents acting on them:
\eqn\Jdes{J^{a_1}(z_1)J^{a_2}(z_2)\cdots J^{a_n}(z_n)\bar J^{\bar a_1}(\zb_1) 
J^{\bar a_2}(\zb_2)\cdots J^{\bar a_{\bar n}}(\zb_{\bar
n})\,\Phi_{\la,\mu}(z,\zb)\ .} 
They are therefore called {\it descendant} fields. More usually, the basis
elements are written as
\eqn\Jndes{J_{-n_1}^{a_1}\cdots J_{-n_N}^{a_N}\bar J_{-\bar n_1}^{\bar a_1} 
\cdots \bar J_{-\bar n_{\bar N}}^{\bar a_{\bar N}}\,\Phi_{\la,\mu}(z,\zb)\ .}

\subsec{Field-state correspondence}

$|0\>$ is the vacuum of the WZW model. $t^a_\la\phi_\la$ means $\sum_{v\in
L(\la)}(t^a_\la)_{u,v}\phi_{\la,v}$. If $\phi_{\la,v}=\delta_{v,v_\la}$, where
$v_\la$ denotes the highest-weight vector of $L(\la)$, and $v\in L(\la)$, we
have  \eqn\philv{\boxEq{\phi_\lambda(0)|0\>\ =\ |v_\la\>\ .}}
This is the basis of the field-state correspondence. More generally,
defining   \eqn\phiv{|\phi_\la\>\ :=\ \sum_{v\in L(\la)}\,\phi_{\la,v}|v\>\ ,}
we can write
\eqn\fstgen{\phi_\lambda(0)|0\>\ =\ |\phi_\la\>\ .} We can also consistently
write
\eqn\uzfs{\phi_\la(z)\ =\ \sum_{u\in L(\la)}\,\phi_{\la,u}\,u(z)\ ,\ \  {\rm
with}\  u(0)\,|0\>\ =\ |u\>\ .}

In terms of $|v_\la\>$, the primary-field conditions read
\eqn\prvl{J_0^a|v_\la\>\ =\ t_\la^a|v_\la\>\ ,\ \ \ J_n^a|v_\la\>\ =\ 0\
(n>0)\ \ ,}
in agreement with \Lnphi. 
The affine algebras are examples of triangularisable algebras (just like the
simple Lie algebras) \ref\moopia{R.V. Moody, A. Pianzola, {\it Lie Algebras
with Triangular Decomposition} (Wiley, 1995)}. This means their generators can
be written as a disjoint sum of three sets, with corresponding decomposition
\eqn\Cdec{\hat g\ =\ \hat g_-\ \oplus\ \hat g_0\ \oplus\ \hat g_+\ .} $\hat
g_0$ is the Cartan subalgebra, while $\hat g_\pm\oplus\hat g_0$ are Borel
subalgebras. $\hat g_+(\hat g_-)$ correspond to positive (negative) roots, and
so contain raising (lowering) operators. Now, in the basis used,  $\hat g_+$ is
generated by $\{J^a_{n>0}\}\oplus g_+$, where $g_+\subset\{J_0^a\}$ contains
the raising operators of $g\subset\hat g$. But since $t^a_\la$ are the
generators of $g$ in a representation $L(\la)$ of highest weight $\la$, we know
$g_+|v_\la\>=0$. So by \prvl, $\hat g_+|v_\la\>=0$, \ie\ $|v_\la\>$ is the
highest-weight state (highest state/vector) of the affine representation
$L(\h\la)$ of $\hat g$. 

The rest of the states in the representation $L(\h\la)$ can be obtained as
descendant states, \ie\ as linear combinations of states of the form 
\eqn\Llhdes{J_{-n_1}^{a_1}\cdots J_{-n_N}^{a_N}\phi_\la(0)|0\>\ \ .} 
Now, there is still an infinite number of possible highest weights. But we'll
find that for fixed $k\in\Z_{> 0}$, only a finite number of inequivalent
highest weights are possible. These are the (unitary) {\it integrable}
highest weights; they generate representations of $\hat g$ that can be
integrated to representations of $\widehat{LG}$. By the $G(z)\otimes G(\zb)$
invariance of the WZW model, these representations are precisely the relevant
ones. We will arrive at this result from an algebraic perspective, however. 

To do this, we first need to discuss $g,\hat g$ and their relation. This
justifies an interesting digression on Kac-Moody algebras \ref\Kac{V. Kac,
{\it Infinite-dimensional Lie Algebras} (Cambridge U. Press, 1990)}
\ref\kmps{S. Kass, R.V. Moody, J. Patera, R. Slansky, {\it Affine Lie
Algebras, Weight Multiplicities, and Branching Rules} (U. California Press,
1990)} \ref\fuchs{J. Fuchs, {\it Affine Lie Algebras and Quantum Groups}
(Cambridge U. Press, 1992)} \ref\godoli{P. Goddard, D. Olive, Int. J. Mod.
Phys. {\bf A1} (1986) 303}. 

\newsec{Affine Kac-Moody Algebras}

\subsec{Kac-Moody algebras: simple Lie algebras}

$g,\hat g$ are examples of Kac-Moody algebras, which can be presented in terms
of a {\it Cartan matrix} $A=(A_{i,j})$, with integer entries (\ie\ Kac-Moody
algebras are generalised Cartan matrix Lie algebras). Let's first define $g$
this way. If $X$ is generated by $x_1$ and $x_2$, for example, we use the
notation
$X= \<x_1, x_2\>$. 

Recall $g=g_+\oplus g_0\oplus g_-$. Now 
\eqn\ggg{g_+=\<e_i\ :\ i=1,\ldots,r\>\ ,\ \ g_0=\<h_i\ :\ i=1,\ldots,r\>\ ,\ \ 
g_-=\<f_i\ :\ i=1,\ldots,r\>\ ,}
where $r$ is the rank of $g$, and $\{e_i,h_i,f_i\ :\ i=1,\ldots,r\}$ are the
{\it Chevalley generators} of $g$. The commutation relations of the generators 
can be expressed in terms of the Cartan matrix:
\eqn\chcrs{\boxEq{\eqalign{[h_i,h_j]\ =&\ 0\cr [h_i,e_j]\ =&\ A_{j,i}e_j\cr
[h_i,f_j]\ =&\ -A_{j,i}f_j\cr [e_i,f_j]\ =&\ \delta_{i,j}h_j\ .\cr
}}}
The Chevalley presentation of the algebra $g$ is completed by the {\it Serre
relations}:
\eqn\serre{\boxEq{\eqalign{\left[\ad(e_i)\right]^{1-A_{j,i}}\ e_j\ =&\ 0\ \
,\cr
\left[\ad(f_i)\right]^{1-A_{j,i}}\ f_j\ =&\ 0\ \ .\cr}}}
The $r\times r$ Cartan matrix has diagonal entries $A_{i,i}=2$, so that $\<
e_i,h_i,f_i \>\cong s\ell(2)$ for all $i=1,\ldots,r$. For simple $g$,  $g\not\cong
\oplus^r_{i=1} s\ell(2)$, so $A_{i,j}\not=0$ for at least one pair $i\not=j$, if
$r>1$. For all Kac-Moody algebras (including semi-simple Lie, affine,
hyperbolic, etc. algebras)\foot{For a discussion of generalised Kac-Moody
algebras, or Borcherds-Kac-Moody algebras, see \gann.}, $A_{i,i}=2\
\forall i$, as just mentioned; $A_{i,j}\in-\Z_{\geq 0}\ \forall\ i\not=j$; and
$A_{i,j}=0\ \Leftrightarrow\ A_{j,i}=0$. In addition, the Cartan matrices are
{\it symmetrisable}: there exist positive rational numbers $q_j$ such that
$AD'$ is a symmetric matrix, where $D'={\rm diag}(q_j)$. 

For $g$ a semi-simple Lie algebra, $A_{i,j}\in\{0,-1,-2,-3\}$  for $i\not=j$,
and most importantly, $\det\, A>0$, \ie\ the Cartan matrix is invertible. For
simple $g$, $A$ must be indecomposable. 

The information contained in the Cartan matrix can be encoded in a so-called
{\it (Coxeter-)Dynkin diagram}. $r$ nodes are drawn, each associated with a
row (or column) of $A$. Node $i$ and node $j$ ($j\not=i$) are joined by a number
$A_{i,j}A_{j,i}$ of lines; and if $A_{i,j}\not= A_{j,i}$, so that there are
more than one lines, an arrow is drawn from node $i$ to node $j$ if
$|A_{i,j}|>|A_{j,i}|$. The Coxeter-Dynkin diagrams of the simple Lie 
algebras are drawn in Figure 6. 

\midinsert
\vskip0cm
\epsfxsize6cm
\centerline{\epsfbox{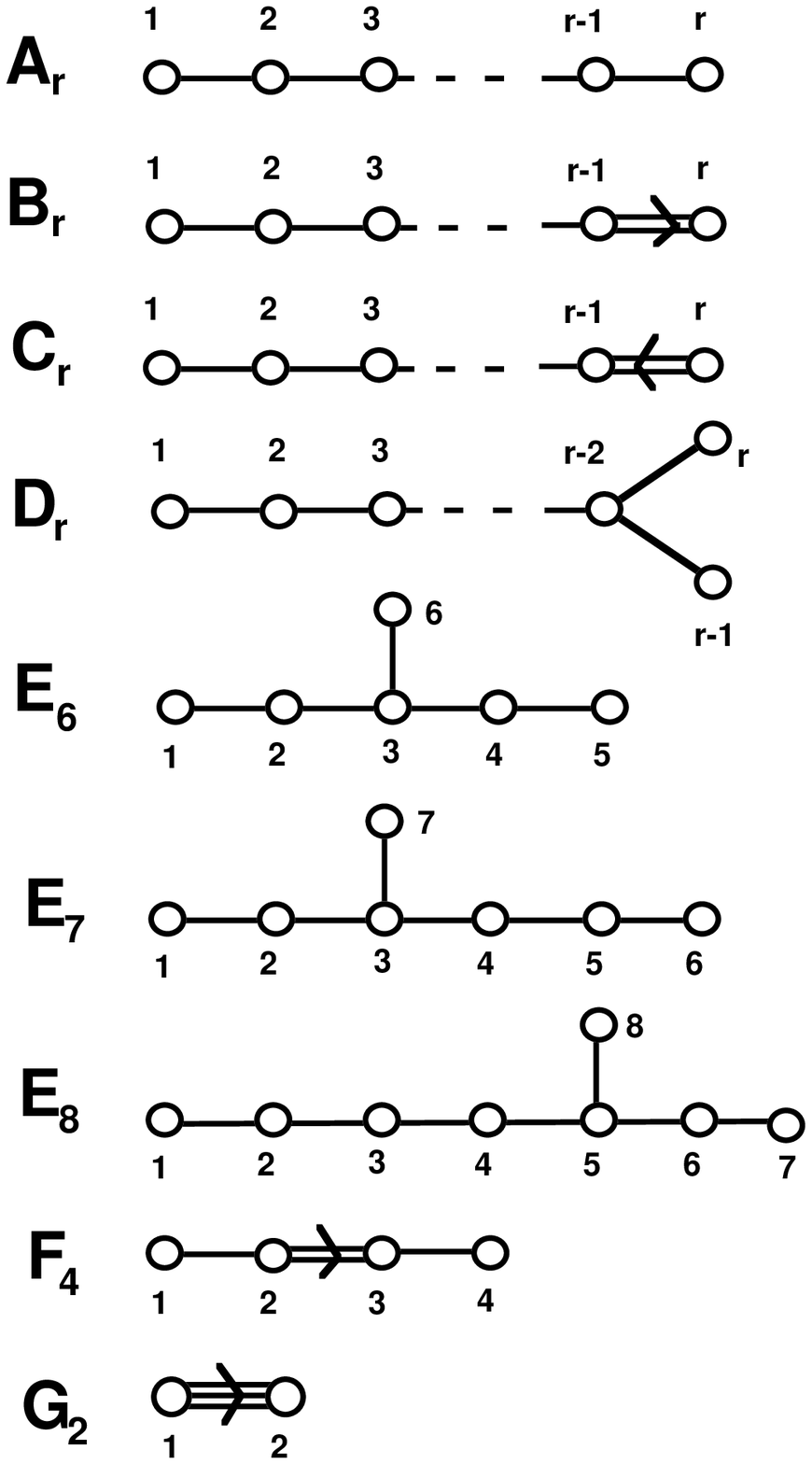}}
\smallskip\smallskip
\leftskip=1cm
\rightskip=1cm
\noindent
\baselineskip=12pt
{\it Figure 6}. The (Coxeter-)Dynkin diagrams of the simple Lie
algebras.\bigskip \leftskip=0cm\rightskip=0cm
\baselineskip=15pt  
\endinsert

A more direct significance, in terms of the roots of the algebra, can be given
to the Cartan matrix and Coxeter-Dynkin diagrams. As we do this, we'll
introduce  another presentation of $g$, the {\it Cartan-Weyl presentation}. 

First, find a maximal set of commuting Hermitian generators $H^i$,
($i=1,\ldots,r$):
\eqn\HH{\boxEq{[H^i,H^j]\ =\ 0\ \ (1\leq\,i,j\,\leq r)\ \ .
}}
Any two such maximal Abelian subalgebras $g_0$ (Cartan subalgebras) are
conjugate under the action of $\exp g$ (the covering group of a group with Lie
algebra $g$). Fix a choice of Cartan subalgebra $g_0$.

Since the $H^i$ mutually commute, they can be diagonalised simultaneously in
any representation of $g$. The states of such a representation are then
eigenstates of the $H^i$, $1\leq i\leq r$. We write
\eqn\wtH{H^i\, |\mu;\,\ell\>\ =\ \mu^i\, |\mu;\,\ell\>\ .} 
Here $\mu$ is called a {\it weight vector} (or just a weight), with $r$
components $\mu^i$.  The corresponding  $r$-dimensional space is known as {\it
weight space}. As we'll discuss shortly, it is dual to the Cartan subalgebra
$g_0$. The $\ell$ of the kets $|\mu;\,\ell\>$ is meant to indicate any
additional labels. 

A basis for the whole of $g$ can be
constructed by appending $\{E^\alpha\}$, obeying 
\eqn\root{\boxEq{[H^i,E^\alpha]\ =\ \alpha^iE^\alpha\ \ (1\,\leq i\leq\, r)\
.}} $\alpha$ is a $r$-dimensional vector, called a {\it root}, and $E^\alpha$
is a step operator (raising or lowering operator, depending). For simple Lie
algebras, $E^\alpha$ is determined by $\alpha$, up to normalisation. If
$\alpha$ is a non-zero root, the only multiple of $\alpha$ that is also a root
is $-\alpha$, and we can take   \eqn\Edag{E^{-\alpha}\ =\
\big(E^\alpha\big)^\dagger\ \ .} The set of roots of $g$ will be denoted
$\Delta$. 

To complete the presentation of $g$ in the Cartan-Weyl basis, we also need:
\eqn\EaEb{\boxEq{[E^\alpha,E^\beta]\ =\
\left\{\matrix{N(\alpha,\beta)E^{\alpha+\beta}\ \ \ &\ \ {\rm if\
}\alpha+\beta\in\Delta\cr
2\alpha\cdot H/\alpha^2\ \ &\ \ {\rm if\ }\alpha+\beta=0\cr
0\ \ \ \ \ &\ \ {\rm otherwise\ }\cr}\right. }}
where $\alpha\cdot H$ means $\sum_{i=1}^r\alpha^iH^i=:H^\alpha$,
$\alpha^2:=(\alpha,\alpha)$ (see below), and the $N(\alpha,\beta)$ are
constants.\foot{\Edag\ implies $N(\alpha,\beta)=-N(-\alpha,-\beta)$. Then 
for $\beta+\ell\alpha\in\Delta$ with $p\leq \ell\leq q$, we can set 
$N(\alpha,\beta)^2=q(1-p)(\alpha,\alpha)/2$.} Here we use the scalar product
$(\alpha,\beta)$, defined through the Killing form \killing:
\eqn\Kab{(\alpha,\beta)\ =\ \K(H^\alpha,H^\beta)\ =\ {\rm Tr\,}\big(
\ad(H^\alpha),\ad(H^\beta)\big)/2h^\vee\ .}
This completes the Cartan-Weyl presentation of $g$. 

The Killing form also establishes an isomorphism between the Cartan subalgebra
$g_0$ and its dual $g_0^*$, weight space: for every weight
$\la\in g_0^*$, there corresponds an element $H^\la\in g_0$ by
$\la(\cdot)=\K(H^\la,\cdot)$ (\ie\
$\la(H^\beta)=\K(H^\la,H^\beta)=(\la,\beta)$ for $\beta\in\Delta$). This inner
product can be extended, by symmetry, $(\alpha,\beta)=(\beta,\alpha)$, to all
weights $\alpha,\beta\in g_0^*$. By \EaEb, the rescaled root
$2\alpha/\alpha^2$ has importance; it is called the {\it coroot}
$\alpha^\vee$. 

If we choose a fixed basis for the root lattice ($\subset$ weight lattice), we
call $\alpha$ positive, $\alpha\in\Delta_+$, iff its first nonzero component
in this basis is positive. Otherwise, $\alpha\in\Delta_-$, \ie\ $\alpha$ is a
negative root. $E^\alpha$ is considered a raising (lowering) operator if
$\alpha\in\Delta_+$ ($\alpha\in\Delta_-$). 

A {\it simple root} is a positive root that cannot be written as a linear
$\Z_{\geq 0}$-combination of other positive roots. The set of simple roots will be
denoted $\Pi = \{\alpha_i\ :\ i=1,\ldots,r\}$. The set of simple coroots is 
$\Pi^\vee = \{\alpha^\vee_i\ :\ i=1,\ldots,r\}$. The basis dual to $\Pi^\vee$ 
is the Dynkin basis of {\it fundamental weights}:
\eqn\fwts{\big(\Pi^\vee\big)^*\ =\ \{ \omega^i\ :\ j=1,\ldots,r\ \}\ .}
That is, $(\omega^i,\alpha^\vee_j)=\delta^i_j$. 

Let us now compare the Chevalley and Cartan-Weyl presentations of $g$.
The Chevalley presentation emphasises the $r$ subalgebras of type 
$s\ell(2)\cong A_1$ that are associated with each simple root (or
fundamental weight). It is the more economical presentation,
since it is written in terms of just $3r$ generators, those listed in \ggg. This
economy allowed the discovery of the Kac-Moody algebras: it was natural to
wonder whether loosening the constraints on the Cartan matrix would lead to
other interesting types of algebras. The price to be paid  is the
imposition of the  more complicated Serre relations \serre. But these
relations are what ensure that (among other things) a finite-dimensional algebra
is generated. 

In contrast, the Cartan-Weyl presentation makes use of the $A_1$-subalgebras
associated with every positive root. For every positive root we get a raising
and lowering operator, and the finite-dimensionality of the algebra is built
in. Of course, the cost is the use of more generators, a total of $\dim g$ of
them. 

More concretely, it is not difficult to make the identifications
\eqn\ccwid{e_i\ =\ E^{\alpha_i}\ ,\ \ \ f_i\ =\ E^{-\alpha_i}\ ,\ \ \ 
h_i={{2\alpha_i\cdot H}\over{\alpha_i^2}}\ =\ \alpha_i^\vee\cdot H\ ,}
where $H=\sum_{i=1}^r \ome^i h_i$, and finally
\eqn\Aroots{\boxEq{A_{i,j}\ =\
{{2\big(\alpha_i,\alpha_j\big)}\over{\alpha^2_j}}\ =\
(\alpha_i,\alpha_j^\vee)\ \ . }}
So, the Cartan matrix encodes the scalar products of simple roots with
simple coroots.

Now, $\det\, A>0$ guarantees that weight space is Euclidean. Consider the
hyperplanes in weight space with normals $\alpha_i$. The {\it primitive
reflection} $r_{\alpha_i}=r_i$ of a weight $\la=\sum_{i=1}^r\la_i\ome^i$ across
such a hyperplane is given by
\eqn\rali{r_{\alpha_i}\la\ =\ r_i\la\ =\ \la-(\la,\al^\vee_i)\alpha_i\ .}
Being reflections, the $r_i$ have order 2, and they generate a {\it Coxeter}
group $W$, which can be presented as 
\eqn\Wpres{W\ =\ \<\,r_i\ :\ i=1,\ldots,r\,\>\ ,}
with the relations
\eqn\WCox{(r_ir_j)^{m_{ij}}\ =\ \id\ .}
Clearly, $m_{ii}=1$ for all $i$, and it turns out that
all $m_{ij}\in\{2,3,4,6\}$, when $i\not=j$. This Coxeter presentation can
be encoded in a {\it Coxeter diagram}: nodes are drawn for each primitive
reflection, and $\{0,1,2,3\}$ lines between nodes for $m_{ij}\in\{2,3,4,6\}$,
respectively ($i\not=j$). For simple $g$, we find the Coxeter diagrams are just
the corresponding Dynkin diagrams (see Fig. 6), with the arrows omitted.
In fact, the Coxeter group so obtained is the {\it Weyl group} of $g$. 

The possible weights of integrable representations will lie on the {\it weight
lattice} $P:=\Z(\Pi^\vee)^*$, the points in weight space that are integer linear
combinations of the fundamental weights. Of course, this lattice is periodic and
``fills''  weight space. So we can think of it as an infinite 
crystal. It has a point group isomorphic to the Weyl group, which explains the
restriction $m_{ij}\in\{2,3,4,6\}$, familiar from crystallography. 

Still, it is remarkable that these Coxeter-Weyl groups almost determine
the algebra $g$ completely. More accurately ($B_r$ and $C_r$ have isomorphic
Weyl groups), the simple Lie algebras are essentially those whose weight
lattices can exist in a Euclidean weight space of dimension equal to the rank.

What is the geometry of the Weyl hyperplanes in weight space? There is a
Weyl hyperplane for each root, not just for the simple roots, and they partition
the $r$-dimensional weight space into a finite number of sectors. Each sector
is of infinite hypervolume, and $W$ acts simply transitively on them.
The example of $g=A_2$ is pictured in Fig. 7. 

\topinsert
\vskip-0cm
\epsfxsize=8cm
\centerline{\epsfbox{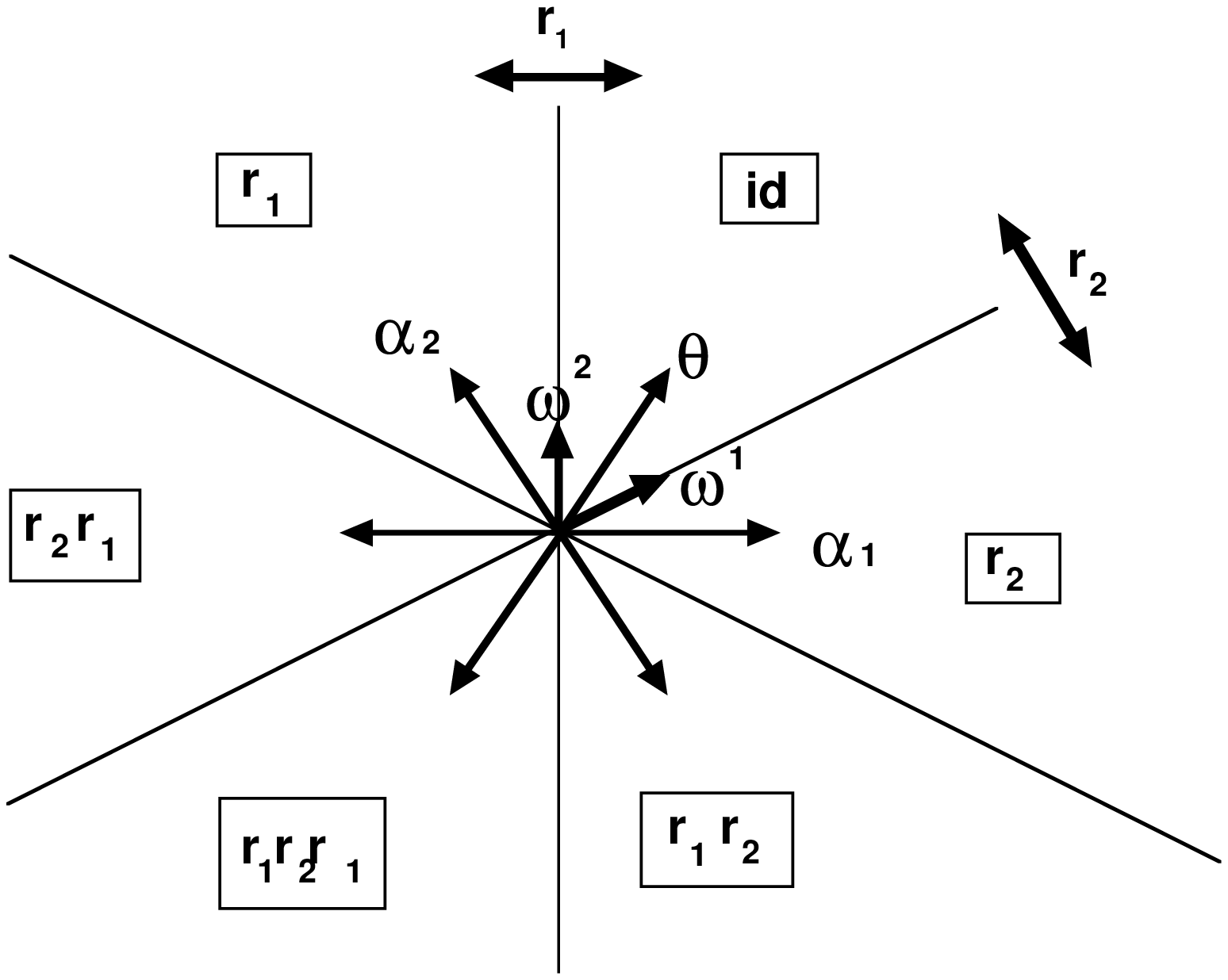}}
\smallskip\smallskip
\leftskip=1cm
\rightskip=1cm
\noindent
\baselineskip=12pt
{\it Figure 7}. The Weyl sectors of $A_2$ weight space. Each sector is
labelled by the Weyl element that maps it to the identity ($\id$) sector. The
identity sector is also known as the dominant sector. Also shown are the
fundamental weights $\ome^1,\ome^2$, and the roots, including the simple roots
$\al_1,\al_2$, and the highest root $\theta$. \bigskip
\leftskip=0cm\rightskip=0cm \baselineskip=15pt   \endinsert

We use the notation $\la = \sum_{i=1}^r\la_i\,\ome^i =
(\la_1,\ldots,\la_r)$, and $L(\la_1,\ldots,\la_r)=L(\la)$. 
Fig. 8  is the weight diagram for the $A_2$ representation $L(2,1)$.
Notice it is symmetric under the action of the Weyl
group $W\cong S_3$ for $A_2$. Let
$\mult(\la;\mu)$ denote the multiplicity of a weight $\mu$ in the
representation $L(\la)$. Then this Weyl symmetry can be written as 
\eqn\Wmult{\mult(\la;\mu)\ =\ \mult(\la;w\mu)\ ,\ \ \forall w\in W\ .} 

\midinsert
\vskip-0cm
\epsfxsize=8cm
\centerline{\epsfbox{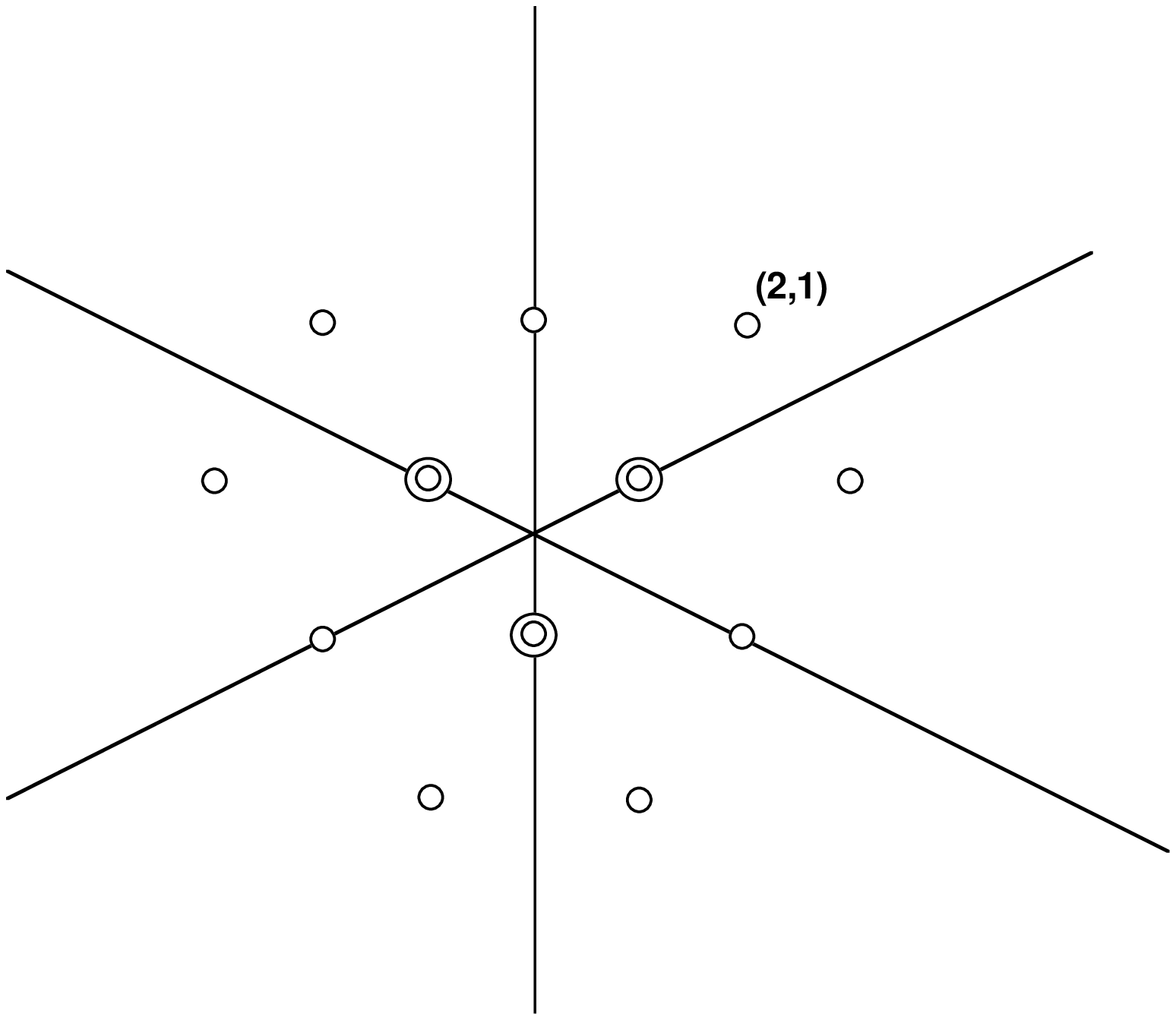}}
\smallskip\smallskip
\leftskip=1cm
\rightskip=1cm
\noindent
\baselineskip=12pt
{\it Figure 8}. The weight diagram of $L(2,1)$, the $A_2$
representation of highest weight $(2,1)$.\bigskip \leftskip=0cm\rightskip=0cm
\baselineskip=15pt   \endinsert

The Weyl symmetry can also be expressed in terms of {\it characters}.
Characters are to representations what weights are to states (vectors). They 
are simpler than the representations themselves, yet still contain sufficient
information to be useful in many contexts. Precisely, the {\it formal
character} of the $g$-representation $L(\la)$ is 
\eqn\fch{\ch_{\la}\ :=\ \Tr_{L(\la)}\, e^{H}\  .}
Equivalently, if we define 
\eqn\Pla{P(\la)\ :=\
\{\, \mu \in  g^{*}_0\ :\ \mult(\la;\mu)\not=0 \,\}\ ,} 
we can write \eqn\fchi{\ch_{\la}\ =\
\sum_{\mu\in P(\la)}\, \mult(\la;\mu)\, e^{\mu}\ \ .} In \fch\ and \fchi,
$e^{H}$ and $e^{\mu}$ are formal exponentials, with the additive property
$e^{\la}e^{\mu} = e^{\la +\mu}$, for example.  

The Weyl symmetry is made manifest in the celebrated {\it Weyl character
formula}:
\eqn\wcf{\boxEq{\ch_\la\ =\ {{\sum_{w\in W}\, (\det\, w)\, e^{w.\la} } \over 
{\prod_{\al\in\Delta_+}\, (1-e^{-\al}) }}\ ,}} 
where the {\it shifted Weyl action} is $w.\la := w(\la +\rho)-\rho$. 
Here $\det\, w = 1$ if $w$ can be written as a composition of an even number of
primitive reflections, and $\det\, w = -1$ for an odd number. 

Since the character of the singlet representation $L(0)$ is $\ch_0=1$, \wcf\
gives the {\it denominator identity}
\eqn\denom{{\prod_{\al\in\Delta_+}\, (1-e^{-\al}) } \ 
=\ {\sum_{w\in W}\, (\det\, w)\, e^{w.0} }\ .}
So the Weyl character formula  can also be written as
\eqn\wcfi{\ch_\la\ =\ {{\sum_{w\in W}\, (\det\, w)\, e^{w(\la +\rho)} } \over 
{\sum_{w\in W}\, (\det\, w)\, e^{w\rho} } }\ .}

If we continue
this relation to weights $\la \not\in P_+$, we can also derive 
\eqn\altw{\boxEq{\ch_\la\ =\ (\det\, w)\, \ch_{w.\la}\ \ .}} 
This relation will be important later. 

We can ``informalise'' the formal character in the following way:
\eqn\ich{\ch_\la(\sigma)\ :=\  
\sum_{\mu\in P(\la)}\, \mult(\la;\mu)\, e^{(\mu,\sigma)}\ \ .} The character
$\ch_\la(\sigma)$ is then a polynomial in the $r$ indeterminates $e^{\sigma_j}$,
$j=1,\ldots,r$.

\subsec{Kac-Moody algebras: affine algebras}

As discussed above, the affine algebras $\hat g$ relevant to WZW models are
central extensions of the loop algebras of $g$, for $g$ semi-simple; we
restrict to $g$ simple here for simplicity. They are known as {\it untwisted
affine} algebras. For such $\hat g$, $g\subset \h g$ is known as the {\it
horizontal subalgebra} of $\h g$. 

The Chevalley presentation for $\hat g$ is
identical to that for $g$ except that the $r\times r$ Cartan matrix
$A=(A_{i,j})$ is replaced by the $(r+1)\times(r+1)$ Cartan matrix $\hat A=(\hat
A_{i,j})_{i,j\in\{0,1,\ldots,r\}}$, with $\hat A_{i,j}=A_{i,j}$ for
$i,j\not=0$. As for $g$, the elements of the Cartan matrix are determined by
scalar products of simple roots and coroots: \eqn\Ahal{\hat A_{i,j}\ =\
(\h\al_i,\h\al^\vee_j)\ \ .} Because of this structure, there is an
intimate relation between the simple roots of $g$ and those of $\hat g$.

An affine Kac-Moody Cartan matrix obeys all the conditions mentioned above
that the simple Lie algebras obey, except that the $\det\, A>0$ condition is
loosened. Let $\hat A_{(i)}$ denote the submatrix of $(r+1)\times(r+1)$  
matrix $\h A$ obtained by deleting the $i$-th row and column. Then we must have 
\eqn\affdet{\det\,\h A\ =\ 0\ ,\ \ {\rm but}\ \ \det\,\hat A_{(i)}\ >\ 0\ \ \
\forall i\in\{0,1,\ldots,r\}\ \ ,}  if $\h A$ is to be an affine Cartan matrix.
This means that the submatrices $\h A_{(i)}$ must be Cartan matrices for
semi-simple Lie algebras. Besides the untwisted affine algebras, twisted affine
algebras also exist, but they are not so directly useful in conformal field
theory. 

\affdet\ guarantees that $\h A$ has no negative eigenvalues, and exactly one
zero eigenvector. For all affine Cartan matrices $\hat A$, there exist positive
integers $a_0,a_1,\ldots,a_r$, called {\it marks}, such that  
$\sum_{i=0}^ra_i\hat A_{i,j}=0$.  If $\hat A$ is affine, meaning it is the
Cartan matrix of an affine algebra, then so is $\hat A^T$ (their Dynkin
diagrams are obtained from each other by reversing their arrows). Because of
this, we also have $\sum_{j=0}^r\hat A_{i,j}a^\vee_j=0$, where the $a^\vee_j$
are known as {\it co-marks} (notice this is consistent with the
symmetrisability of $\hat g$). For untwisted $\hat g$, the marks and co-marks
are determined by the {\it highest root} $\theta$ of $g$, which is its own
co-root $\theta^\vee=\theta$ (here we use the normalisation convention
$\alpha^2=2$ for the longest roots $\alpha$). So we can
expand  \eqn\hroot{\theta\ =\ \sum_{i=1}^r\,a_i\alpha_i\ =\
\sum_{i=1}^r\,a_i^\vee\alpha_i^\vee\ ,} with the expansion coefficients
equalling the (co-)marks. $a_0=a_0^\vee=1$ completes their specification. 

Equivalently, the Dynkin diagrams of the untwisted affine $\hat g$
are simply the {\it extended} Dynkin diagrams  of the corresponding simple 
algebra $g$, obtained by augmenting the set of simple roots $\Pi$ of $g$ by
$-\theta$. The Dynkin diagrams of the untwisted affine algebras are drawn in
Fig. 9. So, the set of affine simple roots 
\eqn\Piaff{\hat \Pi\ =\ \{\,\hat \al_i\ :\ i\in\{0,1,\ldots,r\}\,\}\ } is 
simply related to $\{-\theta,\al_1,\ldots,\al_r\}$. 

\midinsert
\vskip1cm
\epsfxsize=10cm
\centerline{\epsfbox{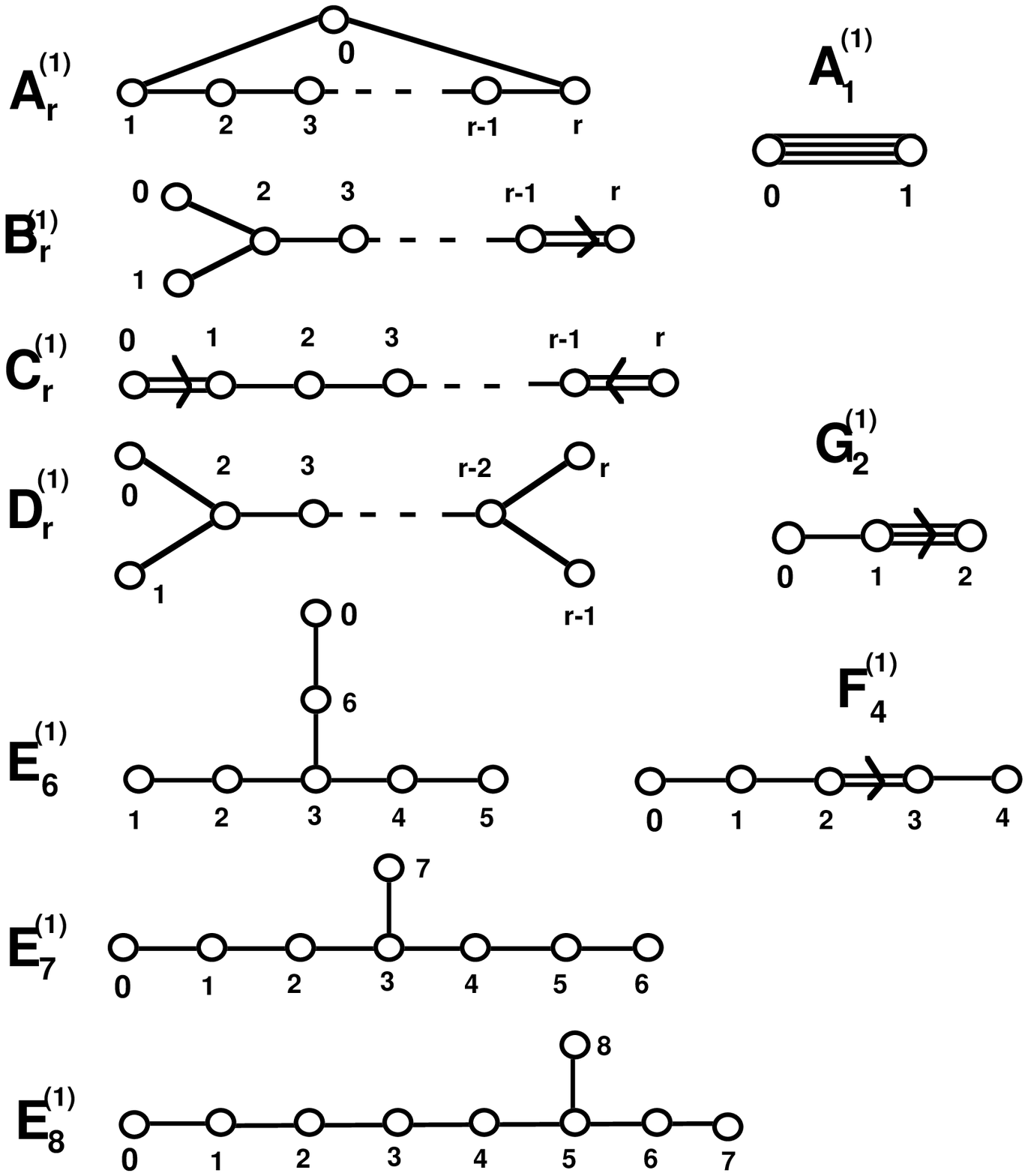}}
\smallskip\smallskip
\leftskip=1cm
\rightskip=1cm
\noindent
\baselineskip=12pt
{\it Figure 9}. The (Coxeter-)Dynkin diagrams of the affine untwisted 
Kac-Moody algebras.\bigskip \leftskip=0cm\rightskip=0cm \baselineskip=15pt  
\endinsert 

Notice that if any node of any of the Dynkin diagrams of Fig. 9 
is omitted, one obtains the Dynkin diagram of a semi-simple Lie algebra. This
is also true of the twisted affine Dynkin diagrams (not drawn here), and is in
agreement with the condition \affdet. It is only the untwisted Dynkin
diagrams, however, that are isomorphic to the extended diagrams of simple Lie
algebras.  

Before making this precise, we must introduce another operator. For an
arbitrary symmetrisable Kac-Moody algebra, one can define the inner product of
simple roots directly from the Cartan matrix $A$ by $(\al_i,\al_j)=A_{i,j}q_j$
($\,$recall $AD'$ is symmetric, with $D'={\rm diag}(q_i)\,$). This inner product
is positive (semi-)definite iff $A$ is of (affine) finite type. Now, many of the
important results for $\hat g$ are obtained as straightforward generalisations
of those for $g$, but these latter rely on having a non-degenerate bilinear
form $(\cdot,\cdot)$. To make the bilinear form non-degenerate in the affine
case,  one needs to enlarge the Cartan subalgebra $\hat g_0$ of $\hat g$, to
$\hat g_0^e$, by adding a derivation $d$. We will denote the {\it
enlarged affine algebra} similarly, by $\hat g^e$. 

The problem is the {\it canonical central element} 
\eqn\ccel{K\ =\ \sum_{i=0}^r\,a_i^\vee \h H^i\ .}
Clearly, $[K,\h H^j]=0$, and furthermore, 
\eqn\KEpmal{[K,E^{\pm\h\al_i}]\ =\ \pm\,\sum_{j=0}^r\,a_j^\vee\hat
A_{i,j}E^{\pm\h\al_i}\ =\ 0\ ,}
showing that $K$ is indeed central. Actually, the coefficient $k$ of the WZ term
\wz\ in the WZW action \action\ is to be identified with the eigenvalue of
$K$, which is fixed in the current algebra of a given WZW model. 

If we extend the bilinear form by choosing $\K(K,d)=1$, $\K(K,K) = \K(d,d) =
0$, the resulting form is non-degenerate. The operator $d$ is
very natural in WZW models: $-d$ will be identified with the Virasoro zero mode
$L_0$. 

If a step operator $E^{\hat\al}$ is an element of $\hat g$, with
\eqn\Ehal{\eqalign{[\h H^i,E^{\hat\al}]\ =\ \al^i E^{\hat\al}&\ \ \ 
(i\in\{1,\ldots,r\})\ ,\ \cr [K,E^{\hat\al}]\ =\ 0\ ,&\ \  \ [d,E^{\hat\al}]\
=\ n\ \ ,\cr}} we denote $\hat\al=(\al,0,n)$, and write $E^{\hat\al}=:E_n^\al$.
So one can think of affine roots as vectors with $r+2$ components, $r$ of which
describe a root of $g$, and the other two correspond to the elements $K,d\in
\hat g^e$.  The inner product on $\h g^{e*}_0$ is 
\eqn\iphg{\boxEq{(\hat\al,\h\beta)\ =\
\big(\,(\al,k_\al,n_\al),(\beta,k_\be,n_\be)\,\big)\ =\ (\al,\be)\
+\ k_\al n_\be\ +\ n_\al k_\be\ \ .}} 
It is determined by the symmetry and invariance:
\eqn\Kinv{\K(x,[z,y])\ +\ \K([z,x],y)\ =\ 0\ \ \ (\forall x,y,z\in\h g^e_0)\ ,}
of the corresponding bilinear form on $\h g_0^e$. Notice that $k_\al,n_\al$
behave like light-cone coordinates $x_\pm=(t\pm x)/\sqrt{2}$ in a Minkowski
metric, so the signature of the inner product on $\h g_0^{e*}$ is
Lorentzian. 

Notice that $\delta=\sum_{i=0}^r
a_i\h\al_i = (0,0,1)$, so that $\delta$ is the root corresponding to $d=-L_0\in
\h g_0^e$. The dual weight is  denoted $\Lambda_0=(0,1,0)$. 

The affine simple roots are 
\eqn\affsr{\boxEq{\h\al_0\ =\ (-\theta,0,1)\ ,\ \ \ \h\al_{i\not=0}\ =\
(\al_i,0,0)\ \ .}}
This explains why the extended Dynkin diagram of $g$ is
identical to the Dynkin diagram for $\h g$.

The fundamental weights are
\eqn\affwts{\boxEq{\h\ome^0\ =\ (0,1,0)\ =\ (0,a_0^\vee,0)\ ,\ \ \ \h
\ome^{i\not=0}\ =\ (\ome^{i\not=0},a^\vee_{i\not=0},0)\ \ .}}
For an arbitrary affine weight $\h\la=(\mu,\ell,n)$, $\ell$ is called the
level of the weight, and $n$ is called its grade. In the WZW context, the
level of weight vectors will usually be fixed by the WZ coefficient $k$, and
the grade is directly related to the eigenvalue of the Virasoro zero mode
$L_0$. For $\h\la$ as just written, we will adopt the notational convention
that $\la=\mu$; if the ``hat'' is removed from an affine weight, the result is
the horizontal projection, or ``finite part'' of it. This is consistent with
\affsr\ and \affwts, and also allows us to write $\ome^0=0$, $\al_0=-\theta$,
for examples. We also use $\phi_\la$ to denote $\phi_{\h\la}$ (and have so
already); so that $\phi_{k\h\ome^0}=\phi_0$, for instance. 

This notational convention also allows us to write
\eqn\lahla{\h\la_i\ =\ \la_i\ ,\ \ \ \forall\ i\in\{1,\ldots,r\}\ .}

These are the affine roots and weights. What about the affine Weyl group $\h
W$? It is also the Coxeter group associated with the corresponding Dynkin
diagram, generated by the primitive Weyl reflections
\eqn\affpw{\h r_i\h\la\ =\ \h\la\ -\ (\h\la,\h\al^\vee_i)\,\al^\vee_i\ .}
Suppose $\h\la=(\la,k,n)$, then this gives
\eqn\rhla{\eqalign{\h r_i\h\la\ =&\ (r_i\la,k,n)\ ,\ \ \ i\not=0\ ;\cr
\h r_0\h\la\ =&\ \h\la-\big[k-(\la,\theta)\big]\,\h\al_0\ =\
\big(\,\la+\big[\,k-(\la,\theta)
\,\big]\,\theta,\, k,\, n-\big[\,k -(\la,\theta)\,\big]\, \big)\ .}}
Notice that $k-(\la,\theta)$ plays the role of $\h \la_0$. This is justified
by $(\delta,\h\la)=\sum_{i=0}^ra_i^\vee\h\la_i =
\h\la_0+\sum_{i=1}^ra^\vee_i\la_i = \h\la_0 + (\la,\theta)$, which should be the
eigenvalue of $K$, \ie\ the level. 

Consequently, we sometimes use $\la_0:= k- (\la,\theta)$. So \lahla\ can be
extended to include $i=0$, once the level $k$ of an affine weight has been
fixed, as it is in WZW models. 

The relation between $\h W$ and $W\subset\h W$ is found by calculating
$r_{\h\al}\h\la$ for $\h\al=(\al,0,m)$. One gets $r_{\h\al}=r_\al(t_\al)^m$,
where
\eqn\tall{t_\al\h\la\ =\
\big(\,\la+k\al^\vee,k,n+\la^2-(\la+k\al^\vee)^2/2k\,\big)\ .}
$t_\al t_\be=t_\be t_\al$, so $\<t_\al\>=T_{kQ^\vee}$, the translation group
in the (scaled) co-root lattice $Q^\vee$ of $g$. Furthermore, $r_\be t_\al
r_\be^{-1} = r_\be t_\al r_\be = t_{r_\be(\al)}$, so $T_{kQ^\vee}$ is a normal
subgroup of $\h W$, and 
\eqn\WhWT{\boxEq{\h W\ =\ W\,\ltimes\,T_{kQ^\vee}\ .}}
This relation has important implications for the modular properties of
affine characters, as we'll see.

The geometry of affine Weyl hyperplanes can be compared to that for the Weyl
hyperplanes of $g$, at least after the horizontal projection 
$\h\la=(\la,k,n)\,\mapsto\,\la$. The situation is analogous, with sectors of
weight space labelled by elements of $\h W$. But this time the sectors are of
finite volume, and there  is an infinite number of them, since $|\h W|=\infty$.
See Fig. 10 for a depiction of the case $g=A_2$. 

\midinsert
\vskip-0cm
\epsfxsize=8cm
\centerline{\epsfbox{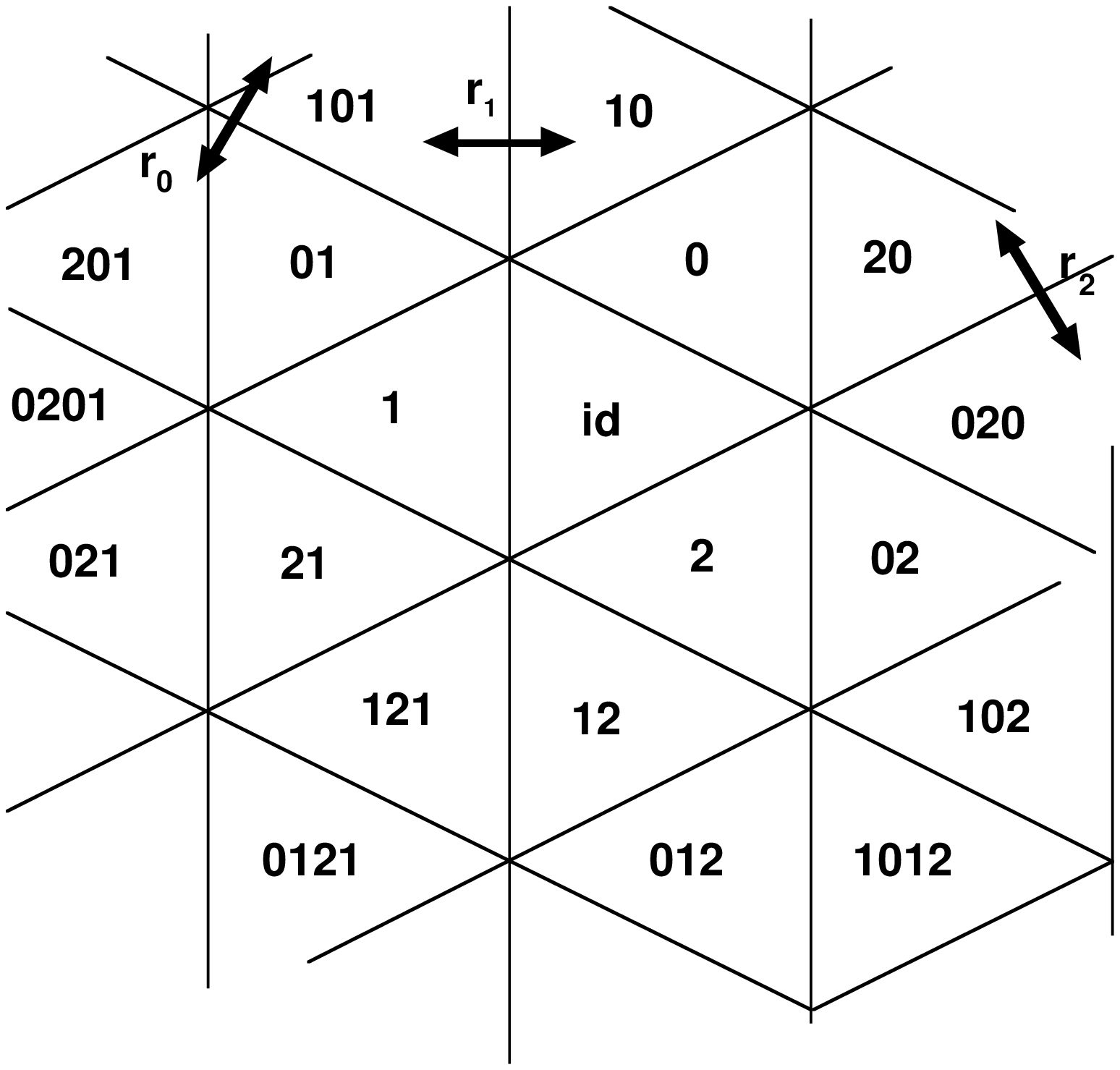}}
\smallskip\smallskip
\leftskip=1cm
\rightskip=1cm
\noindent
\baselineskip=12pt
{\it Figure 10}. The affine Weyl sectors of the horizontal projection of
$\h A_2$ (affine $A_2$) weight space.\bigskip \leftskip=0cm\rightskip=0cm
\vskip0cm\baselineskip=15pt   \endinsert 

This fact is highly suggestive: the integrable highest weights for $g$ are
those integral weights ($\la=\sum_{i=1}^r\la_i\ome^i$ with all $\la_i\in\Z$)
contained in the sector labelled by $\id\in W$, so we expect the integrable
affine highest weights to be finite in number. How does this happen?

First, if $\h\la = (\la,k,n)$ is to be a highest weight for an integrable
representation of $\h g$, then $g\subset\h g$ implies $\la$ must be one for
$g$. So $\la\in P_+=\{\mu=\sum_{i=1}^r\mu_i\ome^i\ :\ \mu_i\in\Z_{\geq 0}\}$, 
ensuring
that each $A_1$ subalgebra $\< e_i,h_i,f_i \> = \<
E_0^{\al_i},H^i_0,E_0^{-\al_i} \>$ ($i\in\{1,\ldots,r\}$) is represented
integrably, \ie\ has $\la_i=2j$ with ``isospin'' $j\in\Z_{\geq 0}/2$. The extra
condition is simply that the $A_1$ subalgebra corresponding to the simple root
$\al_0$ be represented integrably. This just means $\la_0\in\Z_{\geq 0}$ is 
required,
\ie\ $k-(\la,\theta) = k - \sum_{i=1}^r\la_ia^\vee_i \in\Z_{\geq 0}$. In 
other words, 
\eqn\Ppk{\h\la\ \in\ P_+^k\ =\ \bigg\{\,\h\la=\sum_{i=0}^r\la_i\h\omega^i\ :\
\la_i\in\Z_{\geq 0},\ \sum_{i=0}^r\la_ia_i^\vee = k \,\bigg\}\ ,}
explaining why there is a finite number of affine integrable highest weights at
fixed level $k$. If the $r+1$ simple-root $A_1$ subalgebras of $\h g$ are all
represented integrably, that turns out to be sufficient to guarantee that the
whole of $\h g$ is so represented. We also write 
\eqn\Ppkbar{{\overline{P_+^k}}\ =\
\bigg\{\, \la=\sum_{i=1}^r\la_i\omega^i\ :\ \la_i\in\Z_{\geq 0},\
\sum_{i=1}^r\la_ia_i^\vee \leq k \,\bigg\}\ \subset\ P_+\ \ }
for the set of horizontal projections of integrable affine highest weights at
fixed level $k$.

Integrability is signalled by the presence of null vectors, vectors (states)
of zero norm. For example, with $g=A_1$, and highest state $|v_\la\>$ with
$\la=\la_1\ome^1=2j\ome^1$, one finds the null states $e_1|v_\la\>$ (from the
highest-state condition) and $f_1^{\la_1+1}|v_\la\>=f_1^{2j+1}|v_\la\>$. See
Fig. 11 for the example of $j=3/2$, \ie\ the $A_1$ representation
$L(3)$. The existence of null vectors (and so integrability) goes hand-in-hand
with the Weyl symmetry of representations. 

\midinsert
\vskip.5cm
\epsfxsize=8cm
\centerline{\hskip-1cm\epsfbox{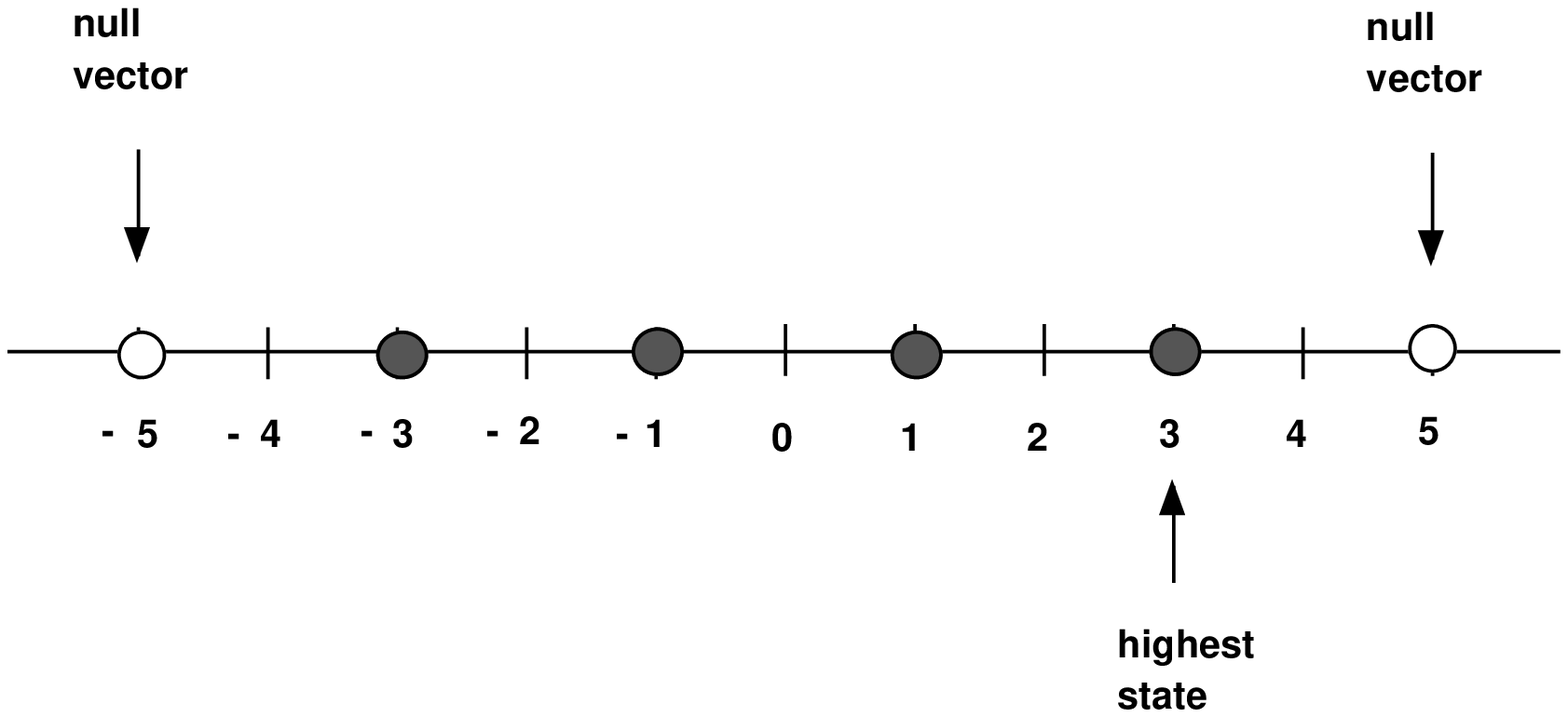}}
\smallskip\smallskip
\leftskip=1cm
\rightskip=1cm
\noindent
\baselineskip=12pt
{\it Figure 11}. The weight diagram for the $A_1$ representation
$L(3\ome^1)$, corresponding to angular momentum $j=3/2$. The weights of the null
vectors are shown. \bigskip \leftskip=0cm\rightskip=0cm \baselineskip=15pt  
\endinsert

For the integrable highest-weight representations of $\h g$, with highest
weight state satisfying $e_i|v_{\h\la}\>=0$ for all $i\in\{0,1,\ldots,r\}$,
there are $r+1$ {\it primitive null vectors}  
$|\eta_i\>:=f_i^{1+\la_i}|v_{\h\la}\>$. Primitive here means that all other
null vectors (an infinite number of them) can be obtained as descendants of
these ones. Because of \affsr, these are of exactly the same form as the
primitive null vectors of the integrable $g$ representation of highest weight
$\la$, for $i\not=0$. The additional ($i=0$) primitive null vector has
interesting consequences in the WZW model, as we'll see. 

The presence of the ``extra'' null vector is also consistent with the
enlargement of the Weyl symmetry $W \rightarrow \h W$. Consider the {\it affine
formal character}  \eqn\affch{\eqalign{\ch_{\h\la}\ :=&\ \Tr_{L(\h\la)}\big(
e^{\h H}  \big)\ \cr =&\
\sum_{\h\mu\in P(\h\la)}\, \mult(\h\la;\h\mu)\, e^{\h\mu}\ \ ,\cr}}
where, $\h H=\sum_{i=0}^r \h\ome^i h_i$, 
$\mult(\h\la; \h\mu)$ denotes the multiplicity of the weight $\h\mu$ in
$L(\h\la)$, and we define  \eqn\Phla{P(\h\la)\ :=\ \{\, \mu \in \h g^{e*}_0\
:\ \mult(\h\la; \h\mu)\not=0 \,\}\ .} 
Then it is the {\it Weyl-Kac formula} that makes manifest the affine Weyl
symmetry of affine characters:
\eqn\wkcf{\boxEq{\ch_{\h\la}\ =\ {{\sum_{\h w\in \h  W}\, (\det\, \h w)\,
e^{\h w.\h\la} } \over  {\prod_{\h\al\in\h \Delta_+}\, (1-e^{-\h \al}) }}\ ,}} 
where $\h w.\h\la$ indicates the shifted action of $\h w$: $\h w.\h\la 
:= \h w(\h\la+ \h\rho) -\h\rho$, with $\h\rho = \sum_{j=0}^r
\h\ome^j$. $\hat\Delta_+$ indicates the set of positive roots of $\h g$, to be
specified shortly. $\ch_{\hat 0}=1$ leads to the affine  denominator formula
\eqn\afdenom{ {\prod_{\h\al\in\h \Delta_+}\, (1-e^{-\h \al}) }\ =\  {\sum_{\h
w\in \h  W}\, (\det\, \h w)\, e^{\h w \h\rho}}\ ,} so that \wkcf\ can also be
written as  \eqn\wkcfi{\ch_{\h\la}\ =\ {{ \sum_{\h w\in \h  W}\, (\det\, \h w)\,
e^{\h w(\h\la+\h\rho)}  }\over{ \sum_{\h w\in \h  W}\, (\det\, \h w)\, e^{\h w
\h\rho}  }}\ .}

One integrable affine highest weight is special: $\h\la=k\h\ome^0$ has
horizontal projection $\la=0$. This indicates that it corresponds to a state
that is $G$-invariant; this state is the vacuum $|0\>$. The corresponding field
$\phi_{k\h\ome^0}=\phi_0$ is known as the {\it identity primary field}, because
of its action on the vacuum:
\eqn\idpf{\phi_{k\h\ome^0}(0)\,|0\>\ =\ \phi_0(0)\,|0\>\ =\ |v_{k\h\ome^0}\>\ =\
|v_0\>\ =\ |0\>} (see \philv).  Now, more on the integrable highest-weight
representations of $\h g$  (they are sometimes called the {\it standard
representations}, for short). 

In terms of
Chevalley generators, the highest state $|v_{\h\la}\>$ is defined by
$e_i|v_{\h\la}\>=0$, for all $i\in\{0,1,\ldots,r\}$. But the highest state is
annihilated by the raising operator corresponding to any positive root. So,
using the Cartan-Weyl presentation, we have 
\eqn\affhwcw{E_0^\al|v_{\h\la}\>\ =\ 0\ ,\ \forall\ \al\in\Delta_+\ \ ;\ \ \ 
J^\al_n|v_{\h\la}\>\ =\ 0\ ,\ \forall\ \al\in\Delta\ ,\ n\in\Z_{> 0}\ ,}
where $J^\al_n\in\{E^{\pm\al}_n,H^\al_n\}$. This then points to the appropriate
choice of the set $\h\Delta_+$ of positive roots of $\h g$:
\eqn\hDpl{\h\Delta_+\ =\ \Delta_+\ \cup\ \{n\delta\ :\ n\in\Z_{> 0}\}\ \cup\
\{\al+n\delta\ :\ \al\in\Delta,n\in\Z_{> 0}\}\ .}
The full set of roots is
$\h\Delta=\h\Delta_+\cup\h\Delta_-=\h\Delta_+\cup(-\h\Delta_+)$. All roots
except 0 and the {\it imaginary} ones $\{n\delta\ :\ 0\not=n\in\Z\}$ have unit 
multiplicity, and each relates to a single element of $\h g$:
$E^{\al+n\delta}=E_{-n}^\al$. $n\delta$ (including $0$) has multiplicity $r$,
relating to the existence of the $r$ elements $H_{-n}^i$. 

So, the generators of $\h g$ can be written to emphasise their similarity with
those for $g$: one simply adds ``mode numbers'' as subscripts to the symbols
for the generators of $g$. Then their commutation relations also take a form
that is simply related to those  for $g$, written in \HH,\root,\EaEb:
\eqn\affcwm{\boxEq{\eqalign{&[H^i_{m},H^j_{n}]\ =\
km\delta_{m+n,0}\delta^{i,j}\ \cr &[H^i_{m},E^\al_n]\ =\
\al^i\,E^\al_{m+n}\ \cr [E^\al_m,E^\be_n]\ =\
&\left\{\matrix{\al^\vee\cdot H_{m+n}+km(2/{\al^2})\delta_{m+n,0}\ \ ,&\ \
\al+\be=0\cr N(\al,\be)E^{\al+\be}_{m+n}\ \ , &\ \ \al+\be\in\Delta\ \cr
0\ \ \ , &\ \ \al+\be\not\in\Delta\ .\cr }\right.
}}}
Here of course, $\al,\be\in\Delta$, and $m,n\in\Z$. 

By the Sugawara construction, $|v_{\h\la}\>$ is also the highest weight of a
representation of $Vir$:
\eqn\Lnvla{L_n|v_{\h\la}\>\ =\ \sum_{m\in\Z}\,
N(J_m^aJ^a_{n-m})\,|v_{\h\la}\>\ = 0\ , \ \ \forall n\in\Z_{> 0}\ .}
We also have $L_0|v_{\h\la}\>=h_\la|v_{\h\la}\>$, as noted above, with conformal
weight $h_\la$ given in \hla. Such an irreducible representation is not
irreducible as a representation of $Vir$; rather, it decomposes into an
infinite number of such representations. 

The highest state is nevertheless the highest state of an irreducible
representation of $Vir$. So, by the state-field correspondence, the $\h
g$-primary field also transforms as a $Vir$-primary field: under the conformal
transformation $z\rightarrow w=w(z)$, an analytic function of $z$, and  
\eqn\virphi{\phi_\la(z)\ \rightarrow\ \phi_\la(w)\ =\ \left({{dw}\over{dz}} 
\right)^{-h_\la}\,\phi_\la(z)\ .} So a $Vir$-primary
field transforms in a tensorial way under conformal transformations. 

Of particular use to us are the so-called {\it projective transformations},
where \eqn\projtr{w\ =\ {{az+b}\over{cz+d}}\ ,\ \ {\rm with}\  ad-bc=1\
.} Writing $a,b,c,d$ as the elements of a $2\times 2$ matrix shows that these
transformations form a group isomorphic to $PSL(2,\C)$: $P$ stands for
projective, meaning the matrix and its negative describe equivalent
transformations \projtr; $S$ stands for special, \ie\ the matrix has
determinant one; and $L$ means linear. The projective transformations are 
the only (invertible) conformal transformations that map the entire complex
plane plus the point at $\infty$ to itself. They
leave the vacuum invariant:  \eqn\projvac{L_{\pm 1}\,|0\>\ =\ L_0\,|0\>\ =\ 0\
,} since the $L_{\pm 1},L_0$ generate the $s\ell(2,\C)$ algebra of the
projective group. 

For more details, consider infinitesimal conformal transformations, \ie\ 
$w=z+\epsilon(z)$, with $|\epsilon(z)|\ll 1$. \virphi\ then yields
\eqn\dephie{\delta\phi_\la(z)\ =\ \big(\epsilon(z)\di_z\ +\
h_\la\epsilon'(z)\big) \phi_\la(z)\ .} If we don't restrict $\epsilon(z)$ 
further, we are considering general infinitesimal conformal transformations.
From their general form \projtr, one can see that
infinitesimal projective transformations give 
\eqn\epsqu{\epsilon(z)\ =\ c_{-1}\ +\ c_0z\ +\ c_1z^2\ ,}
where $c_{\pm1},c_0$ are constants. We write  
\eqn\depeL{\delta\,\phi_{\la}(z)\ =\
\sum_{m=-1}^1\,c_m\,[L_m,\phi_\la(z)]\ ,}
and find 
\eqn\Lmphi{[L_m,\phi_\la(z)]\ =\ \big(z^{m+1}\di_z\ +\ (m+1)h_\la
z^m\big)\phi_\la(z)\ ,}
with $m=\pm 1,0$. This last formula is consistent with the commutation 
relations \Vir\ of $Vir$, for the modes $L_{\pm 1,0}$, and shows they do
generate  $s\ell(2,\C)\subset Vir$. 

Note that $L_{-1}$ acts as the $z$-translation operator and $L_0$ as the
generator of dilations: \eqn\Lmitr{e^{aL_{-1}}\,\phi_\la(z)\, e^{-aL_{-1}}\ =\
\phi_\la(z+a)\ ,\ \ \  \ e^{aL_{0}}\,\phi_\la(z)\, e^{-aL_{0}}\ =\
\phi_\la(e^az)\ .} Including both the holomorphic and antiholomorphic parts
of the primary field, this last equation gives
\eqn\LzLzb{e^{aL_0+\bar a\bar L_0}\,\Phi_{\la,\mu}(z,\zb)\, 
e^{-aL_0-\bar a\bar L_0}\ =\ \Phi_{\la,\mu}(e^az,e^{\bar a}\zb)\ =\
e^{-ah_\la-\bar a h_\mu}\,\Phi_{\la,\mu}(z,\zb)\ ,}
 using \virphi, where $\bar a$ denotes the complex conjugate of $a$. Putting
$a=\alpha+i\theta$, with $\al,\theta\in\R$, we confirm that $L_0+\bar L_0$ is
the generator of dilations (radial Hamiltonian) and $L_0-\bar
L_0$ is the generator of rotations. Furthermore, 
$h_\la+h_\mu$ is the scaling dimension of $\Phi_{\la,\mu}$ and
$h_\la-h_\mu$ is its spin. 

The third generator $L_1$ generates what are known as special
conformal transformations. All three types of transformations (translations,
rotations and special conformal transformations) are conformal in any number 
$N$ of dimensions. If we restrict the base field to $\R$, instead of $\C$, we
have an algebra $s\ell(2,\R)$. The antiholomorphic counterparts $\bar
L_{\pm 1},\bar L_0$ generate another copy of this algebra, and the direct
sum of the two copies is isomorphic to a real form of $so(4)$. In $N$
dimensions, the translations, rotations and special conformal transformations
generate a real form of $so(N+2)$. In $N=2$ the symmetry extends to an
infinite-dimensional one, with infinite-dimensional algebra \Witt. After
central extension, we find $Vir$.  

A simple example of a standard representation of $\h g=\h A_1=A_1^{(1)}$ is
depicted in Fig. 12. There the weights in $\h g_0^{e*}$ are drawn
(except that the fixed eigenvalue $k=2$ of $K$ is not indicated as a
coordinate). Note that the ``horizontal'' weight spaces are those for the
simple Lie algebra $g=A_1$; in general, the horizontal subspaces of a
representation $L(\h\la)$ of $\h g$ will be (reducible) representations of $g$.
(This is where the term horizontal subalgebra $g\subset\h g$ comes from.)
In particular, for the standard representation $L(\h\la)$, the horizontal
representation of lowest $L_0$ eigenvalue is the irreducible representation
$L(\la)$ of $g$. Notice also that the weight diagram is enclosed by a 
parabolic envelope: a parabola passes through all weights $\h \mu$ such that
$\h \mu-\delta$ is not also in the diagram. Its curvature
decreases with increasing level. The parabola becomes a paraboloid for
higher rank algebras. The simple roots are indicated, as well as the weights of
the primitive null vectors. The multiplicities of the weights rise
rapidly with increasing $L_0$ eigenvalue $n$; asymptotically $\mult\big(\la;\,
(\mu,k,-n)\,\big)\sim n^{-3/4}\exp({\rm const.}n^{1/2})$. 

\midinsert
\vskip.5cm
\epsfxsize=8cm
\centerline{\epsfbox{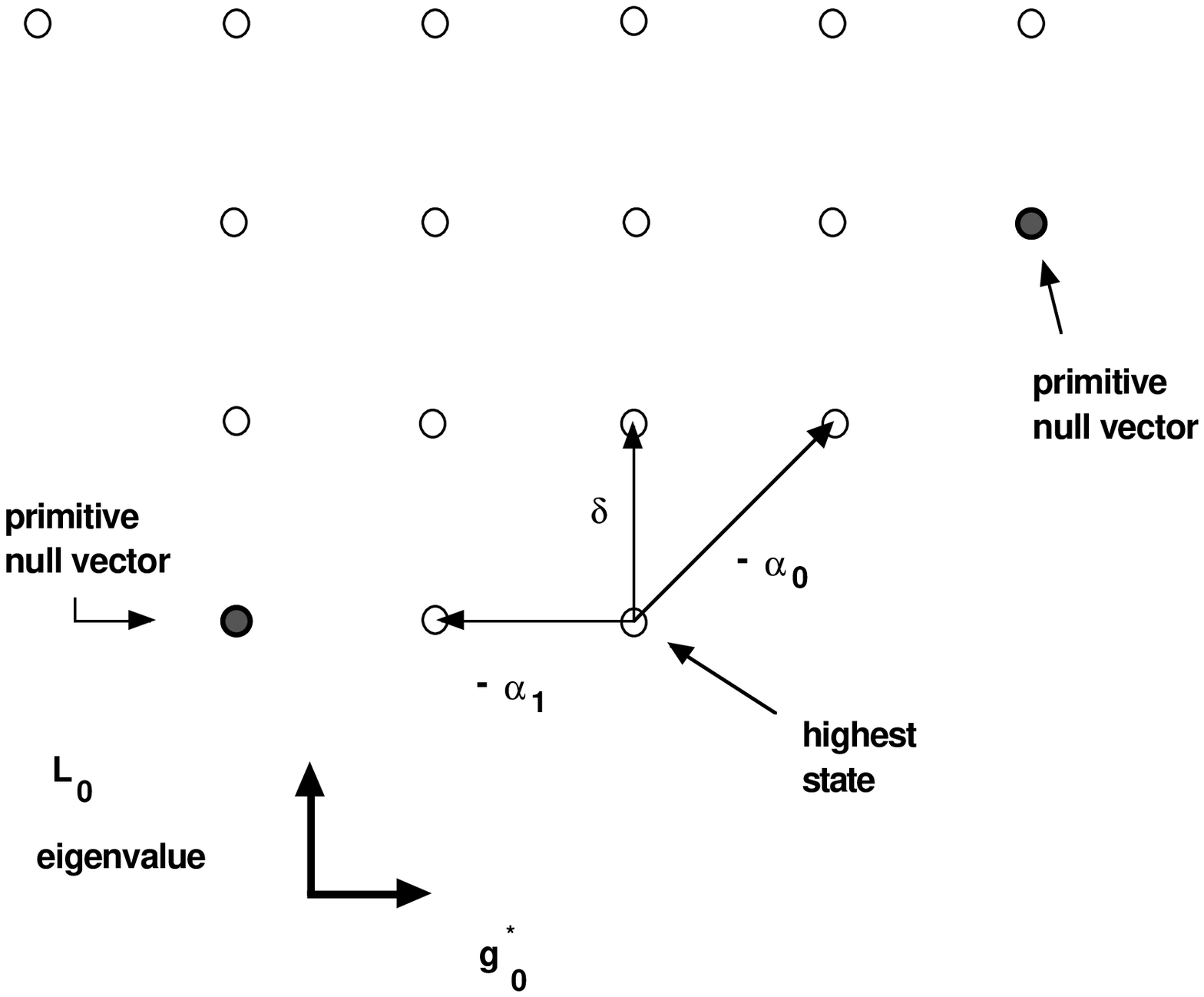}}
\smallskip\smallskip
\leftskip=1cm
\rightskip=1cm
\noindent
\baselineskip=12pt
{\it Figure 12}. The weight diagram of the $\h A_1$ representation
$L(\h\ome^0+\h\ome^1)$. \bigskip \leftskip=0cm\rightskip=0cm
\baselineskip=15pt  
\endinsert 

Fig. 13 shows the standard representations for arbitrary affine
algebras in very schematic fashion. There is a finite number (the case of 
$\card P_+^k = 3$ is drawn) of such representations, each to be associated with
a primary field in the corresponding WZW model. As mentioned above, the
representation $L(k\h\ome^0)$ is special among them: its lowest horizontal
representation is $L(0)$, the scalar representation, and its lowest $L_0$
eigenvalue is the lowest of the low. That's because the single state in the
representation $L(0)$ is to be identified with the vacuum of the WZW model. The
corresponding primary field is called the {\it identity primary field}. The
other standard representations have lowest horizontal representations of higher
dimensions, and lowest $L_0$ eigenvalues that are higher than that of the
vacuum; after all, ${\bf H}=L_0+\bar L_0$, so the vacuum should have lowest
energy. These last two effects go hand-in-hand, as the diagram is meant to
indicate. 

\midinsert
\vskip.5cm
\epsfxsize=8cm
\centerline{\hskip-2cm\epsfbox{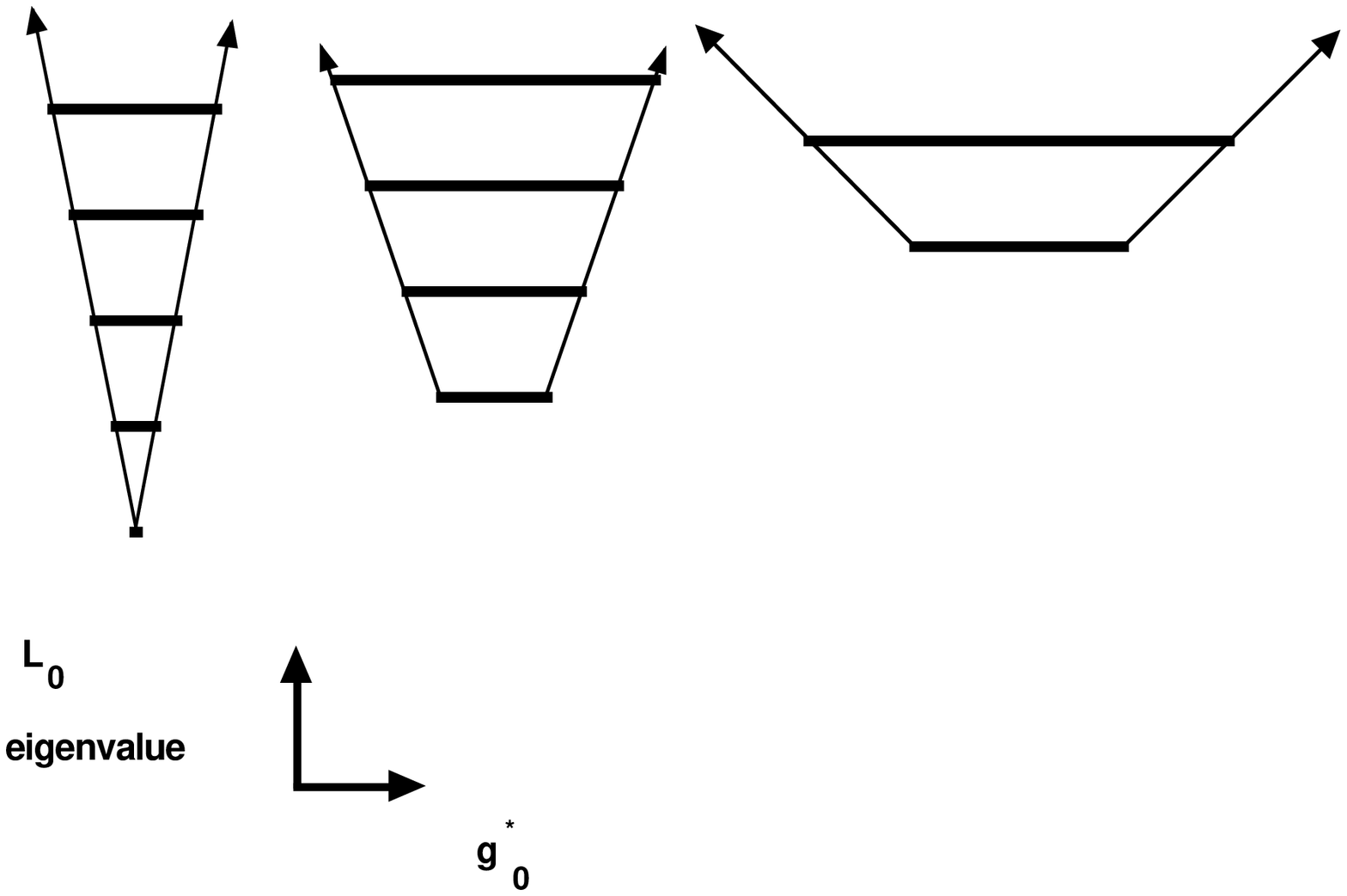}}
\smallskip\smallskip
\leftskip=1cm
\rightskip=1cm
\noindent
\baselineskip=12pt
{\it Figure 13}. Schematic drawing indicating the standard affine
representations that are relevant to a WZW model.\bigskip
\leftskip=0cm\rightskip=0cm \baselineskip=15pt   \endinsert

\LmJn\ shows that elements of $g$ commute with $L_0$. Since ${\bf H}=L_0+\bar
L_0$, this means that $g\oplus g$ is a true symmetry algebra of the WZW model.
The full affine algebra $\h g\oplus \h g$ plays the role of a
spectrum-generating algebra in the theory, generating all the states in the
towers corresponding to the primary fields.  

The remarkable thing is that the states in 
these $\card P_+^k$ primary towers span the space of states of the WZW model!
$\h g\oplus \h g$ generates the full spectrum of the model from the $\card
P_+^k$ primary highest states. We can therefore say that the the
infinite-dimensional affine algebra effectively ``finitises'' the WZW field
theory: one need only study the finite number of primary fields.

\newsec{Affine Algebra Representations and WZW Models}

In the last section we laid the basis for the application of the
representation theory of untwisted affine algebras to WZW models. In this
section we'll describe some specific results in detail. 

\subsec{Gepner-Witten equation}

Null vectors constrain the possible couplings between WZW fields. Consider a
primary field realising a standard representation of $\h g$, with highest
weight $\h\nu$. That is, the primary field has holomorphic part
$\phi_{\h\nu}(z)$. This implies $f_i^{1+\nu_i}\phi_{\h\nu}=0$, for all
$i\in\{0,1,\ldots,r\}$. For $i=0$, this can be rewritten as 
\eqn\nullz{\big(E_{-1}^\theta\big)^{1+\h\nu_0}\,\phi_{\h\nu}(z)\ =\ 0\ .}
Now, suppose this null field appears in a correlation function with the
primary fields $\phi_1(z_1),\ldots, \phi_n(z_n)$; the correlator must then
vanish: \eqn\ncorr{\<\,
\big[\big(E_{-1}^\theta\big)^{1+\h\nu_0}\phi_{\h\nu}(z)\big] \phi_1(z_1) 
\cdots \phi_n(z_n)\,\>\ =\ 0\ .}
After using \JintJ\ to rewrite last equation as
\eqn\ncorrw{\<\, {1\over{2\pi i}}
\oint_z{{dw}\over{w-z}}\left(E^\theta(w)\,
\big[\big(E_{-1}^\theta\big)^{\h\nu_0}\phi_{\h\nu}(z)\big]\right) \phi_1(z_1) 
\cdots \phi_n(z_n)\,\>\ =\ 0\ ,} we can deform the contour of integration in the
manner indicated in Fig. 14  to get 
\eqn\ncors{\eqalign{0\ =\ \sum_{j=1}^n\, {1\over{2\pi i}}
\oint_{z_j}{{dw}\over{w-z}}&\,  \<\,
\big[\big(E_{-1}^\theta\big)^{\h\nu_0}\phi_{\h\nu}(z)\big] \phi_1(z_1)  \cdots
\cr &\cdots\big[E^\theta(w)\phi_j(z_j)\big] \cdots \phi_n(z_n) \phi_n(z_n)\,\>\
.\cr}} Now since the $\phi_j$ are primary, by \JPhi\ we can write
\eqn\ncorsi{0\ =\ \sum_{j=1}^n\, {{t_j^\theta}\over{z-z_j}}\,  \<\,
\big[\big(E_{-1}^\theta\big)^{\h\nu_0}\phi_{\h\nu}(z)\big] \phi_1(z_1)  \cdots
\phi_n(z_n)\,\>\ , } where the $j$ on $t^\theta_j$ indicates that the generator
should act on $\phi_j$. If this process is repeated, we find
\eqn\gepwit{\boxEq{0\ =\ \sum_{{{\{\ell_1,\ldots,\ell_n\}}\atop{\sum \ell_i =
1+\h\nu_0}}}\ 
{{(t_1^\theta)^{\ell_1}/\ell_1!}\over{(z-z_1)^{\ell_1}}}\,\cdots\, 
{{(t_n^\theta)^{\ell_n}/\ell_n!}\over{(z-z_n)^{\ell_n}}}\, 
\<\,\phi_{\h\nu}(z)\phi_1(z_1)\cdots \phi_n(z_n)\,\>\ .}}
This is the {\it Gepner-Witten equation} \ref\gw{D. Gepner, E. Witten, 
Nucl. Phys. {\bf B278} (1986) 493}. Notice that it also holds
if we replace $1+\h\nu_0$ with any $p\geq 1+\h\nu_0$. 

\midinsert
\vskip0cm
\epsfxsize=8cm
\centerline{\epsfbox{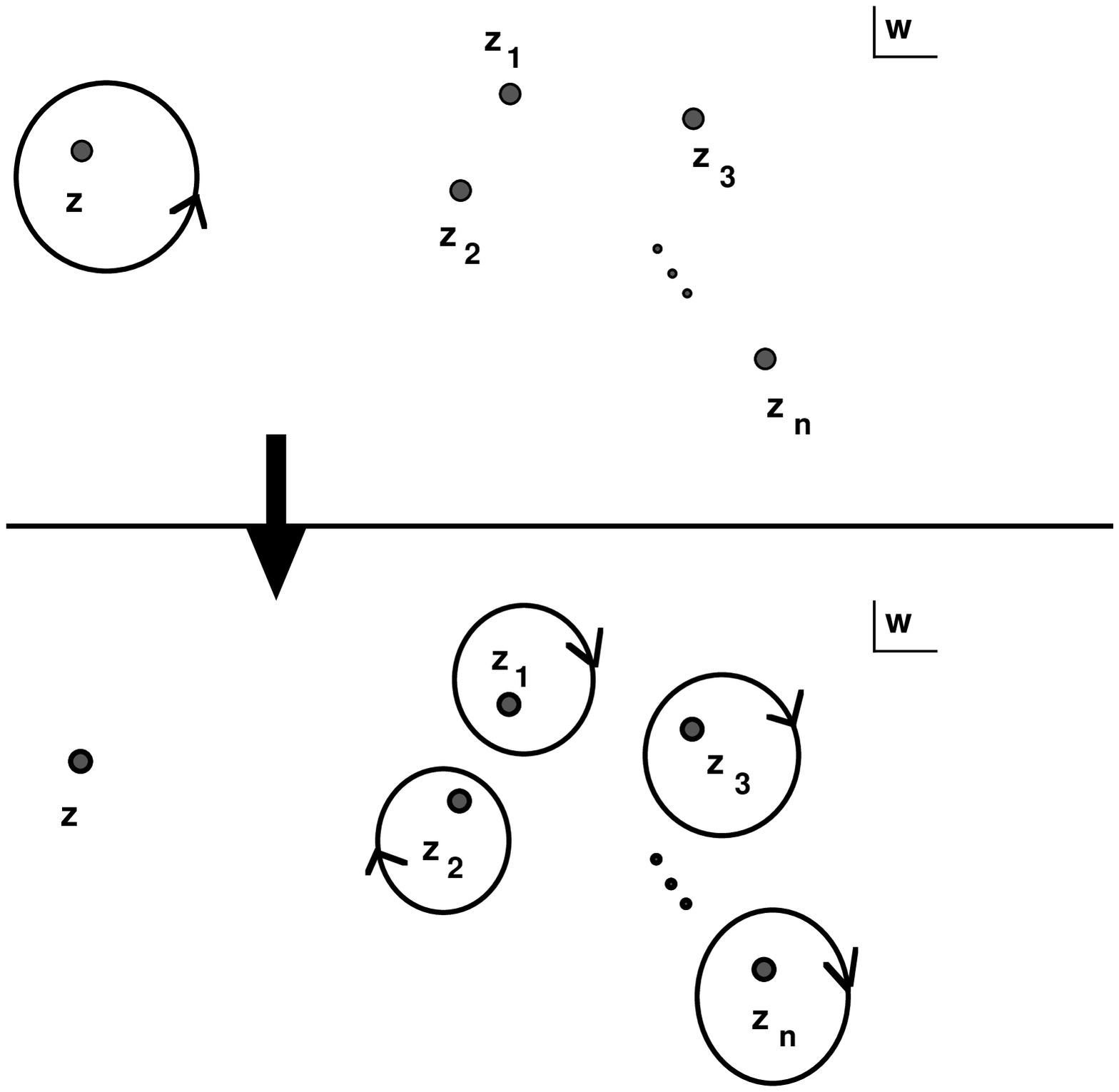}}
\bigskip
\leftskip=1cm
\rightskip=1cm
\noindent
\baselineskip=12pt
{\it Figure 14}. Contour deformation for \ncors. \bigskip
\leftskip=0cm\rightskip=0cm
\baselineskip=15pt  
\endinsert

The first consequence of \gepwit\ is that a non-integrable field  has
vanishing correlators with integrable fields, \ie\ a non-integrable field
decouples from integrable ones. To see this, let $\phi_{\h\nu}$ be the
identity field in \gepwit, that is, set $\h\nu=k\h\ome^0$. Then 
\eqn\phivac{\phi_{\h\nu}(z)|0\>\ =\ e^{zL_{-1}}\phi_{k\h\ome^0}(0)
 e^{-zL_{-1}} |0\>\ =\ |0\>\
,}
using \Lmitr,\projvac,\idpf. 
So we can remove $\phi_{\h\nu}(z)$ from \gepwit. Multiply by $(z-z_n)^{p-1}$
and integrate over $z$ to get
\eqn\gwi{\eqalign{0\ =&\ 
\<\,\phi_1(z_1)\cdots\phi_{n-1}(z_{n-1})
(t_n^\theta)^p\phi_n(z_n)\,\>\cr\ =&\
\<\,\phi_1(z_1)\cdots\phi_{n-1}(z_{n-1})
\big[\big(E_0^\theta\big)^p\phi_n(z_n)\big]\,\>\ ,\cr}} 
for all $p\geq1+\h\nu_0$. Now suppose $\phi_n$ is non-integrable, so that
there is no $p\in\Z_{> 0}$ such that $[\big(E_0^\theta\big)^p\phi_n]=0$. Then \gwi\
will only be satisfied if $\<\phi_1(z_1)\cdots\phi_n(z_n)\>=0$, \ie\ if the
non-integrable field $\phi_n$ decouples from the integrable ones
$\phi_1,\ldots,\phi_{n-1}$. 

\subsec{3-point functions} 

The second application of \gepwit\ will be to the 3-point
correlation functions $\<\phi_\la\phi_\mu\phi_\nu\>$ of primary fields. The 
3-point functions encode the  structure constants of the operator
product algebra, the OPE coefficients (see below). 
So the 3-point functions are arguably the most important, since in
principle, any $n$-point correlation function can be constructed using the
operator product algebra. 

The 3-point functions are highly constrained by global
($z$-independent) $G\otimes G$ invariance, and invariance under the projective
transformations with $s\ell(2,\C)$ algebra generated by $L_{\pm 1},L_0$. We'll
first examine these constraints, before applying the Gepner-Witten equation to
arrive at a quite powerful result. 

Let's work with the holomorphic parts of primary fields, for simplicity. 
Projective invariance yields
\eqn\projn{0\ =\ \sum_{j=1}^n\,\left[ z_j^{m+1}{\di\over{\di z_j}}\,+\,
(m+1)h_jz_j^m \right]\, \<\,\phi_1(z_1)\cdots\phi_n(z_n)\,\>\ ,\ \ {\rm for}\ 
m\in\{-1,0,1\}\ }
(compare to \Lmphi). The general solution to \projn\ is 
\eqn\projgen{\<\,\phi_1(z_1)\cdots\phi_n(z_n)\,\>\ =\ 
F\big(\,\{z_{pqrs}\}\,\big)\ 
\prod_{{i,j=1}\atop{i<j}}^n\, (z_i-z_j)^{h_{ij}}\  ,}
where $h_{ij}=h_{ji}$, $\sum_{i\not=j} h_{ij} =
2h_j$, and $F$ is an arbitrary function of the {\it anharmonic ratios} 
\eqn\anhr{z_{pqrs}\ :=\ {{(z_p-z_q)(z_r-z_s)}\over{(z_q-z_r)(z_s-z_p)}}\ \ . }
These ratios are invariant under the projective transformations $z_i\rightarrow 
(az_i+b)/(cz_i+d)$, $ad-bc=1$. 

Only $n-3$ of these anharmonic ratios are independent, so for the 3-point
function, $F$ is simply a constant. Before writing the explicit form, let us
change notation somewhat. Replace $\phi_1,\phi_2,\phi_3$ with
$\phi_{\h\la},\phi_{\h\mu},\phi_{\h\nu}$, and we'll drop the hats on the
affine weights. We can then write
\eqn\projiii{\<\,\phi_\la(x)\phi_\mu(y)\phi_\nu(z)\,\>\ =\
(x-y)^{h_\nu-h_\la-h_\mu}(x-z)^{h_\mu-h_\nu-h_\la}(y-z)^{h_\la-h_\mu-h_\nu}\
\h C_{\la,\mu,\nu}\ .}
So the computation of the 3-point function boils down to a computation of a
constant $\h C_{\la,\mu,\nu}$. This constant is related to the {\it OPE of
primary fields}
\eqn\opepf{\phi_\la(z)\phi_\mu(0)\ \sim\ {{\h C_{\la,\mu,\nu}\phi_{\nu^t}}
\over{z^{h_\la+h_\mu-h_\nu}}}\ ,}
and so is called an {\it operator product coefficient}. Here $\nu^t$ indicates
the highest weight of the representation contragredient (charge-conjugate) to
$L(\nu)$. (The corresponding field is the unique one with a non-vanishing
2-point function with the primary field $\phi_\nu$.) 

The
global $G$ invariance of a correlation function of $n$ primary
fields imposes
\eqn\Gnpt{0\ =\
\sum_{j=1}^n\,{t_j^a}\,\<\,\phi_1(z_1)\cdots\phi_n(z_n)\,\>\ .} 
That is, the $n$-point function must be a $G$-singlet. For this to be
possible, the tensor product $L(\la)\otimes L(\mu)\otimes L(\nu)$ must contain
the singlet representation $L(0)$. Alternatively, in the tensor-product
decomposition
\eqn\tpdec{L(\la)\otimes L(\mu)\ =\ \sum_{\vp \in
P_+}\,T_{\la,\mu}^\varphi L(\varphi)\ ,} 
we require that the {\it tensor-product coefficients} obey 
\eqn\tpcoeff{T_{\la,\mu}^{\nu^t}\ =\ T_{\la,\mu,\nu}\ \not=\ 0\ .}
This implies that the corresponding {\it Clebsch-Gordan coefficient}
$C_{\la,\mu,\nu} \not= 0$. In summary, then, the global $G$-invariance gives
\eqn\chc{\h C_{\la,\mu,\nu}\ \not=\ 0\ \ \ 
\Rightarrow\ \ \ C_{\la,\mu,\nu}\ \not=\ 0\ .}

We will henceforth concentrate on 3-point functions.

\subsec{Depth rule}

Let's apply the Gepner-Witten equation \gepwit\ to the 3-point function:
\eqn\gwthp{0\ =\ \sum_{{\ell_1,\ell_2}\atop{\ell_1+\ell_2\geq 1+\h\nu_0}}\ 
{{(t_\mu^\theta)^{\ell_1} (t_\la^\theta)^{\ell_2}}\over{\ell_1!\ell_2!
(z-z_1)^{\ell_1} (z-z_2)^{\ell_2}}}\,\<\, \phi_\nu(z) \phi_\mu(z_1)
\phi_\la(z_2) \,\>\ .} 
Since the $z,z_1,z_2$ dependence of $\<\phi_\nu(z) \phi_\mu(z_1)
\phi_\la(z_2)\>$ is fixed, the terms in the summation are independent, and so 
\eqn\gwthpi{(t_\mu^\theta)^{\ell_1} (t_\la^\theta)^{\ell_2} \, 
\<\, \phi_\nu(z) \phi_\mu(z_1)
\phi_\la(z_2) \,\>\ =\ 0\ \ \ \ \forall\ 
\ell_1+\ell_2 > \h\nu_0 = k-(\nu,\theta)\ . } 
$t^\theta$ is the raising operator for the $A_1\subset g$ subalgebra in the
direction of the highest root $\theta$ of $g$. The maximum number of times it
can be applied to a state (vector) $v$ with non-vanishing result, is called the
{\it depth} $d(\theta,v)$ of that state, or sometimes the $\theta$-depth. 

To understand the name, look at the example pictured in Fig. 8. 
There is
drawn the weight diagram of the $A_2$-representation $L(2,1)$.
Consider as the ``upward'' direction that of the highest root
$\theta$ (see Figure 7). Since $E^\theta$ adds the root $\theta$ to the weight
of a state, the depth of a state tells us how far ``down'' it is
from its ``top'', roughly speaking. Concentrating on the 2-dimensional subspace
of weight $(0,-1)$, we see that it breaks up into two one-dimensional subspaces
of depths $d(\theta,v_1)=1$ and $d(\theta,v_2)=2$. An important point, however,
is that $d(\theta,c_1v_1+c_2v_2)=2$, as long as $c_2\not=0$. 

Now, if we write \eqn\uvw{\eqalign{ \<\, \phi_\nu(z) \phi_\mu(z_1)
\phi_\la(z_2)& \,\>\ =\cr\ \sum_{u\in L(\nu)}& \sum_{v\in L(\mu)} \sum_{w\in
L(\la)}\, \phi_{\nu,u}\phi_{\mu,v}\phi_{\la,w}\,\<\, u(z)v(z_1)w(z_2)\,\>\
,\cr}} following \uzfs, we can conclude from \gwthpi\ that even if $\<u|w\otimes
v\>$ is non-zero, the 3-point function $\<u(z)v(z_1)w(z_2)\>$ will vanish
unless  \eqn\gwdr{\boxEq{d(\theta,v)\ +\ d(\theta,w)\ <\ k+1-(\theta,\nu)\ =\ 1
+\h\nu_0.}}  But if $\<u(z)v(z_1)w(z_2)\>$ vanishes, then so does
$\<\phi_\nu(z)\phi_\mu(z_1)\phi_\la(z_2)\>$. Therefore, a necessary condition 
for $\<\phi_\nu(z)\phi_\mu(z_1)\phi_\la(z_2)\>\not=0$ is: whenever
$\<u|v\otimes w\>\not=0$, \gwdr\ must be obeyed. This is the
Gepner-Witten {\it depth rule}.

\subsec{Tensor products and refined depth rule} 

Can we get additional constraints from other null vectors? No: the primitive
null vectors are $|\eta_j\>=f_j^{1+\nu_j}\,|v_\nu\>$ for
$i\in\{0,1,\ldots,r\}$. We have just used the first ($j=0$), and all others
are present at the lowest grade in $L(\nu)$, and so are isomorphic to the
primitive null vectors of the $g$-representation $L(\nu)$. These latter
determine the Clebsch-Gordan coefficients for $g$. So, the depth rule should
allow the determination of the operator product coefficients $\h
C_{\la,\mu,\nu}$ from a knowledge of the Clebsch-Gordan coefficients 
$C_{\la,\mu,\nu}$. This turns out to be less straightforward than one might
hope, however. 

To see why, we consider the simpler problem of computing the operator
product multiplicities (called {\it fusion coefficients}) from the
corresponding tensor-product multiplicities (we'll call them tensor-product
coefficients). Being multiplicities, these coefficients are non-negative
integers. A general tensor-product decomposition is written in \tpdec; there the
$T^\vp_{\la,\mu}\in\Z_{\geq 0}$. The simplest example of such a decomposition is for
$g=A_1$: \eqn\aitpd{L(\la_1)\otimes L(\mu_1)\ =\ L(\la_1+\mu_1)\oplus
L(\la_1+\mu_1-2) \oplus \cdots\oplus L(|\la_1-\mu_1|)\ ,}
where we write $L(\la_1)$ for $L(\la_1\ome^1)$, e.g. 
If we change notation using $\la_1=2j_1$, $\mu_1=2j_2$, and $\nu_1=2j$, we
recognise the rule for the addition of quantum angular momenta. 

Reasoning similar to that given above leads to the following rule. If 
$w,v,u \in L(\la), L(\mu), L(\nu)$, respectively, then $C^\nu_{\la,\mu}=0$
(and so $T^\nu_{\la,\mu}=0$) unless when 
$\<u|w\otimes v\>\not=0$, we have 
\eqn\tpdr{\big(E^{-\alpha_i}\big)^{\ell_1}\,|w\>\,\otimes\, 
\big(E^{-\al_i}\big)^{\ell_2}\,|u\>\ =\ 0\ ,}
for all $\ell_1+\ell_2\geq 1+\nu_i$, for $i\in\{1,\ldots,r\}$. If we generalise
the definition of depth to:
\eqn\gdepth{d(\al,v)\ =\ {\rm min}\{\,\ell\in\Z_{\geq 0}\ :\ (E^\al)^{\ell+1}v=0\
\}\ ,}
then we require
\eqn\tpddr{d(-\al_i,v)\ +\ d(-\al_i,w)\ <\ 1
+\h\nu_i\ ,\ \ \forall\ i\in\{1,\ldots,r\}\ .}
Compare this to \gwdr.

Before writing a more useful version of this rule, let's look again at the
example already mentioned above \uvw. Suppose we have a space $V$ spanned by two
independent states $v_1,v_2$, of depths $d(\theta, v_1)=1$, $d(\theta, v_2)=2$,
respectively. Choosing a different basis, $(v_1\pm v_2)/\sqrt{2}$, say, gives
depths 2,2. The set of depths is a basis-dependent object. So, we should
instead be concerned with the {\it dimensions} of the spaces $V_i:=\{v\in V\ :\
(E^\theta)^{1+i}v=0\}$, for $i=1,2$, which are 1 and 2 in this example, 
respectively. Of course, a particular choice of basis may help; $\{v_1,v_2\}$
is a basis of $V$ that is good for the computation of the required dimensions,
while the other basis is not. 

Now, the highest-weight state $|v_\nu\>$ must appear in $L(\la)\otimes
L(\mu)$ (as well as all others). So, we must be able to write 
\eqn\vltp{|v_\nu\>\ =\ \sum_{w\in L(\la)}\, \sum_{u\in L(\mu)}\, 
C_{w,u}^{v_\nu}\,|w\>\otimes|u\>\ ,}
for some coefficients $C_{w,u}^{v_\la}$. It is not difficult to show that
$|v_\nu\>$ must contain a non-zero component $\propto |w_\la\>\otimes |u\>$,
where $|w_\la\>$ is the highest state of $L(\la)$, and $u\in L(\mu)$.
Otherwise,  $C_{{w_\la},u}^{v_\nu}=0$ would imply that $C_{w,u}^{v_\nu}=0$ for
all $w\in L(\la)$. Of course, so that the component $|w_\la\> \otimes 
|u\>$ has the correct weight $\nu$, we must have 
\eqn\Hunl{H\,|u\>\ =\ (\sum_{j=1}^r\,\ome^j h_i)\,|u\>\ =\ (\nu-\la)\,|u\>\
.}  Therefore, we can write (see \ref\zelbk{D.P. Zelobenko, {\it Compact Lie
Groups and Their Representations} (Am. Math. Soc., 1973)}, for example) 
\eqn\zelo{\boxEq{\eqalign{T^\nu_{\la,\mu}\ =&\ \dim\big\{\,u\in L(\mu;\nu-\la)\
:\ (E^{-\al_i})^{1+\nu_i} u = 0\ \forall\,i=1,\ldots,r\,\big\}\cr\ =&:\ \dim
V_{\la,\mu}^\nu\ ,\cr}}} where we have used 
\eqn\Lmnml{L(\mu; \nu-\la)\ :=\ \big\{\, u\in L(\mu)\ :\ H u = (\nu-\la)
u\,\big\}\ .} 

For the operator product numbers, or fusion coefficients $\Nk_{\la,\mu}^\nu$,
we write a truncated tensor product, or {\it fusion product}:
\eqn\LlkLm{L(\la)\otimes_k L(\mu)\ =\ \oplus_{\nu\in \Ppkb}\,
\Nk_{\la,\mu}^\nu\, L(\nu)\ .} 
Of course, since $\Ppkb\subset P_+$ (see \Ppkbar), $\Nk_{\la,\mu}^\nu$ are
undefined if any of $\la,\mu,\nu\in P_+$ are not in $\Ppkb$. But truncation here
means that  \eqn\truncT{\Nk_{\la,\mu}^\nu\ \leq\ T_{\la,\mu}^\nu\ .}  

An argument similar to that
above leads to the following conjecture \ref\kmsw{A.N. Kirillov, P.
Mathieu, D. S\'en\'echal, M.A. Walton, Nucl. Phys. {\bf B391} (1993) 651; p.
215, vol. 1 in M.A. del Olmo et al (editors), {\it Group Theoretical Methods
in Physics, Proceedings of the XIXth International Colloquium, Salamanca, Spain,
1992} (CIEMAT, Madrid, 1993) } (see also \ref\mwcjp{M. A. Walton, Can. J.
Phys. {\bf 72} (1994) 527}): \eqn\rdr{\boxEq{\eqalign{\Nk^\nu_{\la,\mu}\ =&\
\dim\big\{\,u\in L(\mu;\nu-\la)\ :\ (E^{-\al_i})^{1+\nu_i} u = 0\
\forall\,i=0,\ldots,r\,\big\}\cr\ =&:\ \dim \Vk_{\la,\mu}^\nu\ .\cr}}} Here, a special case of the ``extra'' conditions \gwdr\ has been
incorporated into a formula similar to \zelo. (Recall
that since $\h\al_0 = (-\theta,0,1)$, $\al_0= -\theta$.) 

We'll call \rdr\  the
{\it refined depth rule}. Notice it explains \truncT. In fact, \rdr\ implies
the following stronger relations  \eqn\Nkpi{\Nk_{\la,\mu}^\nu\ \leq\
{}^{(k+1)}N_{\la,\mu}^\nu\ ,\ \ \ \lim_{k\rightarrow\infty}\,\Nk_{\la,\mu}^\nu\
=\ T_{\la,\mu}^\nu\ .}

All this can be encoded in the concept of a {\it threshold level}
\ref\cmw{C.J. Cummins, P. Mathieu, M.A. Walton, Phys. Lett. {\bf 254B} (1991)
386}. When $T_{\la,\mu}^\nu>1$, we say that there are more than one
different ``couplings''  $L(\nu)\subset L(\la)\otimes L(\mu)$. That is, there
is more than one way to assemble the states of $L(\nu)$ in the tensor product
$L(\la)\otimes L(\mu)$. If in addition, $L(\nu)\subset L(\la)\otimes_k L(\mu)$,
we say that the coupling is also a ``fusion coupling'' at level $k$. 

For each of the $T_{\la,\mu}^\nu$ couplings $L(\nu)\subset L(\la)\otimes 
L(\mu)$, there exists a threshold level $k_t$, such that the coupling is not a
fusion coupling at levels $k<k_t$, and is for all levels  $k\geq k_t$. 

The threshold level allows a convenient notation: the fusion products for
all levels can be written as the tensor product with threshold levels as
subscripts:
\eqn\tptl{L(\la)\otimes L(\mu)\ =\ \ \oplus_{k_t}\,\oplus_\nu\,
{}^{(k_t)}n^\nu_{\la,\mu}\, L(\nu)_{k_t}\ .} 
Then 
\eqn\Nknk{\Nk_{\la,\mu}^\nu\ =\ \sum_{k_t}^k\, {}^{(k_t)}n_{\la,\mu}^\nu\ .}
In \Nknk\ the sum is over all couplings, and any coupling with $k_t\leq k$
contributes once. The $A_1$ example \aitpd\ becomes
\eqn\aifpd{\eqalign{L(\la_1)\otimes L(\mu_1)\ =\
L(\la_1+\mu_1&)_{\la_1+\mu_1}\oplus L(\la_1+\mu_1-2)_{\la_1+\mu_1-1}\cr &\oplus
\cdots\oplus L(|\la_1-\mu_1|)_{{\rm max}(\la_1,\mu_1)}\ .\cr}} This can be
derived from the depth rule, by considering the coupling $L(\nu_1) \subset
L(\la_1)\otimes L(\mu_1)$ and corresponding $u\in L(\mu_1)$, with $Hu =
[(\nu_1-\la_1)\ome^1] u$. The depth is easily seen to be
$d(\theta,u)=(\mu_1-\nu_1+\la_1)/2$, so that   \eqn\aikt{k_t\ =\
(\mu_1-\nu_1+\la_1)/2\ +\ (\nu,\theta)\ =\ (\la_1+\mu_1+\nu_1)/2\ .}
Let us rewrite the $A_1$ fusion product one more way:
\eqn\aifp{L(\la_1)\otimes L(\mu_1)\ =\
\oplus_{\nu_1=|\la_1-\mu_1|}^{{\rm min}(\la_1+\mu_1,2k-\la_1-\mu_1)}\
L(\nu_1)\ .} This the original form found by Gepner and Witten, and it makes
clear the level-truncation of the tensor product. 

An $A_2$ example is
\eqn\aiifp{\eqalign{L(1,1)^{\otimes 2}\ =\ &L(0,0)_2\oplus L(1,1)_2 \oplus\cr
&L(3,0)_3 \oplus L(1,1)_3 \oplus L(0,3)_3\ \oplus\cr
&L(2,2)_4\ .
}}
This is perhaps the simplest example with a $T_{\la,\mu}^\nu>1$: we have
${}^{(2)}N_{(1,1),(1,1)}^{(1,1)} = 1$, ${}^{(k>2)}N_{(1,1),(1,1)}^{(1,1)} =
2$. This phenomenon occurs because algebras of rank greater than one (\ie\
$g\not=A_1$) have most $\mult(\mu; \vp):=\dim L(\mu;\vp)>1,$ the spaces of
fixed weight $\vp$ in a representation $L(\mu)$ are typically not 
one-dimensional.\foot{\small Except for a $G_2$ exception, it turns out
that $\mult(\mu; \mu')=1$ iff $\mu-\mu'$ has a unique expression as a
$\Z_{\geq 0}$-linear combination of those positive roots $\alpha$ of $g$ also
obeying $(\mu,\al)>0$ \ref\berzel{A.D. Berenstein, A.V. Zelevinsky, Funkt.
Anal. Pril. {\bf 24} (1990) 1}.} 

\subsec{Good bases and the \LR\ rule}

Now let's try to use the refined depth rule \rdr\ to compute fusion
products like \aiifp. Since the level $k$ will figure more prominently
henceforth, we'll let $g_k$ indicate the affine algebra $\h g$ at fixed level
$k$. 

The problem is to find a good choice of basis of
$L(\mu;\nu-\la)$ to simplify (as much as possible) the computation of
$\Nk_{\la,\mu}^\nu = \dim{}^{(k)}V_{\la,\mu}^\nu \subset V_{\la,\mu}^\nu
\subset L(\mu;\nu-\la)$. Let's first back up and consider $T_{\la,\mu}^\nu =
\dim V_{\la,\mu}^\nu$. After all, the problem of a good choice of basis
already exists in the computation of $T_{\la,\mu}^\nu$ (for the
$\al_i$-depths, if not for the $\theta$-depth). Suppose a basis
$B(\mu;\nu-\la) = \{u^a\ :\ a=1,\ldots,\dim L(\mu;\nu-\la)\}$ of
$L(\mu;\nu-\la)$ were to be good; what should that mean? Ideally, one could
test the $u^a$  one-by-one, and those passing would form a basis of
$V_{\la,\mu}^\nu$. That is, the subspace $V_{\la,\mu}^\nu$ would be spanned by
the subset of $B(\mu;\nu-\la)$ that are elements of $V_{\la,\mu}^\nu$. It
turns out that such {\it good bases} (or proper bases)
$B^{(p)}(\mu;\nu-\la)$ exist \ref\goddb{O. Mathieu, Geom. Ded. {\bf 36} (1990)
51;\hfill\break
I.M. Gelfand, A.V. Zelevinsky, Funkt. Anal. Pril. {\bf 19} (1985) 72}:
\eqn\pbases{V_{\la,\mu}^\nu\ =\ {\rm Span}\{\, u^a\in B^{(p)}(\mu;\nu-\la)\ :\
u^a\in V_{\la,\mu}^\nu \,\}\ ,} for all possible highest weights $\la,\mu,\nu
\in P_+$. 

For the case $g=A_r$, the basis elements are indexed by {\it standard
tableaux}, for example 
\eqn\ateg{ \ST{\STrow{\b1\b2\b2\b4}\STrow{\b3\b3\b5}\STrow{\b4}\STrow{\b5}}\ \
.} The tableaux with numbers removed are called {\it Young tableaux}, and their
boxes are arranged in left-justified rows, of non-increasing length going from
top to bottom. To obtain a standard tableau relevant to the $A_r$ case,
the numbers added must be from the set $\{1,2,\ldots,r+1\}$, and they
must appear in non-decreasing order from left to right in the rows, and
in increasing order from top to bottom in the columns. Notice this implies
that a single column of a standard tableau can contain no more than $r+1$
boxes, for the $A_r$ case. 

In
combination with \zelo, the good bases (with elements indexed by standard
tableaux) lead to a simple rule for the computation of tensor-product
coefficients, the {\it Littlewood-Richardson rule}, to be explained below. Each
box $\ST{\STrow{\b{$i$}}}\ $ stands for a weight in the basic representation
$L(\ome^1)$ of $A_r$, so the weight $wt(\ \ST{ \STrow{\b{$i$}} }\ )$ of $\
\ST{\STrow{\b{$i$}}}\ $ is  \eqn\wtbi{ wt(\ \ST{\STrow{\b{$i$}} }\ )\ =\ \ome^1
-\al_1 - \al_2 - \ldots - \al_{i-1}\  =\  \left\{ \matrix{\ome^1\ \ \ ,&\ \ \
i=1\cr -\ome^{i-1}+\ome^i\ ;&\ i=2,\ldots,r\cr -\ome^r\ \ \ ,&\ \ \ i=r+1 \cr}
\right. } Notice that $\sum_{i=1}^{r+1} wt(\,\ST{\STrow{\b{$i$}}}\,) = 0$;
therefore a  column of length $r+1$ can be dropped from a standard tableau. 
The {\it weight} $wt({\cal T}_\#)$ of a standard tableau ${\cal T}_\#$ is just
the sum of the weights of its component boxes. The {\it shape}  $sh({\cal
T}_\#)$ of a standard tableau ${\cal T}_\#$ is the weight of the tableau
obtained by replacing all the numbers in the $i$-th row with $i$'s. Notice that
the shape of a standard tableau will always be a dominant weight (\ie\ an
element of $P_+$); it will be the highest weight of the relevant
representation. To restrict to those highest weights relevant to the $A_r$ WZW
model, the number of columns of length less than or equal to $r$ must be less
than or equal to $k$ (see \Ppkbar). The weight of the Young tableau ${\cal T}$
obtained from the standard tableau ${\cal T}_\#$ by removing its numbers,  is
defined by $wt({\cal T}) := sh({\cal T}_\#)$. 

The elements of the good basis $B^{(p)}(\mu;\sigma)$ are indexed by the
elements of 
\eqn\Tmusi{{\cal T}_\#(\mu;\sigma)\ =\ \{\, {\cal T}_\# \in {\cal S}_\#\ :\
sh({\cal T}_\#) = \mu,\ wt({\cal T}_\#) = \sigma\,\}\ ,} 
where ${\cal S}_\#$ just means the space of standard tableaux. 
We'll write $v(\Tn)$
for the state indexed by the standard tableau $\Tn$. The elements of these bases
appropriate for the $A_2$ representation $L(1,1)$ are shown in Fig. 15. There
the tableaux are drawn, roughly at the  positions of their weights in weight
space. 

\midinsert
\vskip1cm
\epsfxsize=8cm
\centerline{\hskip-1.5cm\epsfbox{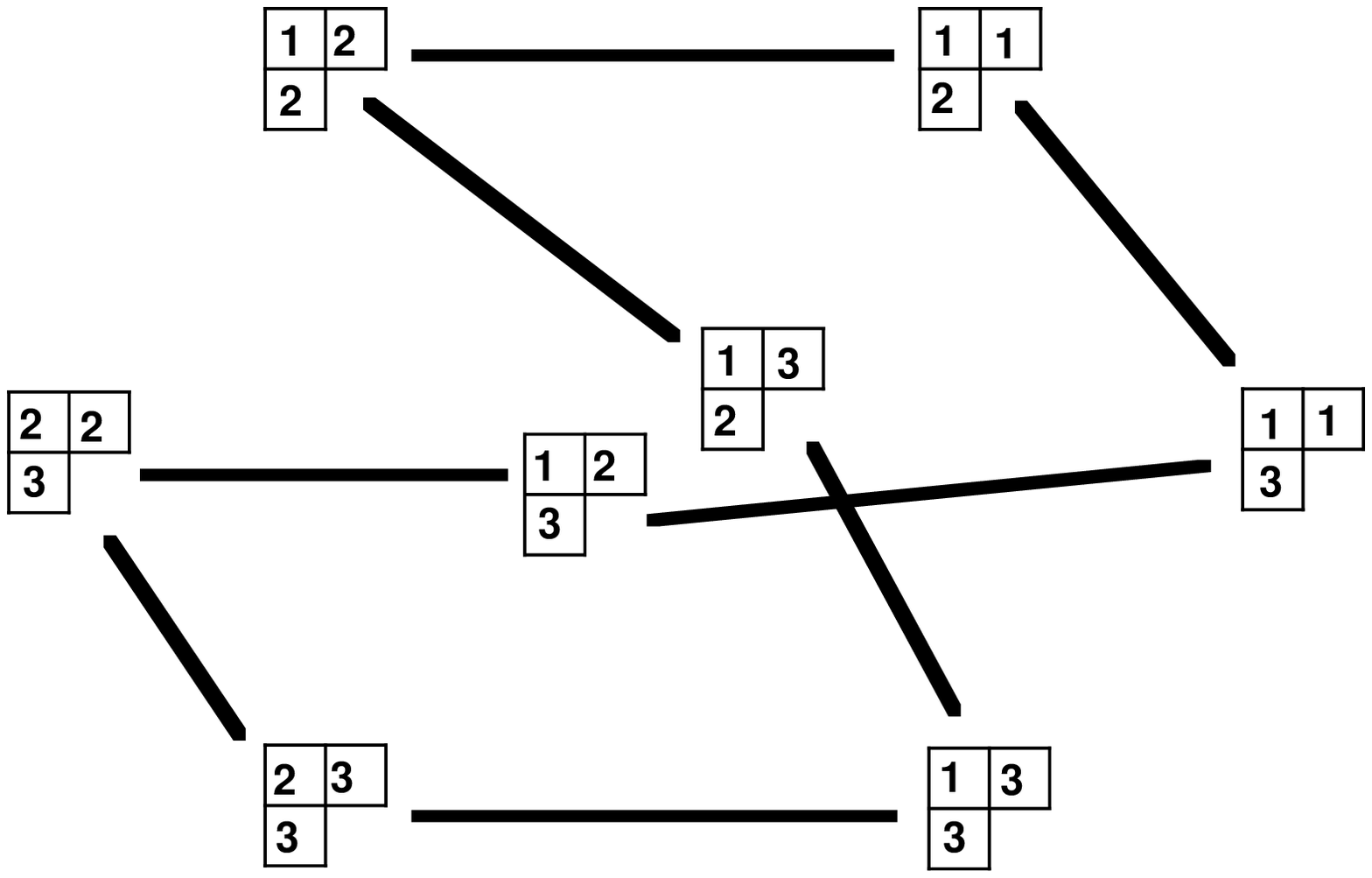}}
\smallskip\smallskip\bigskip\bigskip
\leftskip=1cm
\rightskip=1cm
\noindent
\baselineskip=12pt
{\it Figure 15}. Shown are the standard tableaux that label the states
of the $A_2$ representation $L(1,1)$ in the good basis.\bigskip
\leftskip=0cm\rightskip=0cm \baselineskip=15pt  
\endinsert

The action of the step operators $e_i$ and $f_i$ $(i=1,2)$, can be transcribed
to an action on the standard tableaux. I will describe it later in the more
convenient language of {\it paths}. For the moment, their actions (up to
non-zero multiplicative factors) are indicated in Fig. 15 by the lines drawn.
Notice that a single line specifies the action of both $e_i$ and $f_i$ (fixed
$i$). That's because of the property
\eqn\efT{\eqalign{e_i\Tn\ \not=\ 0\ \ \Rightarrow&\ f_ie_i\Tn\ \propto\ \Tn\ 
\ ,\cr
f_i\Tn\ \not=\ 0\ \ \Rightarrow&\ e_if_i\Tn\ \propto\ \Tn\ 
\ ,\cr }}
which guarantees the ``good''-ness property, as can be seen from the diagram. 

Also, the depths $d(\al_i,\Tn):=d(\al_i,v(\Tn))$ of a standard tableau are
easily found. Notice that the $j$ right-most columns of a standard tableau
form a standard tableau; call it $\Tn^{(j)}$, where $j$ ranges from 0 to the
full width of $\Tn$. It turns out that the weights of these sub-tableaux
determine the depths $d(\al_i,\Tn)$ 
for all  $i=1,2,\ldots,r$. Precisely, we have
\eqn\dmaxTj{d(\al_i,\Tn)\ =\ {\rm max}_j\big(\, wt(\Tn^{(j)}) , -\al_i^\vee
\,\big)\ .}

We can also speak of the {\it height} $h(\al_i,\Tn)$. By 
$d(\al_i,\Tn)$ ($h(\al_i,\Tn)$) is meant the maximal non-negative integer
$\ell$ such that $e_i^\ell\Tn \not= 0$ ($f_i^\ell\Tn \not= 0$). So, 
$h(\al_i,\Tn) = d(-\al_i,\Tn)$, for example.  
They are also simply found, since for a state $u$ in any
$A_1=\<e_i,h_i,f_i\>  \subset g$ representation,  $h(\al_i,u) - d(\al_i,u) = 
(\sigma,\al_i^\vee)$, if $Hu = (\sigma,\al_i)u$. Now, notice that the
tableaux drawn in Fig. 15, can be assigned to a single irreducible
representation of $\<e_i,h_i,f_i\>$ for each of $i=1,2$. This property
generalises to all ranks $r$, and so 
\eqn\hTn{h(\al_i,\Tn)\ =\ \big(\, wt(\Tn) , \al_i^\vee \,\big)\ +\
d(\al_i,\Tn)\ .}
The property just mentioned is another way of describing the ``good''-ness of
the required bases. 

Consider the $g=A_2$ example 
$B^{(p)}(\,(1,1); (0,0)\,)$. Its elements are indexed by the tableaux
\eqn\aiiadz{\ST{\STrow{\b1\b2}\STrow{\b3}}\ ,\ 
\ST{\STrow{\b1\b3}\STrow{\b2}}\ .}
Since
\eqn\Tji{wt\left( \ST{\STrow{\b2}} \right)\ =\ (-1,1)\ ,\ 
wt\left( \ST{\STrow{\b1\b2}\STrow{\b3}} \right)\ =\ (0,0)\ ,}
\dmaxTj\ gives
\eqn\dhTji{\eqalign{d\left( \al_1,\ST{\STrow{\b1\b2}\STrow{\b3}} \right)\ =\ 1\
\ \Rightarrow&\ 
h\left( \al_1,\ST{\STrow{\b1\b2}\STrow{\b3}} \right)\ =\ 1\ ;\cr
d\left( \al_2,\ST{\STrow{\b1\b2}\STrow{\b3}} \right)\ =\ 0\
\ \Rightarrow&\ 
h\left( \al_2,\ST{\STrow{\b1\b2}\STrow{\b3}} \right)\ =\ 0\ .\cr
}}
Similarly, 
\eqn\Tji{wt\left( \ST{\STrow{\b3}} \right)\ =\ (0,-1)\ ,\ 
wt\left( \ST{\STrow{\b1\b3}\STrow{\b2}} \right)\ =\ (0,0)\ ,}
implies
\eqn\dhTjii{\eqalign{d\left( \al_1,\ST{\STrow{\b1\b3}\STrow{\b2}} \right)\ =\ 0\
\ \Rightarrow&\ 
h\left( \al_1,\ST{\STrow{\b1\b3}\STrow{\b2}} \right)\ =\ 0\ ;\cr
d\left( \al_2,\ST{\STrow{\b1\b3}\STrow{\b2}} \right)\ =\ 1\
\ \Rightarrow&\ 
h\left( \al_2,\ST{\STrow{\b1\b3}\STrow{\b2}} \right)\ =\ 1\ .\cr
}}
A glance at Fig. 15 confirms these results. 

When substituted into \zelo, the depths obtained by \dmaxTj\ yield the
Littlewood-Richardson rule, a simple rule for the computation of the $g=A_r$
tensor-product coefficients $T_{\la,\mu}^\nu$:
\eqn\LRrule{\boxEq{T_{\la,\mu}^\nu\ =\ {\rm card}\big\{\, \Tn\in 
\Tn(\mu;\nu-\la)\  :\
d(\al_i,\Tn)\leq \nu_i\ \ \ (\forall
i=1,\ldots,r) \,\big\}\ .  }} 
By \Tmusi, we have 
\eqn\Tnmnl{\Tn(\mu;\nu-\la)\ =\ \big\{\, \Tn\in{\cal S}_\#\
:\  sh(\Tn)=\mu,\ wt(\Tn)=\nu-\la \,\big\}\ .}
We should mention that this is not precisely the form of the rule originally 
given by Littlewood and Richardson. It is, however, related to it by a simple
transformation (see \ref\weym{J. Weyman, Contemp. Math. {\bf 88} (1989) 177},
for example). 

To use the \LR\ rule, one can first draw the Young tableau $\T_\la$ of shape
$\la$. Consider a standard tableau $\Tn$ of shape $\mu$ and weight $\nu-\la$.
Add its sub-tableaux $\Tn^{(j)}$ to $\T_\la$ by placing boxes
$\ST{\STrow{\b{$\ell$}}}$ to the right of the $\ell$-th row (numbered from top
to bottom)of $\T_\la$. An $A_5$ example, with $\la=(3,1,2,1)$, $\mu=(1,1,2,0)$
and $\nu=(2,0,3,2)$ is
\eqn\aexT{\ST{\STrow{\bv\bv\bv\bv\bv\bv\bv}\STrow{\bv\bv\bv\bv}
\STrow{\bv\bv\bv}\STrow{\bv}}\ +\ 
\ST{\STrow{\b1\b2\b2\b3}\STrow{\b3\b3\b4}\STrow{\b4\b5}}\ =\ 
\ST{\STrow{\bv\bv\bv\bv\bv\bv\bv\b1}\STrow{\bv\bv\bv\bv\b2\b2}
\STrow{\bv\bv\bv\b3\b3\b3}\STrow{\bv\b4\b4}\STrow{\b5}}\ .} If the shape of the
resulting {\it mixed tableau} (defined in the obvious way) is dominant (\ie\ in
$P_+$) for all $\Tn^{(j)}$, then $\Tn$ contributes 1 to $T_{\la,\mu}^\nu$.
Also, the corresponding vector $v(\Tn)$ is an element of the basis  of
$V_{\la,\mu}^\nu$. 

To make things clear, consider a simple $A_2$ example: let's verify that
$T_{\la,\mu}^\nu = T_{(1,0),(1,1)}^{(1,0)} = 1$. The tableaux of shape
$\mu=(1,1)$ and weight $\nu-\la = (0,0)$ are those drawn in \aiiadz. Adding
the sub-tableaux $\Tn^{(j)}$ ($j=1$ and 2) of the first one to
$\T_\la=\ST{\STrow{\bv}}$ gives the mixed tableaux
\eqn\mTi{\ST{\STrow{\bv}\STrow{\b2}}\ ,\ \
\ST{\STrow{\bv\b1}\STrow{\b2}\STrow{\b3}}\ ,}
so that the first tableau contributes 1 to the tensor-product coefficient. 
With the second, however, adding $\Tn^{(1)}=\ST{\STrow{\b3}}$ to $\T_\la$ gives
\eqn\mTii{\ST{\STrow{\bv}\vskip.38cm \STrow{\b3}}\ .}
Therefore, the second tableau of \aiiadz\ does not contribute, and we find 
$T_{(1,0),(1,1)}^{(1,0)} = 1$. 

One nice feature of the Littlewood-Richardson rule  is that one simply counts
the number of standard tableaux of a certain type that pass a specific test.
Furthermore, this test can be applied to the candidate tableaux one-by-one,
without referring to the other candidates. For example, there are no
redundancies or cancellations between candidates. For this reason, we'll call
the Littlewood-Richardson rule a {\it combinatorial rule}. 

The Littlewood-Richardson rule can be stated in many different ways. The
version described is well suited to the application of \zelo, however, with its
connection between standard tableaux and the vectors of
$B^{(p)}(\mu;\nu-\la)$. 

More importantly, the standard tableaux can be replaced by {\it universal}
objects, that can be defined for any simple $g$ in a
uniform way. So the \LR\ rule can be adapted from the case of $g=A_r$ to all
simple $g$. It was Littelmann who completed this generalisation, first using 
sequences of Weyl group elements as the universal objects, and then using
sequences of weights \ref\LitWp{P. Littelmann, J. Alg. {\bf 130} (1990) 328;
Invent. Math. {\bf 116} (1994) 329}. In the \LR\ rule, both sequences have to do
with the sub-tableaux  $\Tn^{(j)}$ of a standard tableaux $\Tn$, or
equivalently, the columns of $\Tn$, reading from right to left. 

The more economical generalisation is the one using the sequences of weights.
It is phrased in terms of {\it Littelmann paths} in weight space. We'll look
at the $g=A_r$ case, but generalisation is straightforward. To each standard
tableau, we can associate a piecewise linear path in weight space as follows:
read the weights of the columns of the standard tableau from right to left,
and associate a piece of the path to each. The vector position of the end of
such a piece minus that of the beginning equals the weight of the
corresponding column. So, by adding the pieces, in order (right to left on the
tableau), we obtain the relevant path. 

As an example, Fig. 16 shows the Littelmann paths for the outer 
weights of the $A_2$ representation $L(2,1)$. 

\midinsert
\vskip1cm
\epsfxsize=8cm
\centerline{\hskip-2cm\epsfbox{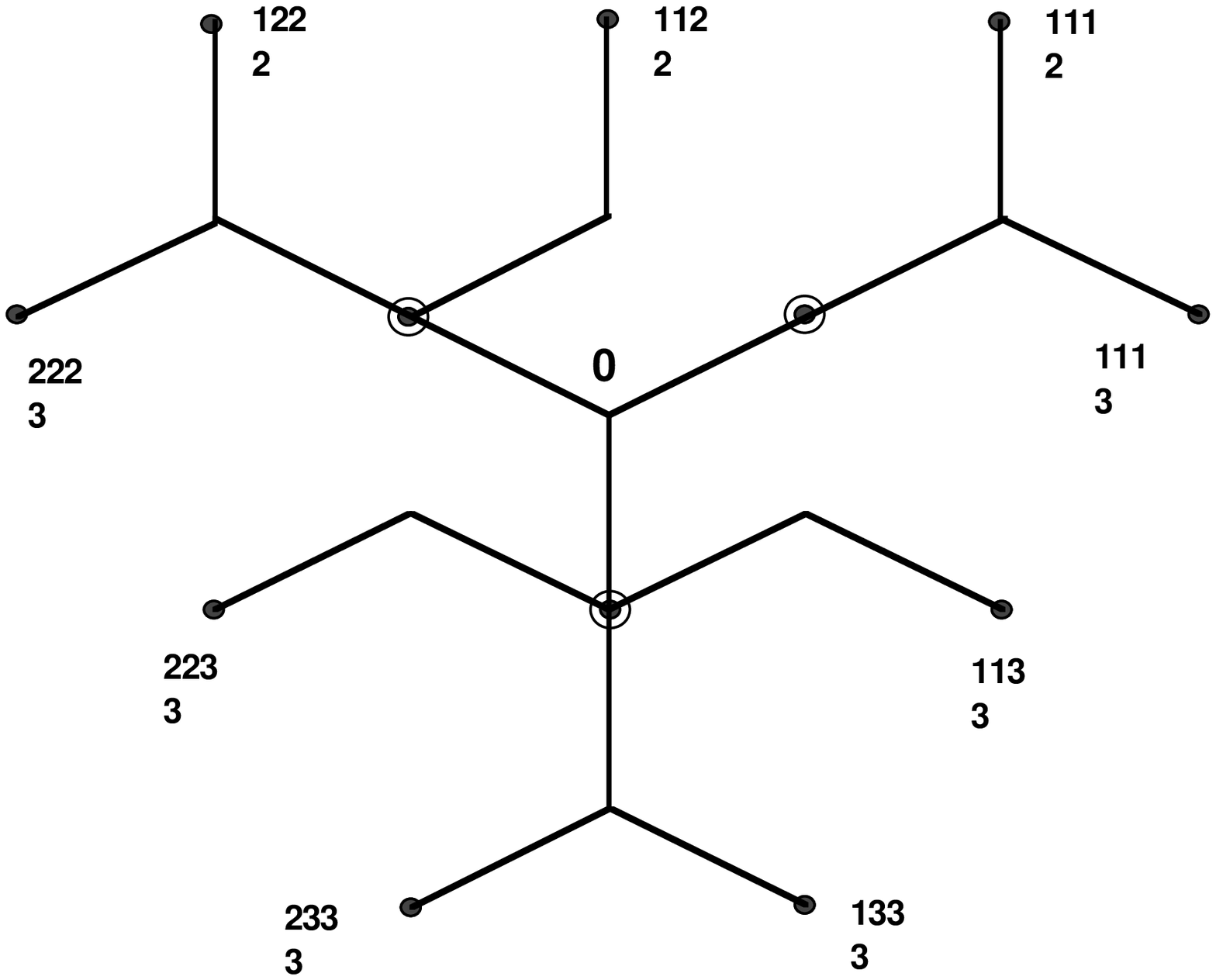}}
\smallskip\smallskip\smallskip
\leftskip=1cm
\rightskip=1cm
\noindent
\baselineskip=12pt
{\it Figure 16}. The Littelmann paths are drawn for the outermost
weights of the representation $L(2,1)$ of $A_2$. For each such weight $\mu$,
the shortest route from 0 to $\mu$ is the relevant path. \bigskip
\leftskip=0cm\rightskip=0cm \baselineskip=15pt   \endinsert

As promised, we'll indicate the action of the step operators $e_i,f_i$ 
$(i=1,\ldots,r)$ on the paths. By the equivalence of 
standard tableaux and Littelmann paths, this will also describe the actions on
the standard tableaux. It is important to realise, however, that what we
describe is only the action of the step operators up to normalisation. Let
$v(\pi)$ denote the state labelled by the path $\pi$. $e_i,f_i$
acting on a path $\pi$ yield other paths, $e_i\pi,f_i\pi$. Then we have $e_i\,
v(\pi)\propto\, v(e_i\pi)$ and  $f_i\, v(\pi)\propto\, v(f_i\pi)$, with 
non-zero multiplicative factors left unspecified.

First consider $e_i$. Parametrise a path $\pi$ with
parameter $t\in[0,1]$. $wt\big( \pi(t)\big)$ is the weight of a point $t$ of
the path.  $wt\big( \pi(0)\big) = 0$, and $wt\big( \pi(1)\big)$ is the weight of
the path $wt(\pi)$, and the weight of the corresponding vector and standard
tableau. Similarly, the shape of the path $sh(\pi)$ equals the shape of the
corresponding Young tableau, or the weight of the highest vector. First find
the minimum non-positive integer value of $\big( wt(\pi(t)),\al_i^\vee \big)$
for $t\in[0,1]$; call it $M_i$. Let $t_2$ be the minimum value of $t$ where
$\big( wt(\pi(t)),\al_i^\vee \big) = M_i$. If $M_i=0$, then $e_i\pi = 0$. If
$M_i\leq -1$, then find the maximum value of $t<t_2$ such that $\big(
wt(\pi(t)),\al_i^\vee \big) = M_i+1$, and call that $t_1$. Now break the path
up into three pieces, corresponding to the intervals: \eqn\tints{0\leq t\leq
t_1\ ,\ \ t_1\leq t \leq t_2\ ,\ \ t_2\leq t \leq 1\ \ .}  Weyl reflect the
middle piece across the hyperplane normal to $\al_i$ at $wt(\pi(t_1))$.
Finally, re-attach the third piece (corresponding to $t_2\leq t\leq 1$) at
$\pi(t_2)+\alpha_i$, to obtain the path $e_i\pi$. 

It is not hard to see that this action yields $d(\al_i,\,v(\pi)\,)=-M_i$ for the
vector $v(\pi)$ that is indexed by the path $\pi$. A diagram sketching the
action of $e_i$ on a path $\pi$ is given in Fig. 17. 

\topinsert
\vskip2cm
\epsfxsize=8cm
\centerline{\hskip-3cm\epsfbox{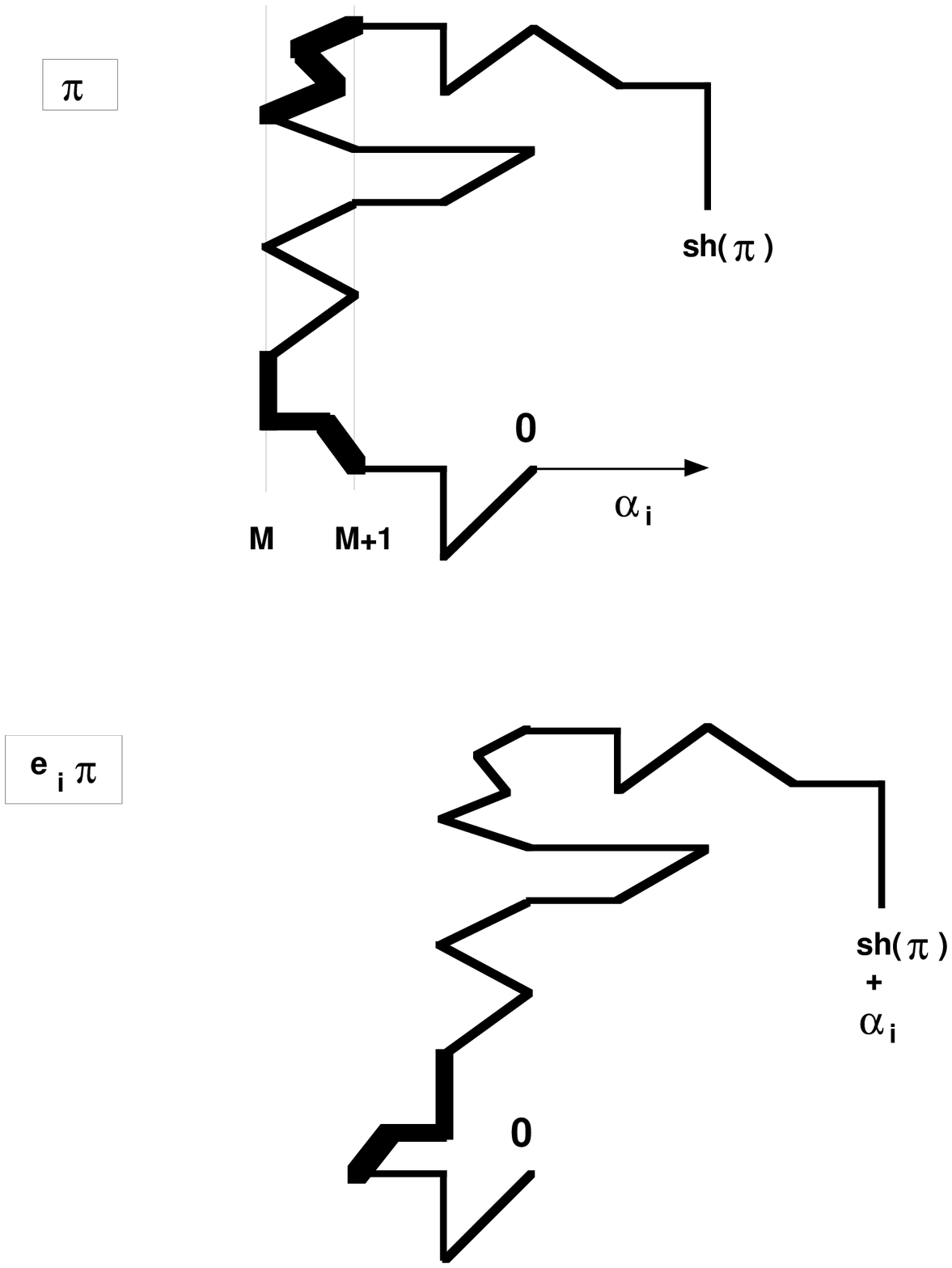}}
\smallskip\smallskip
\leftskip=1cm
\rightskip=1cm
\noindent
\baselineskip=12pt
{\it Figure 17}. The action of $e_i$ on a Littelmann path $\pi$. The
thickened parts of the path $\pi$ are those segments to be Weyl-reflected by
$e_i$ or $f_i$ (see Fig. 18 for the latter). \bigskip
\leftskip=0cm\rightskip=0cm \baselineskip=15pt   \endinsert

The action of the lowering operator
$f_i$ is defined similarly; see Fig. 18. With $M_i$ defined as above, consider
$M'_i= wt(\pi)- M_i$. If $M'_i=0$, then $f_i\pi = 0$. If $M'_i \geq 1$, first
find the maximum value of $t$ such that  $\big( wt(\pi(t)),\al_i^\vee \big) =
M_i$; call it $t_1$. Then obtain the minimum value of $t>t_1$ such that $\big(
wt(\pi(t),\al_i^\vee \big) = M_i+1$; call that $t_2$. Then the three intervals
\tints\ are again relevant, and $f_i\pi$ is found by reflecting the middle
piece across the hyperplane normal to $\al_i$ at $wt(\pi(t_1))$, and
re-attaching the third piece at $\pi(t_2)-\alpha_i$. 

It is not hard to see that $M'_i$ so defined is the height $h(\al_i,\,
v(\pi)\,)$ for the vector $v(\pi)$ that is indexed by the path $\pi$.

\topinsert
\vskip.5cm
\epsfxsize=6cm
\centerline{\epsfbox{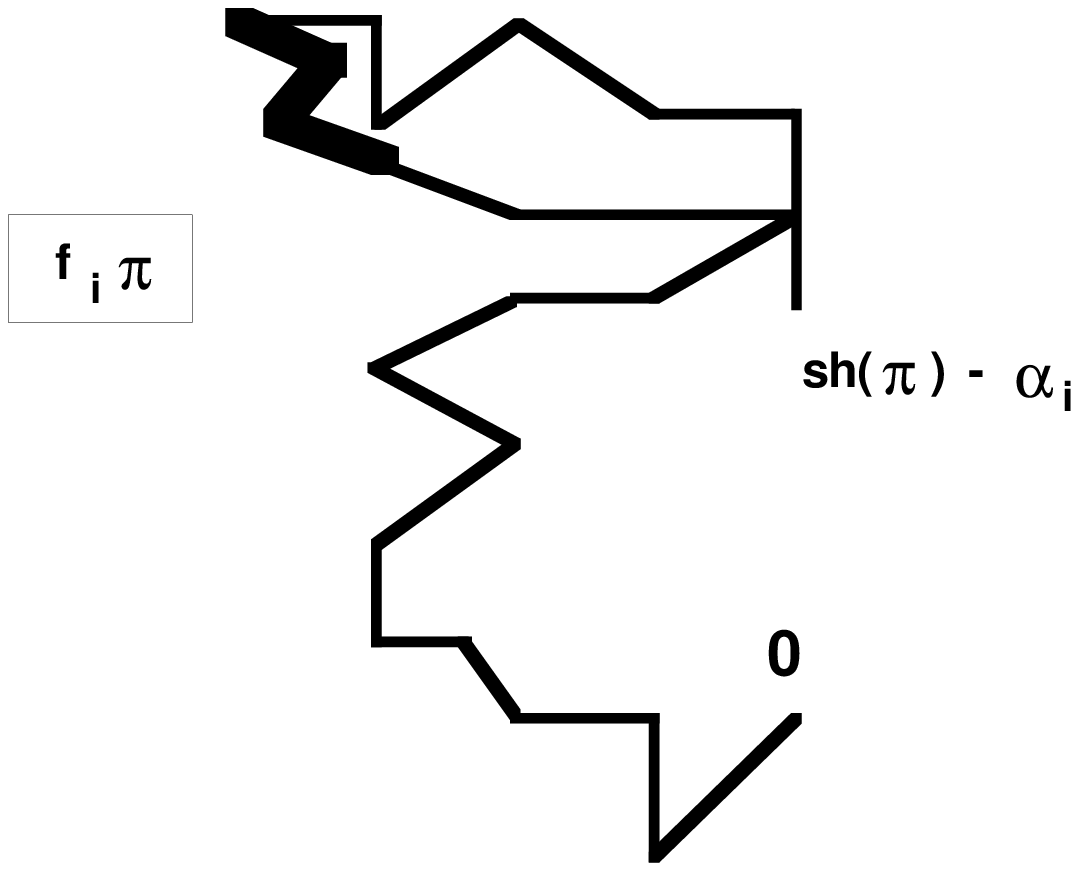}}
\smallskip\smallskip
\leftskip=1cm
\rightskip=1cm
\noindent
\baselineskip=12pt
{\it Figure 18}. The action of $f_i$ on the Littelmann path $\pi$ of Fig. 17.
\bigskip \leftskip=0cm\rightskip=0cm
\baselineskip=15pt  
\endinsert

With the actions so-defined, we've drawn in Fig. 19 the analogue of
Fig. 15 for the $A_2$ representation $L(2,1)$. Again, the lines
indicate the action of the step operators $e_1,e_2,f_1,f_2$. 

Incidentally, the graphs of Figs. 15,19 are (essentially) examples of so-called
{\it crystal graphs} \ref\kash{M. Kashiwara, Commun. Math. Phys.
{\bf 133} (1990) 249}\ref\Licg{P. Littelmann, J. Alg. {\bf 175} (1995) 65}.
These arise in the theory of {\it quantum groups} $U_q(g)$, the
$q$-deformations of the universal enveloping algebra $U(g)=U_1(g)$ of $g$. Such
quantum groups allow the construction of integrable lattice models in
two-dimensions, in the spirit of the 2-dimensional Ising model. The graphs
reflect the simplified representation theory of $U_q(g)$ at $q=0$. (One can also say that the existence of the
canonical bases at $q=1$ is explained by the simplified representation theory
at $q=0$ \ref\lusz{G.
Lusztig, J. Amer. Math. Soc. {\bf 3} (1990) 447; Prog. Theor. Phys. Suppl.
{\bf 102} (1990) 175}.) But physically, $q=0$ corresponds to absolute zero 
temperature, justifying the reference to crystals. 

\midinsert
\vskip0cm
\epsfxsize=8cm
\centerline{\epsfbox{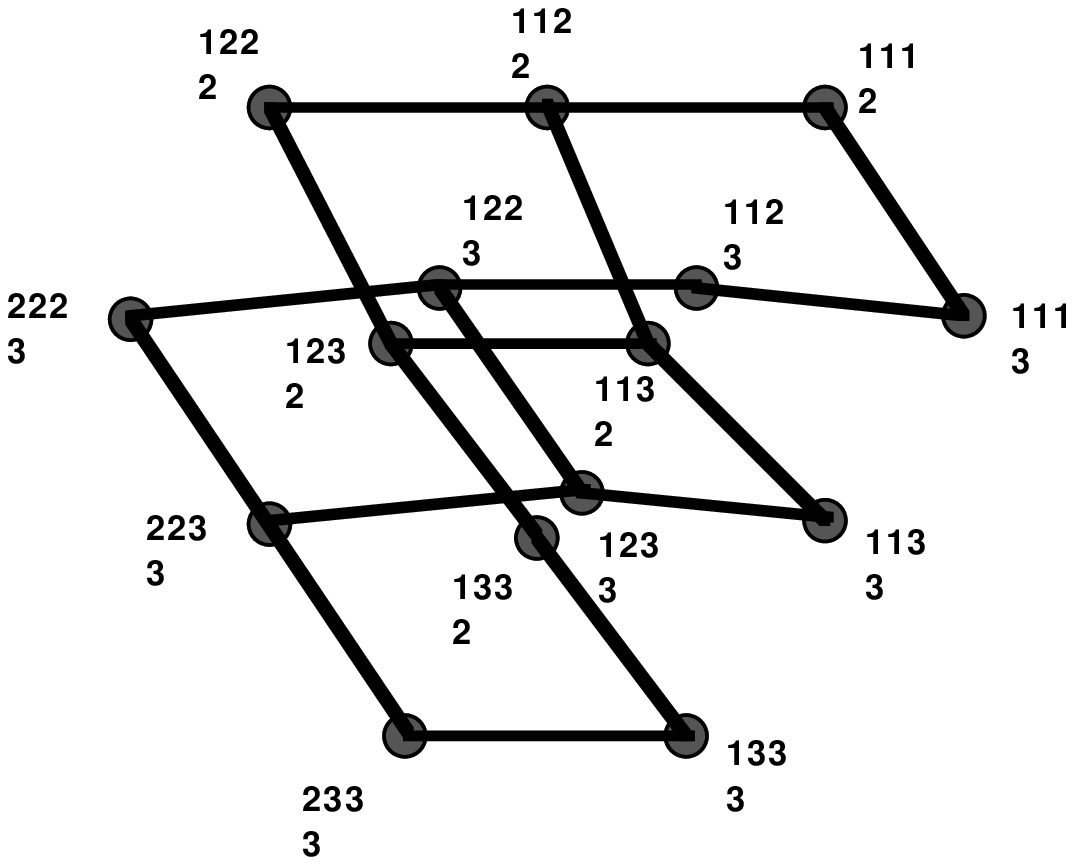}}
\smallskip\smallskip
\leftskip=1cm
\rightskip=1cm
\noindent
\baselineskip=12pt
{\it Figure 19}. The standard tableaux of the $A_2$ representation
$L(2,1)$ are drawn (roughly) at the positions of their weights. The lines
indicate the action of the step operators $e_1,e_2,f_1,f_2$ on the good basis
elements indexed by the tableaux (or by the corresponding Littelmann paths). 
\bigskip \leftskip=0cm\rightskip=0cm \baselineskip=15pt   \endinsert 

To make the actions of the ladder operators on paths completely clear, we'll do
a couple of examples. First consider the path of weight (2,-2) drawn in Fig. 16.
We indicate this path by the weights of its straight segments, written in
sequence: \eqn\piseq{\pi\ =\ \{\,(0,-1), (1,0),  (1,-1)\,\}\ .}
To find  $e_2\pi$, we first need $M_2$, the minimum non-positive value of 
$\big( wt(\pi(t)),\al_2^\vee \big)$; here it is -2. We find $wt(\pi(t_1))=
(1,-1)$ and $wt(\pi(t_2)) = (2,-2)$. (Notice here that the ``third piece'' of
\tints\ corresponds to the single value $t=1$.) The ``middle piece'' consists of
a single, straight segment, of weight $(1,-1)$, and so is reflected to a
similar single, straight segment of weight $r_2(1,-1) = (0,1)$. So we find
\eqn\eiipi{e_2\,\pi\ =\ \{\,(0,-1), (1,0),  (0,1)\,\}\ ,}
a path that is not drawn in Fig. 16. 

Now consider $f_1\pi'$, with $\pi'$ the path 
\eqn\piseqi{\pi'\ =\ \{\,(-1,1), (1,0),  (0,1)\,\}\ ,} 
of weight $(0,2)$. We have $M_1=-1$, and so $M'_1= 0-(-1)=1$. $wt(\pi'(t_1))=
(-1,1)$ and $wt(\pi'(t_2)) = (0,1)$ indicate that the ``middle piece'' is a
single, straight segment of weight $(1,0)$, that must be replaced by one of
weight $r_1(1,0)=(-1,1)$. As a result, we find 
\eqn\fipi{e_1\,\pi'\ =\ \{\,(-1,1), (-1,1),  (0,1)\,\}\ ,} 
a path that is drawn in Fig. 16.

What is the form of the \LR\ rule in terms of paths? To calculate
$T_{\la,\nu}^\nu$, one considers all paths $\pi$ of shape $sh(\pi)=\mu$ and
weight $wt(\pi)= \nu-\la$.  Such a path will contribute 1 to $T_{\la,\mu}^\nu$
iff $\la + wt\big(\pi(t)\big)$ has non-negative Dynkin indices for all
$t\in[0,1]$. That is, we require  $\big(\la + wt(\pi(t)), \al_i^\vee) \geq 0$,
for all $i\in\{1,\ldots,r\}$, and all $0\leq t\leq 1$. In other words, the
$\la$-translated path $\la + wt\big(\pi(t)\big)$ must remain in the dominant
sector; it must be a {\it dominant path}. 

This form of the \LR\ rule is very general, applicable  to all symmetrisable
Kac-Moody algebras. These paths also have some interesting invariances. For
example, one can generate a valid set of paths for the vectors of $L(\mu)$ by
acting successively with lowering operators on the ``highest path'', in all
inequivalent ways. Furthermore, one can use any dominant path as the highest
path \ref\Lipr{P. Littelmann, Ann. Math. {\bf 142} (1995) 499}, even the
straight line from 0 to $\mu$! 

This is all interesting, but what we really want to know is how to adapt this
machinery to the formula \rdr\ for the computation of fusion coefficients
$\Nk^\nu_{\la,\mu}$. Can we find an analogue of the \LR\ rule for fusions: a
combinatorial rule that computes the fusion coefficients $\Nk_{\la,\mu}^\nu$?

For that we also need to consider the action of
$E^\theta$. Restricting to $g=A_r$, we have $\theta =
\al_1+\al_2 +\ldots +\al_r$. By the Cartan-Weyl commutation relations \EaEb, 
\eqn\Eth{E^\theta\ \propto\ \ad(e_1)\ad(e_2)\cdots \ad(e_r)\,e_r\ .}
For example, if $r=2$, $E^\theta\propto [e_1,e_2]$. Figures 15,
19 reveal that the basis $B^{(p)}(\mu; \sigma)$ that is good for the
calculation of tensor-product coefficients, is {\it not} also good for the
calculation of their truncated versions, the fusion coefficients \kmsw
\ref\mwjmp{M.A. Walton, J. Math. Phys. {\bf 39} (1998) 665}. 

For example, consider again $B^{(p)}(\,(1,1); (0,0)\,)$, and the relevant
tableaux \aiiadz. We get
\eqn\Etheg{\eqalign{E^\theta\,\ST{\STrow{\b1\b2}\STrow{\b3}}\ \propto&\ 
[e_1,e_2]\, \ST{\STrow{\b1\b2}\STrow{\b3}}\ =\ -e_2e_1\, 
\ST{\STrow{\b1\b2}\STrow{\b3}}\ \propto\ \ST{\STrow{\b1\b1}\STrow{\b2}}\ ,\cr
E^\theta\,\ST{\STrow{\b1\b3}\STrow{\b2}}\ \propto&\
[e_1,e_2]\, \ST{\STrow{\b1\b3}\STrow{\b2}}\ =\ e_1e_2\, 
\ST{\STrow{\b1\b3}\STrow{\b2}}\ \propto\ \ST{\STrow{\b1\b1}\STrow{\b2}}\ .\cr
}}
On the other hand, we know (by means to be discussed soon) that
${}^{(2)}N_{(1,1),(1,1)}^{(1,1)}$ $= 1$, and ${}^{(\geq
3)}N_{(1,1),(1,1)}^{(1,1)} = 2$. This means that some linear combination $v_0$
of the vectors labelled by $\ST{\STrow{\b1\b2}\STrow{\b3}}$ and 
$\ST{\STrow{\b1\b3}\STrow{\b2}}$ has depth $d(\theta,v_0)=0$ and an
independent vector $v_1$ in their span has $d(\theta,v_1) = 1$. 

Is there another basis that is good for $\Nk^\nu_{\la,\mu}$? No. Such a basis
must be good for the tensor-product coefficients $T_{\la,\mu}^\nu$, and in the
case of $g=A_2$, such a basis is unique. So no such basis exists for $A_2$; 
neither does one for any $A_r\supset A_2$. 

So, to find the fusion coefficients by this route, the appropriate linear
combinations must be found. They would be useful for the computation of the
OPE coefficients $\h C_{\la,\mu}^\nu$. But for the fusion coefficients only,
this is a bit much. Luckily, there is another way to find the fusion
coefficients, due to Verlinde.  

\subsec{Affine characters and modular transformations}

In order to motivate Verlinde's approach, let us first discuss some results on
affine algebras that will turn out to be important for affine fusion
\ref\kacpet{V.G. Kac, D. Peterson, Adv. Math. {\bf 53} (1984) 125}. Consider the
affine character \affch\   made informal:
\eqn\afich{ch_{\h\la}(\h\sigma)\ :=\ 
\sum_{\h\mu\in P(\h\la)}\, \mult(\h\la; \h\mu)\, e^{(\h\mu,\h\sigma)}\ \ .}
We'll set
\eqn\setsi{\sigma\ =\ -2\pi i\,\big( \zeta:=\sum_{j=1}^r z_j\al^\vee_j, \tau, 0
\big)\ ,} and use \wkcfi\ to bring in the affine Weyl group $\h W$. Then one
finds 
\eqn\thth{\ch_{\h\la}(\h\sigma)\ =\ e^{-2\pi
i\tau\left(h_\la-{{c(g,k)}\over{24}}\right)}\, {{\sum_{w\in W}\,(\det\, w)\,
\Theta_{w(\la+\rho)}^{(k+h^\vee)}(\tau, \pzp) } \over  {\sum_{w\in W}\,(\det\, 
w)\, \Theta_{w\rho}^{(h^\vee)}(\tau, \pzp) }} \ ,}
where $\pzp$ stands for $(z_1,z_2,\ldots,z_r)$, and the {\it theta
functions} are  
\eqn\thfn{\Theta_\la^{(k)}(\tau,\pzp)\ :=\ \sum_{\al\in
Q^\vee}\,  e^{-\pi i\big[ 2(\la+k\al,\zeta) - {\tau\over{k^2}}\,|\la
+k\al|^2 \big]}\ .} Recall also that $h_\la$ and $c(g,k)$ are the conformal
weight of $\phi_\la$, and the Virasoro central charge, respectively (see
\hla,\cwzw). 
It is convenient to define the {\it normalised character} 
\eqn\normch{\chi_{\h\la}(\tau,\pzp)\ :=\ 
e^{2\pi
i\tau\left(h_\la-{{c(g,k)}\over{24}}\right)}\, \ch_{\h\la}(\h\sigma)\ .}
Then we have the simple relation
\eqn\nchth{\boxEq{\chi_{\h\la}(\tau,\pzp)\ =\ {{\sum_{w\in W}\,(\det\, w)\,
\Theta_{w(\la+\rho)}^{(k+h^\vee)}(\tau, \pzp) } \over  {\sum_{w\in W}\,(\det\, 
w)\, \Theta_{w\rho}^{(h^\vee)}(\tau, \pzp) }} \ .}}

This last result is remarkable. First, notice that the sums are both over
the finite Weyl group of the simple Lie algebra $g$. Consequently, there is
a striking resemblance to the Weyl character formula \wcfi\ for $g$. 

One can trace the appearance of the $W$-sums to the semi-direct product
structure of $\h W$, \WhWT. When acting on a shifted weight, such as
$\h\la+\h\rho$, $\h\la\in P_+^k$, we have $\h W\ =\
W\,\ltimes\,T_{(k+h^\vee)Q^\vee}$. The other factor $T_{(k+h^\vee)Q^\vee}$ is
responsible for the presence of the theta functions. These latter are
well-known to have remarkable transformation properties under the group
$PSL(2,\Z)$, the so-called {\it modular group} $\Gamma$. The modular group is
generated by the elements $S\,:\, \tau\rightarrow -1/\tau$ and $T\,:\, \tau
\rightarrow \tau+1$, and the general modular transformation has the form
\eqn\modabcd{\tau\ \rightarrow\ {{a\tau+b}\over{c\tau+d}}\ ,\ \ \ \ 
a,b,c,d\in\Z\ ,\ \ \ \ ad-bc=1\ .} So, this transformation can be encoded in a
$2\times 2$ integer  matrix, of determinant one, or the negative of such a
matrix; thus $\Gamma \cong PSL(2,\Z)$. 

As a consequence of the appearance of the theta functions in \nchth, we find
that the normalised characters transform among themselves under modular
transformations. Specifically, Kac and Peterson showed that 
\eqn\nchST{\boxEq{\eqalign{\chi_{\h\la}(-1/\tau,\{z/\tau\})\ =&\ \sum_{\h\mu\in
P_+^k}\ \Sk_{\la,\mu}\, \chi_{\h\mu}(\tau,\pzp)\ \ ,\cr
\chi_{\h\la}(\tau+1,\pzp)\ =&\ \sum_{\h\mu\in
P_+^k}\ \Tk_{\la,\mu}\, \chi_{\h\mu}(\tau,\pzp)\ \ .\cr
  }}}
Consider the $\card P_+^k\times \card P_+^k$ $=\card\Ppkb\times \card
\Ppkb$ (see \Ppkbar) matrices $\Sk$ and $\Tk$, with elements $\Sk_{\la,\mu}$ and
$\Tk_{\la,\mu}$, respectively; they turn out to be unitary, showing that the
normalised characters form a unitary representation of (a subgroup of) the
modular group $\Gamma$. The  explicit forms of their elements are:
\eqn\KPST{\boxEq{\eqalign{\Sk_{\la,\mu}\ =&\ F(g,k)\, \sum_{w\in W}\, (\det\, 
w)\,  e^{-{{2\pi i}\over{k+h^\vee}}\, \big(\, \la+\rho, w(\mu+\rho) \,\big)}\ \
,\cr
\Tk_{\la,\mu}\ =&\ \delta_{\la,\mu}\,  e^{-{2\pi i}
\big( h_\la-{{c(g,k)}\over{24}} \big)}\ \ ,\cr }}} 
where $F(g,k)$ is a constant independent of $\la,\mu$.     

\subsec{Kac-Peterson relation and Verlinde formula} 

The form of the $\Sk_{\la,\mu}$ recalls the numerator of the Weyl character
formula \wcfi. To recover the full character, rather than just the numerator,
we can take a ratio to find  
\eqn\KPrel{\boxEq{ {{\Sk_{\la,\mu}}\over {\Sk_{0,\mu}}}\ =\ \ch_\la\big(\, 
-2\pi i {{\mu+\rho}\over{k+h^\vee}} \,\big)\ .
}}
We'll call this important result the {\it Kac-Peterson} relation. 

Why should there be such an intimate relation between the modular matrix $\Sk$
of the affine characters and the characters of $g$? An answer is provided by
conformal field theory, and more specifically here, by the WZW model. 

First, recall one of the uses of characters in the representation
theory of $g$. Consider the tensor-product
decomposition $L(\la)\otimes L(\mu) = \sum_{\nu\in P_+} T_{\la,\mu}^\nu
L(\nu)$, and the formal element $e^{H}$ in it. Taking traces gives
\eqn\tpch{\ch_\la \ch_\mu\ =\ \sum_{\nu\in P_+}\, T_{\la,\mu}^\nu \ch_\nu\ .} 
One says that the simple Lie characters obey the tensor-product algebra. 

It is therefore natural to wonder if the {\it Kac-Peterson ratios}
\eqn\KPrat{\chi^{(k)}_\la(\mu)\ :=\ {{\Sk_{\la,\mu}}\over {\Sk_{0,\mu}}}\ }
have interesting multiplicative properties. After all, \KPrel\ says that
they are ``discretised'' simple Lie characters. One finds 
\eqn\KPfus{\boxEq{\chi^{(k)}_\la(\sigma)\, \chi^{(k)}_\mu(\sigma)\ =\ 
\sum_{\nu\in \Ppkb}\, \Nk_{\la,\mu}^\nu\, \chi^{(k)}_\nu(\sigma)\ ,}} 
valid $\forall\sigma\in \Ppkb$. That is,
the Kac-Peterson ratios obey the WZW fusion algebra! 

If we rewrite this in terms of the modular $S$ matrix, $\Sk$, we find 
\eqn\Verl{\boxEq{ \Nk_{\la,\mu}^\nu\ =\ \sum_{\sigma\in \Ppkb}\  
{{\Sk_{\la,\sigma}\, \Sk_{\mu,\sigma}\, \Sk_{\nu,\sigma}^*} \over 
{\Sk_{0,\sigma}}}\ \,
}} 
using the unitarity of $\Sk$. This is the celebrated {\it Verlinde formula}
\ref\ver{E. Verlinde, Nucl. Phys. {\bf B300} (1988) 360}. It is valid for all
RCFT's, when the corresponding modular $S$ matrices are used.

\subsec{Duality} 

The Verlinde formula, in the form \KPfus, provides a rationale for the
Kac-Peterson relation \KPrel. As the depth rule makes clear, the WZW
fusion products  should be truncated versions of the $g$ tensor products,
because the horizontal subalgebra $g\subset \h g$ is a true symmetry of the
theory. It is perhaps not too surprising then that the quantities $\{
\chi^{(k)}_\la(\sigma)\ :\ \la\in \Ppkb \}$ that obey the fusion rule algebra
\KPfus\ turn out to be discretised characters of $g$. Roughly, that they are
characters means that the fusion coefficients are intimately related to the
corresponding tensor-product coefficients. Their ``discretisation'' is a
consequence of constraints coming from the spectrum-generating algebra $\h g
\supset g$. It will imply that the fusion coefficients are bounded above by the
tensor-product coefficients. 

But that does not explain the Verlinde formula in any way: why should ratios
of modular $S$ matrix elements have anything to do with fusions, let alone
represent the fusion algebra? 

The answer arises from the powerful concept of {\it duality} in conformal field
theory \ref\moosei{G. Moore, N. Seiberg, Phys. Lett. {\bf 212B} (1988) 451}.
Consider an arbitrary correlation function in a conformal field theory, not
necessarily a WZW model. One finds that such a correlation function factorises
into holomorphic and antiholomorphic parts: \eqn\hahfact{\eqalign{{\cal
C}(\, z_1,\ldots,z_n,\tau_1,\ldots,\tau_m\,;\, 
\zb_1,\ldots,\zb_n,\bar\tau_1,\ldots,\bar\tau_m \,)\ =&\ \cr
\sum_{I,\bar J}\, {\cal C}_{I,\bar J}\,
{\cal B}_I(\, z_1,\ldots,z_n,\tau_1,\ldots,\tau_m \,)\, 
{\bar{\cal B}}_{\bar J}&(\, \zb_1,\ldots,\zb_n, \bar\tau_1,\ldots, 
\bar\tau_m \,)\ \ .\cr}} 
Here $z_1,\ldots, z_n$ are meant to indicate the positions of the $n$ fields
(points) of the correlation function, and the $\tau_j$ are constants
(sometimes moduli) set by the type of correlation function under
consideration. For the general class of conformal field theories known as {\it
rational conformal field theories (RCFT's)}, which includes the WZW models, the
sums over $I, \bar J$ are (discrete and) finite. The functions ${\cal B}_I$,
$\bar{\cal B}_{\bar J}$ are known as {\it conformal blocks}. 

The factorisation \hahfact\ is not unique, however. As was alluded to earlier,
the conformal blocks can (in principle) be calculated using the operator
product algebra. If we symbolise a non-zero operator product coefficient $\h
C_{\la,\mu}^\nu$ by the graph of Figure 20,  we are associating
that graph to the conformal block of a 3-point function. In a similar way,
we can label a choice of conformal blocks by a trivalent graph. Part of the
non-uniquenes of the conformal blocks comes from the non-uniqueness of the
trivalent graph as label. 

\topinsert
\vskip0cm
\epsfxsize=6cm
\centerline{\epsfbox{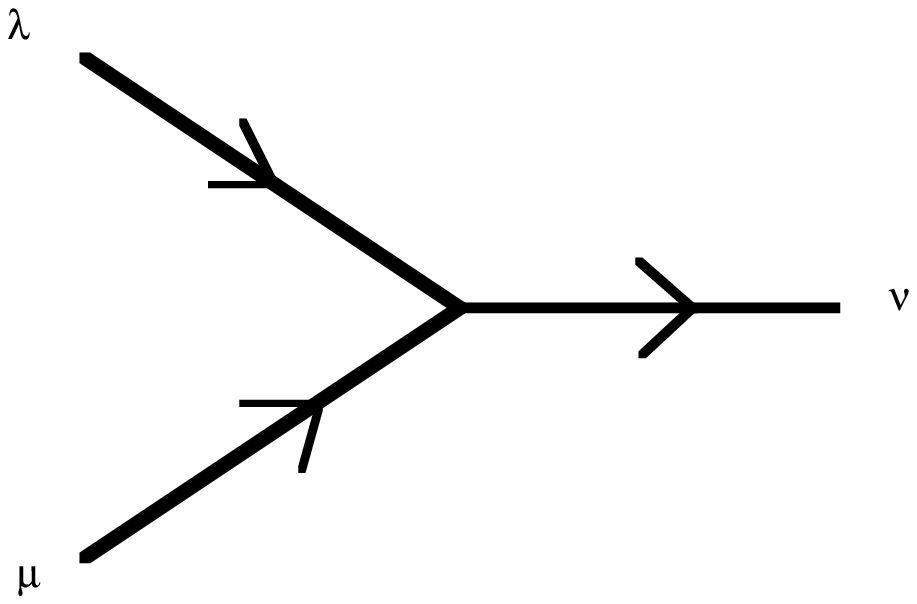}}
\smallskip\smallskip\bigskip
\leftskip=1cm
\rightskip=1cm
\noindent
\baselineskip=12pt
{\it Figure 20}. The graph labelling a conformal block for the 3-point
function. \bigskip \leftskip=0cm\rightskip=0cm
\baselineskip=15pt  
\endinsert 

For example, consider a 4-point function. Its conformal blocks can be
labelled by either of the two trivalent graphs of Fig. 21. The
underlying assumption of duality is that the final physical correlation
function must not depend on the choice of graph. Furthermore, duality states
that there must be a linear relation between the conformal blocks
associated with the two graphs, such that this is true. The matrix encoding
the particular linear relation shown in the Figure, is called the fusing
matrix $F$. 

\midinsert
\vskip1cm
\epsfxsize=10cm
\centerline{\epsfbox{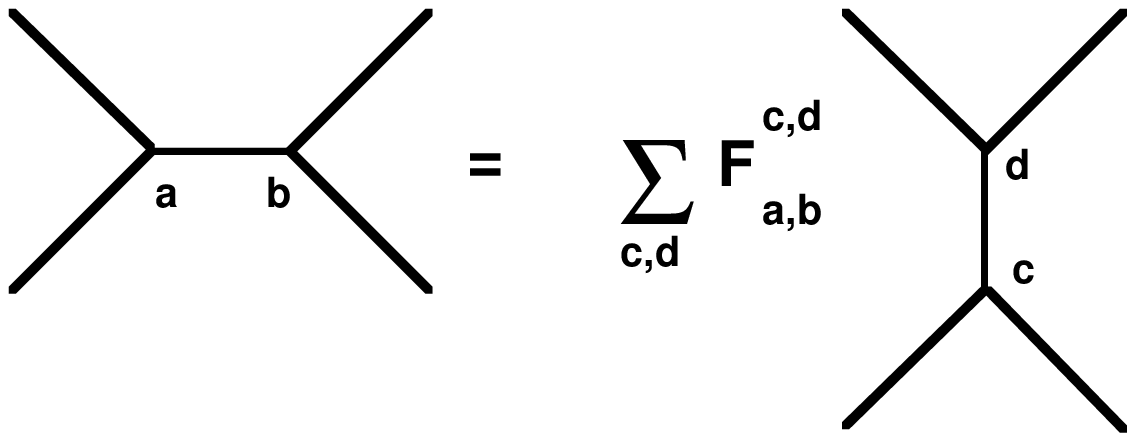}}
\smallskip\smallskip
\leftskip=1cm
\rightskip=1cm
\noindent
\baselineskip=12pt
{\it Figure 21}. Duality for 4-point functions.\bigskip
\leftskip=0cm\rightskip=0cm
\baselineskip=15pt  
\endinsert 

There is a another, related way of thinking of these trivalent graphs. When
radial quantisation was mentioned earlier, the conformal transformation
\cyl\ from the cylinder to the complex plane was discussed. One can think of
the resulting plane as having two special, ``marked'' points, at the origin
and at $\infty$. The inverse conformal transformation maps the plane with
these two marked points (or punctures) to a Riemann surface, the cylinder.
Letting the radius shrink to zero gives a particularly simple trivalent graph, a
straight line. 

If one considers a 4-point function, however, one has four marked points, and
a conformal transformation can be found to map the plane to a Riemann surface.
The Riemann surface will be topologically equivalent to a Riemann sphere with
four marked points. One can then recover a trivalent graph by shrinking the
sphere in different ways. In particular, one can recover the two different
graphs of Fig. 21. 

With this latter picture, duality can tell us why the modular group $\Gamma$
enters consideration. The modular group is intimately connected with Riemann
surfaces (with no marked points) of genus one, \ie\ with tori. Represent a
torus by a parallelogram with opposite sides identified, as in Fig. 22.  We
will consider a conformal field theory on such a torus. By conformal
invariance, the overall scale doesn't matter, so we set the sides of the
paralellogram as shown in the Figure. The conformal class of the torus can thus
be specified by one complex number, its modulus $\tau$. But this is still a
redundant description. It is simple to see that $T\, :\, \tau \rightarrow
\tau+1$ does not change the underlying torus. Furthermore, after a rescaling,
$S\, :\, \tau \rightarrow -1/\tau$ doesn't either. So the conformal class of
the torus is invariant under the full modular group $\Gamma = \< S,T \>$. 

\topinsert
\vskip0cm
\epsfxsize=6cm
\centerline{\epsfbox{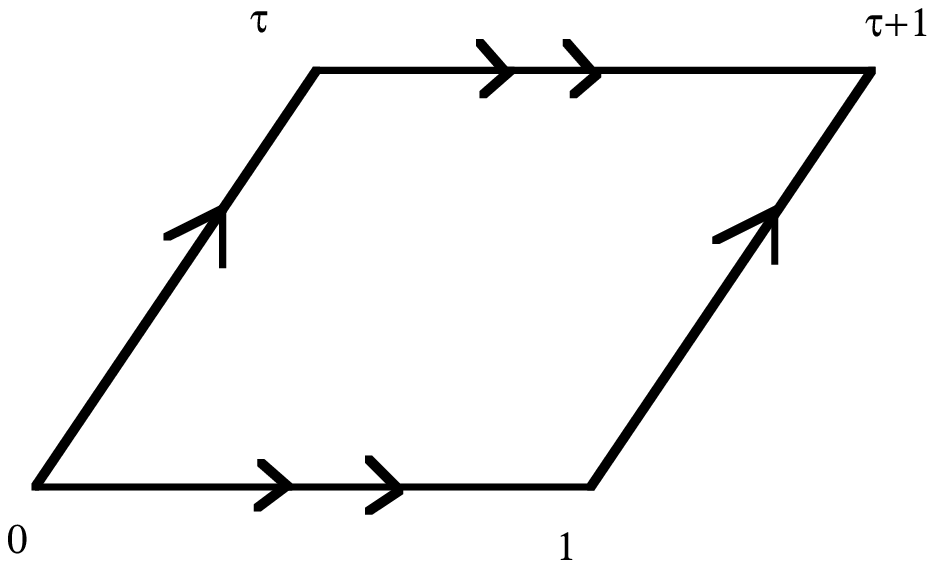}}
\smallskip\smallskip
\leftskip=1cm
\rightskip=1cm
\noindent
\baselineskip=12pt
{\it Figure 22}. Torus, as parallelogram.\bigskip
\leftskip=0cm\rightskip=0cm
\baselineskip=15pt  
\endinsert

Now consider the ``correlation function'' 
\eqn\partfn{\eqalign{ Z(\tau)\ =&\ e^{2\pi i({c\over{24}}\tau-{{\bar
c}\over{24}}\bar\tau)}\,  \Tr_{\cal H}\, 
e^{-2\pi i\big(\, {\bf H} (i{\rm Im}\,\tau) + {\bf P} ({\rm Re}\,\tau) \,\big)}\
\cr =&\ \Tr_{\cal H}\, 
e^{-2\pi i\big(\,(L_0-{c\over{24}})\tau + (\bar L_0-{\bar 
c\over{24}})\bar\tau \,\big)}\ ,\cr
}}
where $\tau$ is the complex conjugate of $\tau$, $c$ and $\bar c$ are the
holomorphic and antiholomorphic central charges, and  ${\cal H}$ is the
Hilbert space of the theory. We have used that the Hamiltonian and rotation
operators are ${\bf H}=L_0+\bar L_0$ and ${\bf P}=L_0-\bar L_0$, respectively
(see after \LzLzb). \partfn\ is known as the torus partition function. 

Let us restrict once more to consideration of WZW models; what we'll say,
however, can also be adapted straightforwardly to all RCFT's. Now, by the
holomorphic-antiholomorphic factorisation of WZW models, the Hilbert space has
the form \eqn\Hilb{ {\cal H} =\ \oplus_{\la,\mu \in \Ppkb}\,
M_{\la,\mu}\, {\cal H}_{\la} \otimes {\cal H}_{\mu}\ ,}
with $M_{\la,\mu}\in\Z_{\geq 0}$. Here ${\cal H}_\la$ ($\la\in \Ppkb$) is the
Hilbert space of states in the conformal tower of $L(\h\la)$ ($\h\la\in
P_+^k$). This factorisation is manifested in the partition function:
\eqn\pfnch{Z(\tau,\bar\tau)\ =\ \sum_{\la,\mu\in \Ppkb}\, 
M_{\la,\mu}\   \chi_{\h\la}(\tau,\{0\})\,  \chi^*_{\h\mu}(\bar\tau,\{0\})\ .}
That is, the conformal blocks for the torus partition function are the
normalised characters. The corresponding trivalent graph is a loop, but the
blocks are also labelled by the torus modulus $\tau$. But $\tau$, $\tau+1$,
$-1/\tau$, etc., are all different ways of labelling the same torus. So duality
implies the modular covariance \nchST\ of the normalised characters. 

Summarising to this point: in a RCFT, (normalised) characters appear naturally 
as conformal blocks for the torus partition function, and duality implies that
they must be modular covariant. 

\midinsert
\vskip1cm
\epsfxsize=8cm
\centerline{\epsfbox{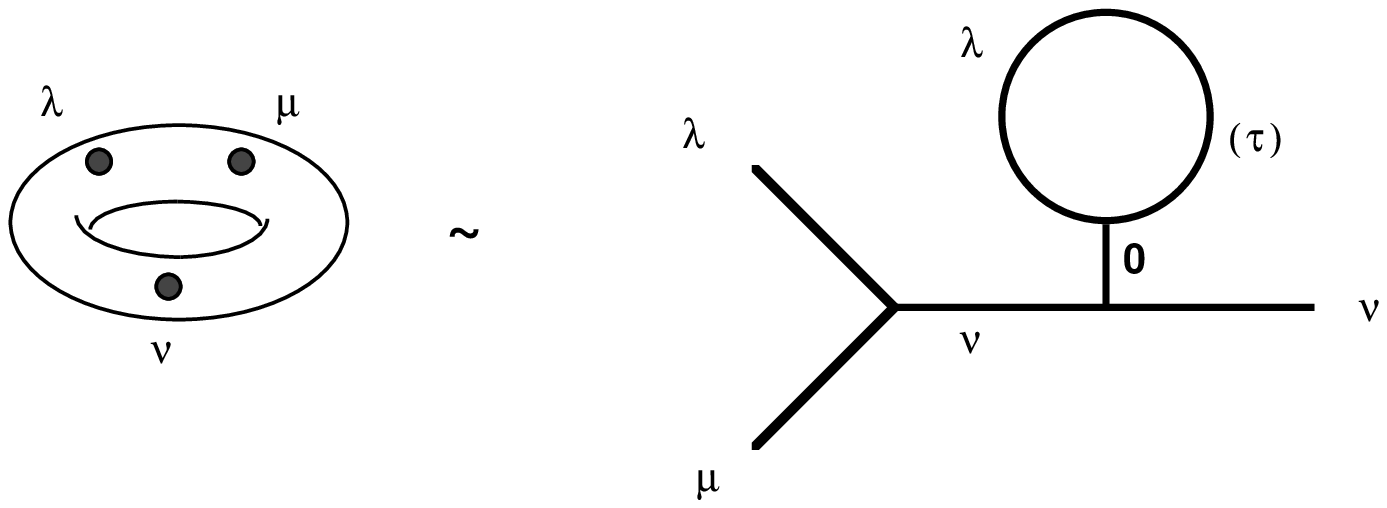}}
\smallskip\smallskip
\leftskip=1cm
\rightskip=1cm
\noindent
\baselineskip=12pt
{\it Figure 23}. A depiction of a 3-point function on a torus, and one
possible conformal block for it. \bigskip \leftskip=0cm\rightskip=0cm
\baselineskip=15pt  
\endinsert

But this is just one simple consequence of duality; much more can be extracted,
including the Verlinde formula. To derive it from duality,
one needs to consider a correlation function that can involve both the
3-point functions (and so the fusion coefficients) and the torus (so that
the modular transformations are involved).  It is  the 3-point functions on
the torus that are pertinent in this context. Fig. 23 shows a
torus with three marked points, and a choice of a trivalent graph to label the
corresponding conformal block. That choice makes it clear how the 3-point
functions and the characters appear. But there are many other choices of
graphs, and one can also replace $\tau$ with $-1/\tau$, for example. These
freedoms are not completely independent, however. They turn out to be sufficient
to prove the Verlinde formula, but we will not provide the detailed argument 
here \moosei. 

Recapping, we have seen that the remarkable modular properties of
affine characters are accounted for in the physical context of  WZW
models. The underlying concept is duality, a property that extends to all
RCFT's. It also implies many other
important relations. The Verlinde formula is just one symptom of duality in
conformal field theory.

\subsec{Fusion coefficients as Weyl sums} 

To close this section, let us return to the problem of computing WZW fusion 
coefficients, and apply the Verlinde formula \Verl. The Kac-Peterson
relation \KPrel\ in \tpch\ implies
\eqn\Kptp{\chi^{(k)}_\la(\sigma)\, \chi^{(k)}_\mu(\sigma)\ =\ 
\sum_{\vp\in P_+}\, T_{\la,\mu}^\vp\, \chi^{(k)}_\vp(\sigma)\ ,}
where we have used the notation \KPrat. Compare this with \KPfus. Both are
valid $\forall\sigma\in P_+^k$. \KPfus\  led to  \Verl\ by the unitarity
of the matrix $\Sk$. A relation between the tensor-product coefficients
$T_{\la,\mu}^\vp$ and the fusion coefficients $\Nk_{\la,\mu}^\nu$ should
result from the same unitarity, except that the  ranges are not
identical. The sum over $P_+$ in \Kptp\ must be restricted to a sum 
over $\Ppkb$, as in \KPfus. 

To do so, we make use of the alternating Weyl symmetry \altw, as applied to
the Kac-Peterson ratios: 
\eqn\altwKP{\chi^{(k)}_{w.\vp}(\sigma)\ =\ (\det\, w)\, \chi^{(k)}_\vp(\sigma)\
, \ \  \forall\ w\in W\ .}
This is not sufficient, however. To restrict the sum over $\vp\in P_+$
to a sum over $\nu\in \Ppkb$, we need to use elements of the affine Weyl
group $\h W$, not just $W\subset\h W$ (see Fig. 10, for example).
More accurately, we need the action of the affine Weyl group projected onto
the weight space of the horizontal subalgebra $g\subset \h g$. 

Consider the affine primitive reflection $\h r_0$, with action on affine
weights given in \rhla. Just as $\la$ is used to denote the horizontal part
of the affine weight $\h \la$, $r_0.\la$ will indicate the horizontal part of
$\h r_0.\h \la$. One finds
\eqn\rzrth{r_0.\la\ =\ r_\theta.\la\ +\ (k+h^\vee)\theta\ ,}
where $r_\theta\in W$ is the Weyl reflection across the hyperplane normal
to the highest root $\theta$. Then 
\eqn\rzrthe{e^{-{{2\pi i}\over{k+h^\vee}}(r_0.\la,\sigma+\rho)}\ =\ 
e^{-{{2\pi i}\over{k+h^\vee}}(r_\theta.\la,\sigma+\rho)}\ ,}
for $\sigma\in P_+$, since then $(\theta,\sigma+\rho)\in \Z$. Using this and
the Weyl character formula \wcfi, we find
\eqn\altrz{\chi^{(k)}_{r_0.\vp}(\sigma)\ =\ (\det\, r_\theta)\,
\chi^{(k)}_\vp(\sigma)\ 
=\ (\det\, r_0)\,
\chi^{(k)}_\vp(\sigma)\ .}
Since the affine Weyl group $\h W$ can be obtained from the Weyl group $W$ of
$g$ by adjoining $\h r_0$ as a generator, we have
\eqn\altwKP{\boxEq{\chi^{(k)}_{w.\vp}(\sigma)\ =\ (\det\, w)\,
\chi^{(k)}_\vp(\sigma)\ , \ \  \forall\ w\in \h W\ .}}

Using this in \Kptp, and comparing with \KPfus, we get
\eqn\TsumN{\sum_{\vp\in \Ppkb}\, \sum_{w\in\h W}\, (\det\, w)\,
T_{\la,\mu}^\vp\, \chi^{(k)}_{\vp}(\sigma)\ =\ 
\sum_{\nu\in \Ppkb}\, \Nk_{\la,\mu}^\nu\, \chi^{(k)}_{\nu}(\sigma)\ ,}
for all $\sigma\in \Ppkb$. The unitarity of $\Sk$ means that the coefficients
of $\chi^{(k)}_{\nu}(\sigma)$ on the left hand side of \TsumN\ can be equated
with those on the right hand side, for all $\nu\in \Ppkb$. Therefore we find 
\Kac\ref\MW{M.A. Walton, Phys.
Lett. {\bf 241B} (1990) 365; 
Nucl. Phys. {\bf B340} (1990) 777}\ref\FGP{P. Furlan, A. Ganchev,
V. Petkova, Nucl.\ Phys.\
{\bf B343} (1990) 205}
\eqn\NeT{\boxEq{\Nk_{\la,\mu}^\nu\ =\ \sum_{w\in \h W}\, (\det\,w)\,
T_{\la,\mu}^{w.\nu}\ \ .}}
The dependence of the right-hand side on the
level $k$ is implicit: for a fixed $\nu\in P_+$, $w.\nu$ can change with
changing level, when $w\in \h W$ (see \rzrth).

This last equation provides a fairly simple way of computing the fusion
coefficients. For example, one can employ the \LR\ rule (or Littelmann's
generalisation) to first find the tensor-product coefficients
$T_{\la,\mu}^\nu$, and then perform the alternating Weyl sum of \NeT.
However, since it is an alternating sum (the signs are inherited from the Weyl
character formula) cancellations occur, and the rule is not a combinatorial
one.  

Another expression can be given for the fusion coefficients as an alternating
affine Weyl sum. This one will allow a connection with the refined depth rule
\rdr. Using \fchi\ brings in the weight multiplicities, and in \tpch\ it
leads to
\eqn\Temult{T_{\la,\mu}^\nu\ =\ \sum_{w\in W}\, (\det\,w)\, 
\mult(\mu; w.\nu-\la)\ .}
In \NeT, this results in
\eqn\Nemult{\boxEq{\Nk_{\la,\mu}^\nu\ =\ \sum_{w\in \h W}\, (\det\,w)\, 
\mult(\mu; w.\nu-\la)\ .}}

This formula encodes a straightforward algorithm for the computation of the
fusion coefficients. First, the weight $\la+\rho$ is added to all the
weights of $P(\mu)$. the resulting weights are regarded as horizontal
projections of affine weights, at level $k+h^\vee$. One then attempts to Weyl
transform the resulting weights into the dominant ($w=\id$) sector, using the
horizontal projection of the affine Weyl group.  Some of the
weights will be fixed by an affine Weyl reflection; that is, they can only lie
on the boundary of the dominant sector after Weyl transformation. These should
be ignored. The others will be transformed into the dominant sector, some by
elements of $\h W$ of determinant +1, and some by elements of determinant -1.
The latter will always cancel some other dominant weights of the first kind.
The final result will be a set of dominant weights $\{ \nu+\rho\}$ with
multiplicities $\Nk_{\la,\mu}^\nu$, the quantities to be calculated. 

The algorithm can be pictured in a weight diagram, for ranks $\leq 2$. A simple
example is drawn in Fig. 24. Recall now the refined depth rule \rdr.
The conditions $(E^{-\al_i})^{1+\nu_i}u=0$ can be interpreted as saying that
the states $u$ that count toward $\Nk_{\la,\mu}^\nu$ are those for which the 
``cancelling states'' $(E^{-\al_i})^{1+\nu_i}u$ do not exist. If one does,
then the alternating Weyl sum formula \Nemult\ tells us that the cancelling
state's weight will be reflected into the dominant sector, to cancel the
weight $\nu$, in the algorithm of the previous paragraph. 

\midinsert
\vskip1cm
\epsfxsize=8cm
\centerline{\epsfbox{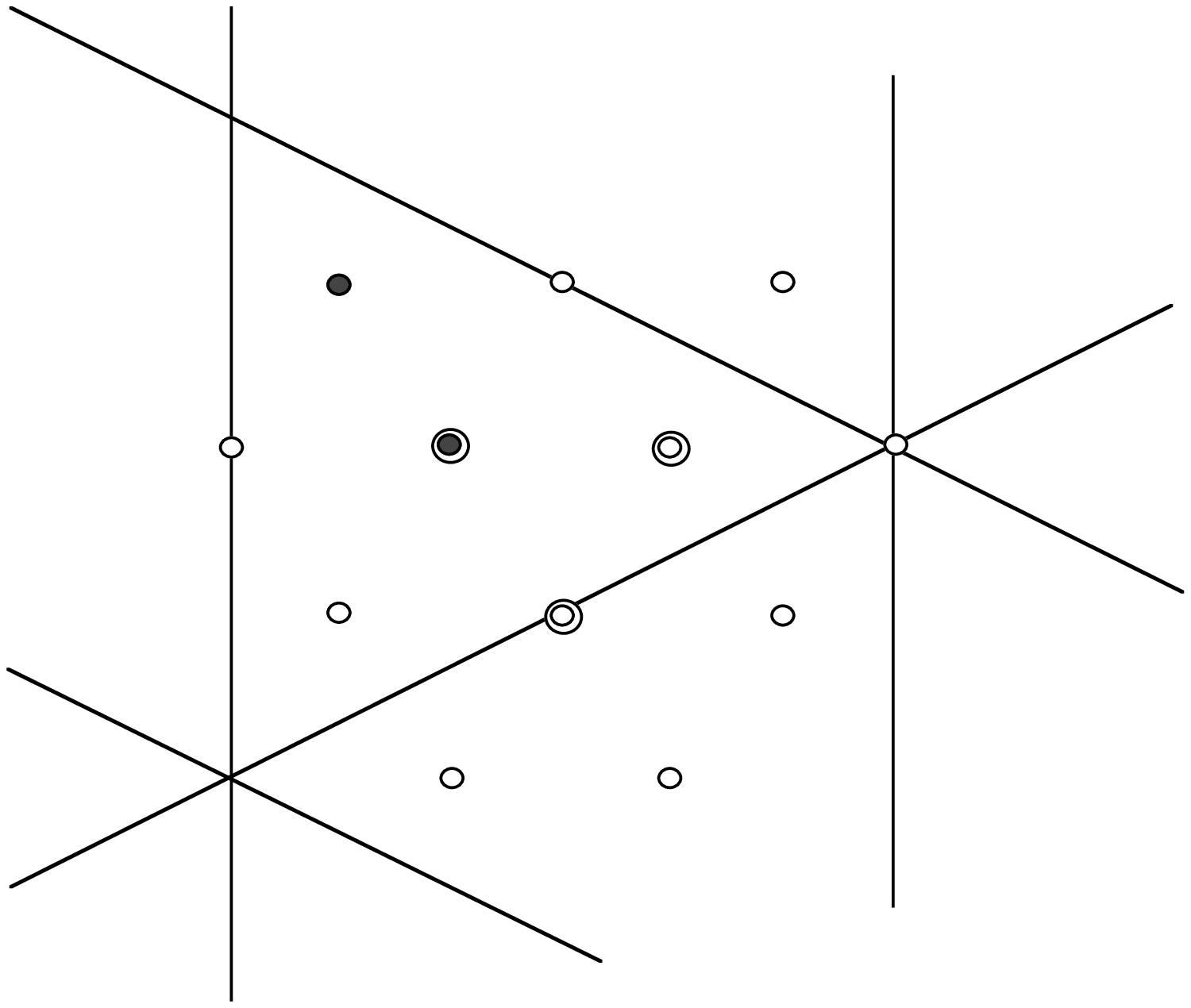}}
\smallskip\smallskip
\leftskip=1cm
\rightskip=1cm
\noindent
\baselineskip=12pt
{\it Figure 24}. Weight diagram illustrating the computation by \Nemult\ of the
$A_2$ fusion $L(2,0)\otimes_3 L(2,1) = L(0,3) \oplus L(1,1)$. Darkened 
circles indicate weights that correspond to the representations $L(0,3)$ and
$L(1,1)$; the others are cancelled. \bigskip \leftskip=0cm\rightskip=0cm
\baselineskip=15pt   \endinsert 

This concludes our discussion of the computation of fusion coefficients, as an
example of an application of the representation theory of affine algebras to 
WZW models.

\newsec{Conclusion}

We hope these lecture notes have given some indication of the beauty of the
subject of conformal field theory, and the associated infinite-dimensional
algebras. 

We'll close with a brief (and incomplete) guide to the review literature on
conformal field theories, affine algebras, and their relation. 
The most comprehensive reference to date is \ref\DMS{P. Di Francesco,
P. Mathieu, D. S\'en\'echal, {\it Conformal Field Theory} (Springer-Verlag,
1996)}. Another monograph is \ref\ket{S.V. Ketov, {\it Conformal Field Theory}
(World Scientific, 1995)}. 

For the beginner in conformal field theory, the reviews \ref\zub{J.-B. Zuber,
Acta Phys. Polon. {\bf B26} (1995) 1785} \ref\gin{P. Ginsparg, in Les Houches
session XILX, Fields, Strings, and Critical Phenomena, eds. E. Br\'ezin, J.
Zinn-Justin (Elsevier, 1989)} \ref\schell{A.N. Schellekens,
Fortschr. Phys. {\bf 44} (1996) 605}  and \ref\sai{Y. Saint-Aubin,  Univ. de
Montr\'eal preprint, CRM-1472 (1987)} (in French) are quite ``user-friendly''.
So is Cardy's review \car, which emphasises the statistical mechanics
applications of conformal field theory. For more applications, see 
\aff. 

For a review of rational conformal field theory, especially its 
duality, see \msrev.

The ultimate reference for affine algebras is \Kac. \kmps\ is somewhat more
accessible, however, but less comprehensive. Also accessible are 
\godoli\ and  \fuchs. \DMS\ reviews the basic facts of affine algebras required
in the study of conformal field theory. 

For another treatment of  affine
algebras in conformal field theory, see \ref\fuchsr{J. Fuchs, lectures  
at the Graduate Course on Conformal Field Theory and Integrable Models
(Budapest, August 1996), to appear in Springer Lecture Notes in Physics}.
\gaw\ is an elegant exposition of functional integral methods applied to the
WZW model. 

\vskip1cm
\noindent {\bf Acknowledgements}

It is a pleasure to thank the members of the Feza Gursey Institute for their
warm hospitality. Thanks also go to the other lecturers and the
participants in the summer school for enjoyable conversations. I am also
grateful to Terry Gannon and Pierre Mathieu  for helpful readings of  the
manuscript. Gannon gets additional thanks  
for encouragement, \ie\ for gloating when he finished his lecture notes first. 

The author acknowledges the support of a research grant from NSERC of Canada.

\listrefs 

\bye